\documentclass[10pt,journal,compsoc]{IEEEtran}

\usepackage{booktabs}
\usepackage{moreverb}
\usepackage{fontenc}
\usepackage{amsmath}
\usepackage{amssymb}
\usepackage{amsthm}
\usepackage{fancybox}
\usepackage{color}
\usepackage{colortbl}
\usepackage{array}
\usepackage{multirow}
\usepackage{multicol}
\usepackage{listings}
\usepackage{graphicx}
\usepackage{makecell}
\usepackage{epigraph}
\usepackage[multiple]{footmisc}

\usepackage{ragged2e}

\usepackage{boxedminipage}

\ifCLASSOPTIONcompsoc
  \usepackage[nocompress]{cite}
\else
  \usepackage{cite}
\fi

\ifCLASSINFOpdf
\else
\fi
\usepackage[ruled,vlined,lined,commentsnumbered]{algorithm2e}
\usepackage{balance}
\usepackage{csvsimple}
\usepackage{longtable}
\usepackage[skip=1pt,font=small,labelfont=bf,textfont=bf]{caption}
\usepackage{url}
\usepackage{lscape}
\usepackage{rotating}
\usepackage{tikz}
\usepackage[normalem]{ulem}
\usepackage{enumitem}

\usepackage[most]{tcolorbox}
\usepackage{threeparttable}

\usepackage{marvosym}
\usepackage{textcomp}

\usepackage{tikz}
\newcommand*\circled[1]{\tikz[baseline=(char.base)]{
            \node[shape=circle,draw,inner sep=0.5pt] (char) {#1};}}

\setlist{noitemsep} %to leave space around whole list

\definecolor{codegreen}{rgb}{0,0.6,0}
\definecolor{codegray}{rgb}{0.5,0.5,0.5}
\definecolor{codepurple}{rgb}{0.58,0,0.82}
\definecolor{backcolour}{rgb}{0.95,0.95,0.92}
 
\definecolor{DarkOrange}{rgb}{0.8,0.3,0.0} 
\definecolor{yellow}{RGB}{255,255,153}
\definecolor{grey}{RGB}{224,224,224}

\newboolean{showcomments}
\setboolean{showcomments}{true}
\ifthenelse{\boolean{showcomments}}
 { \newcommand{\mynote}[2]{
      \fbox{\bfseries\sffamily\scriptsize#1}
        {\small$\blacktriangleright$\textsf{\emph{#2}}$\blacktriangleleft$}}}
        { \newcommand{\mynote}[2]{}}

\theoremstyle{definition}
\newtheorem{definition}{Definition}

\newcommand{\nallproject}{2014\xspace}
\newcommand{\nmavenproject}{730\xspace}
\newcommand{\nviolationtype}{400\xspace}

\newcommand{\nalarms}{16,918,530\xspace}
\newcommand{\nfixedalarms}{88,927\xspace}

\newcommand{\mincommits}{500\xspace}

\newcommand{\findbugs}{{\ttfamily FindBugs}\xspace}
\newcommand{\github}{{\ttfamily \url{Github.com}}\xspace}
\newcommand{\gumtree}{{\ttfamily GumTree}\xspace}

\newcommand{\find}[1]{
\begin{tcolorbox}[tile,size=fbox,boxsep=3mm,boxrule=0pt,top=0pt,bottom=0pt,
borderline west={1.5mm}{0pt}{blue!50!white},colback=blue!5!white]
#1
\end{tcolorbox}

}

\begin{document}
\title{Mining Fix Patterns for FindBugs Violations}

\author{Kui Liu,
        Dongsun Kim,
        Tegawend\'e F. Bissyand\'e, Shin Yoo,
        and~Yves Le Traon% <-this % stops a space
    \IEEEcompsocitemizethanks{%
    \IEEEcompsocthanksitem Kiu Liu, Dongsun Kim, Tegawend\'e F. Bissyand\'e,
        and~Yves Le Traon are with the Interdisciplinary Centre for Security, Reliability and Trust (SnT) at University of Luxembourg, Luxembourg. \protect\\
    E-mail: \{kui.liu, dongsun.kim, tegawende.bissyande, yves.letraon\}@uni .lu
    \IEEEcompsocthanksitem Shin Yoo is with the School of Computing, KAIST, Daejeon, Republic of Korea. \protect\\
    E-mail: shin.yoo@kaist.ac.kr
    }% <-this % stops an unwanted space
% \thanks{Manuscript received April 19, 2005; revised August 26, 2015.}
}

% \markboth{Journal of \LaTeX\ Class Files,~Vol.~14, No.~8, August~2015}%
% {Shell \MakeLowercase{\textit{et al.}}: Bare Demo of IEEEtran.cls for Computer Society Journals}

%\renewcommand{\shortauthors}{}

\IEEEtitleabstractindextext{%
\begin{abstract}
    \justifying
    Several static analysis tools, such as Splint or FindBugs, have been proposed
    to the software development community to help detect security vulnerabilities or bad programming practices. However, 
    the adoption of these tools is hindered by their high false positive rates. If the false positive rate is too high, developers may get acclimated to violation reports from these tools, causing concrete and severe bugs being overlooked.
    Fortunately, some violations are actually addressed and resolved by developers. 
    We claim that those violations that are recurrently fixed are likely to be true positives, and an automated approach can learn to repair similar unseen violations.
    However, there is lack of a systematic way to investigate the distributions on existing violations and fixed ones in the wild, that can provide insights into prioritizing violations for developers, and an effective way to mine code and fix patterns which can help developers easily understand the reasons of leading violations and how to fix them.
    
    In this paper, we first collect and track a large number of fixed and unfixed violations across revisions of software. 
    The empirical analyses reveal that there are discrepancies in the distributions of violations that are detected and those that are fixed, in terms of occurrences, spread and categories, which can provide insights into prioritizing violations. 
    To automatically identify patterns in violations and their fixes, we propose an approach that utilizes convolutional neural networks to learn features and clustering to regroup similar instances. We then evaluate the usefulness of the identified fix patterns by applying them to unfixed violations. 
    The results show that developers will accept and merge a majority (69/116) of fixes generated from the inferred fix patterns. It is also noteworthy that the yielded patterns are applicable to four real bugs in the Defects4J major benchmark for software testing and automated repair.

\end{abstract}

}

\maketitle

\IEEEdisplaynontitleabstractindextext

% \fi
\IEEEpeerreviewmaketitle

%\IEEEpubidadjcol


% \epigraph{``The real power of patterns is not to hand us exotic solutions, but to give us a way to remember the simple, ordinary, basic solutions that we know but forget.''}{ Linda Rising in \underline{The Benefits of Patterns}, IEEE Software 2010 (issue 5)}

\section{Introduction}
\label{sec:introduction}

% SA tools and their adoption
Modern software projects widely use static code analysis tools to assess software quality and identify potential defects. 
Several commercial~\cite{coverity,objectLab,semmle} and open-source%{\color{red}~\cite{coccinelle}}~\dongsun{coccinelle is a bit far from static analysis.}
~\cite{hovemeyer_finding_2004,pmd,splint,cppCheck} tools are integrated into many software projects, including operating system development projects~\cite{koyuncu_impact_2017}. For example, Java-based projects often adopt \findbugs~\cite{hovemeyer_finding_2004} or {\tt PMD}~\cite{pmd} while C projects use {\tt Splint}~\cite{splint}, {\tt cppcheck}~\cite{cppCheck}, or {\tt Clang Static Analyzer}~\cite{clang_static_analyzer}, while Linux driver code are systematically assessed with a battery of static analyzers such as Sparse and the LDV toolkit.
Developers may benefit from the tools before running a program in real environments even though those tools do not guarantee that all identified defects are real bugs~\cite{bessey_few_2010}.

% example of static analysis violation
Static analysis can detect several types of defects such as security vulnerabilities, performance issues, and bad programming practices (so-called code smells)~\cite{engler_bugs_2001}. Recent studies denote those defects as {\bf static analysis violations}~\cite{venkatasubramanyam_automated_2014} or {\bf alerts}~\cite{heckman_model_2009}. In the remainder of this paper, we simply refer to them as {\em violations}.
Fig.~\ref{fig:violationExample} shows a violation instance, detected by \findbugs, % within Commit \texttt{bdf3fe} in \texttt{nbm-maven-plugin} project~\footnote{\url{https://github.com/mojohaus/nbm-maven-plugin}},
which is a violation tagged \texttt{\footnotesize BC\_EQUALS\_METHOD\_SHOULD\_WORK\_FOR\_ALL\_OBJECTS}, 
as it does not comply with the programming rule that the implementation of method \texttt{equals(Object obj)} should not make any assumption about the type of its \texttt{obj} argument~\cite{findbugs:description}. 

\begin{figure}[!h]
    % \centering
\begin{lstlisting}[linewidth={\linewidth},frame=tb,basicstyle=\footnotesize\ttfamily]
    public boolean equals(Object obj) {
        // (@{\bf Violation Type:}@)
        // BC_EQUALS_METHOD_SHOULD_WORK_FOR_ALL_OBJECTS
(@{\color{red}~~~~~~return getModule().equals(}@)
(@{\color{red}~~~~~~~~~~~((ModuleWrapper) obj).getModule());}@)
    }
\end{lstlisting}
    \caption{Example of a detected violation, taken from 
    {\em PopulateRepositoryMojo.java} file
    at revision {\tt bdf3fe} in
    project \protect{\tt nbm-maven-plugin}\protect\footnotemark.}
    \label{fig:violationExample}
    \vspace{-0.5cm}
\end{figure}
~\footnotetext{\url{https://github.com/mojohaus/nbm-maven-plugin}}

As later addressed by developers via a patch represented in Fig.~\ref{fig:expfixchange}, the method should return {\tt false} if \texttt{obj} is not of the same type as the object being compared. In this case, when the type of  \texttt{obj} argument is not the type of \texttt{ModuleWrapper}, a\texttt{ java.lang.ClassCastException} should be thrown.

\begin{figure}[!h]
    % \centering
    \vspace{0.2cm}
\begin{lstlisting}[linewidth={\linewidth},frame=tb,basicstyle=\footnotesize\ttfamily]
    public boolean equals(Object obj) {
(@{\color{red}-~~~~~return getModule().equals(}@)
(@{\color{red}-~~~~~~~~~~((ModuleWrapper) obj).getModule());}@)
(@{\color{codegreen}+~~~~~return obj instanceof ModuleWrapper \&\& }@)
(@{\color{codegreen}+~~~~~~~~~~getModule().equals(}@)
(@{\color{codegreen}+~~~~~~~~~~((ModuleWrapper) obj).getModule());}@)
    }
\end{lstlisting}
    \caption{Example of fixing violation, taken from Commit {\tt 0fd11c} of project {\tt nbm-maven-plugin}.}
    \label{fig:expfixchange}
\end{figure}

Despite wide adoption and popularity of static analysis tools (e.g., FindBugs has more than 270K downloads\footnote{\url{http://findbugs.sourceforge.net/users.html}}), 
accepting the results of the tools is not yet guaranteed. Violations identified by static analysis tools are often ignored by developers~\cite{ayewah_evaluating_2007}, since static analysis tools may yield high rates of \textbf{false positives}. 
Actually, a (false positive) violation might be (1) not a serious enough concern to fix, (2) less likely to occur in a runtime environment, or (3) just incorrectly identified due to the limitations of the tool. 
Depending on the context, developers may simply give up on the use of static analysis tools or they may try to prioritize violations based on their own criteria.

Nevertheless, we can regard a violation as {\bf true positive} if it is recurrently removed by developers through source code changes as in the example of Fig.~\ref{fig:expfixchange}. Otherwise, a violation can be considered as ignored (i.e., not removed during revisions) or disappearing (a file or program entity is removed from a project) instead of being fixed. 
%Figure~\ref{fig:expfixchange} shows a code change for fixing a violation, which is a fixing patch, taken from \texttt{commit 0fd11c}, for the violation in Figure~\ref{fig:violationExample} by adding an {\tt instanceof} check.
We investigate in this study different research questions regarding ({\bf RQ1}) {\em to what extent do violations recur in projects?} ({\bf RQ2}) {\em what types of violations are actually fixed by developers?}(i.e., true positives) ({\bf RQ3}) 
{\em what are the patterns of violations code that are fixed or unfixed by developers?} From this question, we can identify common code patterns of violations that could help better understand static analysis rules.
({\bf RQ4}) 
{\em how are the violations resolved when developers make changes?} 
Based on this question, for each violation type, we can derive fix patterns that may help summarize common violation (or real bug) resolutions and may be applied to fixing similar unfixed violations. 
({\bf RQ5}) {\em can fix patterns help systematize the resolution of similar violations?}
% The first question may provide an insight on what kinds of violations are recurrent in software projects. 
% The second question may highlight which violations are more likely to be {\em true} or {\em false} positive so that developers (or researchers as well) can focus on higher priority violations. 
This question may shed some light on the effectiveness of common fix patterns when applying them to potential defects.

To answer the above questions, we investigate violations and violation fixing changes collected from \nmavenproject open source Java projects. Although the approach is generic to any static bug detection tool, we focus on a single tool, namely \findbugs, applying it to every revision of each project. We thus identify violations in each revision and further enumerate cases where a pair of consecutive revisions involve the resolution of a violation through source code change (i.e., the violation is found in revision $r_1$ and is absent from $r_2$ after a code change can be mapped to the violation location): we refer to such recorded changes as \textit{violation fixing changes}.  
We further conduct empirical analyses on identified violations and fixed violations to investigate their recurrences, their code patterns, etc.
%and answer first two questions. After collecting distinct unfixed violations and fixed violations, we design an approach to identify common code patterns of unfixed violations and fixed ones based on Convolutional Neural Networks (CNNs)~\cite{mou_convolutional_2016} and {\em X-means} clustering algorithm~\cite{pelleg_xmeans_2000}. 
% \sout{Currently, this paper does not justify why CNNs are the best tool for this task.}

After collecting violation fixing changes from a large number of projects using an AST differencing tool~\cite{falleri_fine_2014}, we mine developer fix patterns for static analysis violations. The approach encodes a fixing change into a vector space using Word2Vec~\cite{yoav_word2vec_2014}, extracts discriminating features using Convolutional Neural Networks (CNNs)~\cite{mou_convolutional_2016} and regroups similar changes into a cluster using {\em X-means} clustering algorithm~\cite{pelleg_xmeans_2000}. We then evaluate the suitability of the mined fix patterns by applying them to 1) a subset of unfixed violations in our subjects, to 2) a subset of faults in Defects4J~\cite{just2014defects4j} and to 3) a subset of violations in 10 open source Java projects.
% under their latest versions identified by \findbugs to evaluate the effectiveness of fix patterns.

% research questions

% experiments results.
% \dongsun{actual results will be summarized here.}

Overall, this paper makes the following contributions:
% \dongsun{it is necessary to split contributions and findings from the following list.}
\begin{enumerate}[leftmargin=*]
    \item \textbf{Large-scale dataset of static analysis violations:} we have carefully and systematically tracked static analysis violations across all revisions of a large set of projects. This dataset, which has required substantial effort to build, is available to the community in a labelled format, including the violation fixing change information. 
    \find{We release a dataset of \nalarms unique samples of \findbugs violations across revisions of 730 Java projects, along with \nfixedalarms code changes addressing some of these violations.}
    %our study reports the results of a large-scale collection of static analysis violations. In addition, this paper provides violation tracking results across project revision histories, which helps us figure out fixed/unfixed violations.
    \item {\bf Empirical study on real-world management of \findbugs' violations:} our study explores the nature of violations that are widespread across projects and contrasts the recurrence of developer (non)fixes for specific categories, providing insights for prioritization research to limit deterrence due to overwhelming false positives, thus contributing towards improving tool adoption. 
    \find{Our analyses reveal cases of violations that appear to be systematically ignored by developers, and violation categories that are recurrently addressed. The pattern mining of violation code further provides insights into how violations can be prioritized towards enabling static bug detection tools to be more adopted.}
    \item \textbf{Violation fix pattern mining:} we propose an approach to infer common fix patterns of violations leveraging CNNs and {\em X-means} clustering algorithm. Such patterns can be leveraged in subsequent research directions such as automated refactoring tools (for complying with project rules as done by checkpatch\footnote{\url{http://tuxdiary.com/2015/03/22/check-kernel-code-checkpatch}}\footnote{\url{https://github.com/spotify/linux/blob/master/scripts/checkpatch.pl}} in the Linux kernel development), or automated program repair (by providing fix ingredients to existing tools such as PAR~\cite{kim_automatic_2013}).
    \find{Mined fix patterns can be leveraged to help developers rapidly and systematically address high-priority cases of static violations. In our experiments, we showed that 40\% of a sample set of 500 unfixed violations could be immediately addressed with the inferred fix patterns.}
    \item \textbf{Pattern-based violation patching:} we apply the fix patterns to unfixed violations and actual bugs in real-world programs. Our experiments demonstrate the potential of the approach to infer patterns that are effective  which shows the potential of automated patch generation based on the fix patterns.
    \find{Developers are ready to accept fixes generated based on mined fix patterns. Indeed out of 113 generated patches, 69 were merged in 10 open source projects. It is noteworthy that since static analysis can uncover important bugs, mined patterns can be leveraged for automated repair. Out of the 14 real-bugs in the Defects4J benchmark which can be detected with \findbugs, our mined fix patterns are immediately applicable to produce correct fixes for 4 bugs.}
        % We formalize the definitions of a patch, a fix pattern and a bug fixing process, which is an essential step towards a common understanding of approaches in our community.
    % \item We propose a refined AST tree to present code in a concise and discriminating way. \dongsun{this is a bit confusing. it should be elaborated more.}
    % {\color{red} For example, variable node {\tt a} is represented as ({\em Variable, a}) but not ({\em SimpleName, a}). The details are presented in Section~\ref{sec:definitions} and its example construct is shown in Figure~\ref{fig:trees}.
    % \item We design an approach to identify common code patterns of violations and common fix patterns of fixed violations based on CNNs and {\em X-means}.
    % \item Fix patterns per static analysis violation type: \dongsun{more detail.}
    %Application of fix patterns: \dongsun{program repair results here.}
\end{enumerate}

The remainder of this paper is organized as follows. We propose our study method in Section~\ref{sec:approach}, describing the process of violation tracking, and the approach for mining code patterns based on CNNs. Section~\ref{sec:evaluation} presents the study results in response to the research questions.
Limitations of our study are outlined in Section~\ref{sec:discussion}. Section~\ref{sec:relatedwork} surveys related work. We conclude the paper in Section~\ref{sec:conclusion}with discussions of future work. Several intermediary results, notably w.r.t. the statistics of violations are most detailed in the appendix.
\section{Methodology}
\label{sec:approach}

%\subsection{Overview}
\label{sec:appoverview}

\begin{figure}[!ht]
    \centering
    \includegraphics[width=1\linewidth]{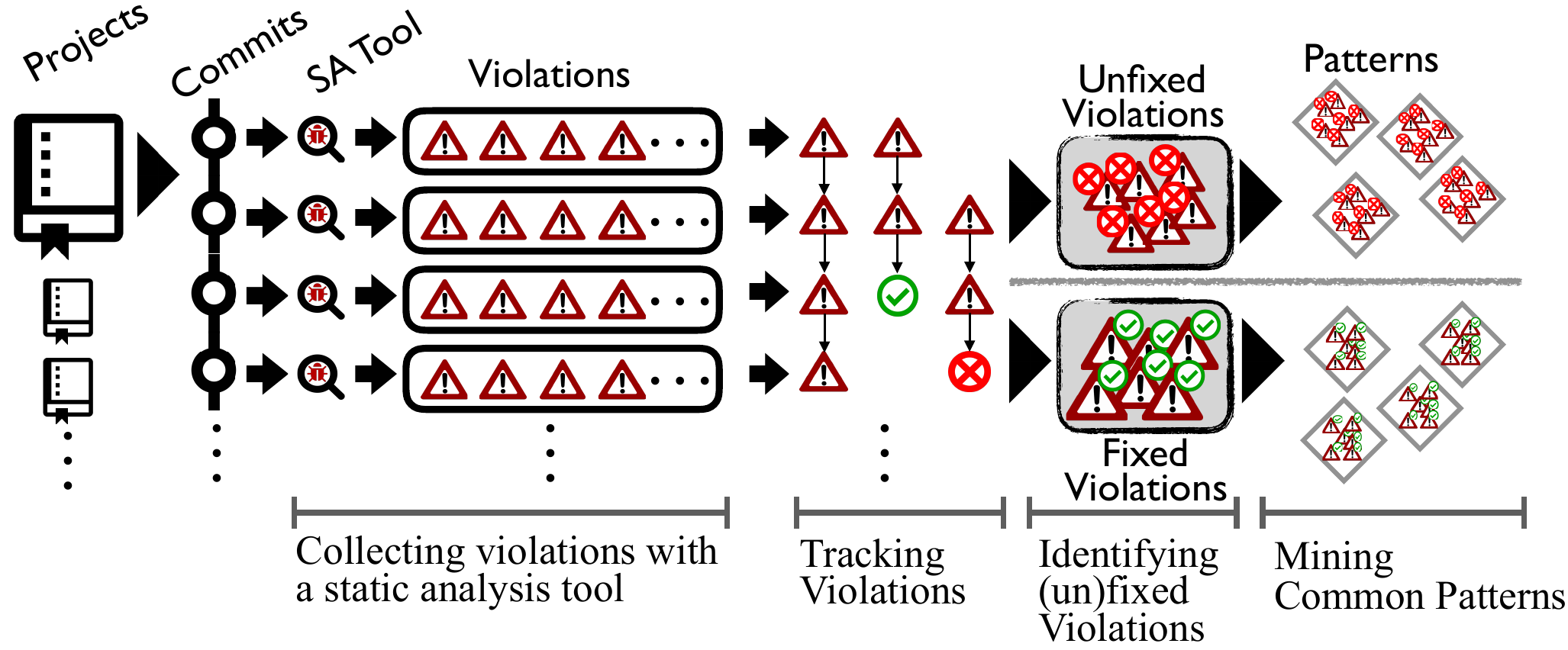}
    \caption{Overview of our study method.}
    \label{fig:overview}
\end{figure}

Our study aims at uncovering common code patterns related to static analysis violations and to developers' fixes.
As shown in Figure~\ref{fig:overview}, our study method unfolds in four steps: (1) applying a static analysis tool to collecting violations from programs, (2) tracking violations across the history of program revisions, (3) identifying fixed and unfixed violations, (4) mining common code patterns in each class of violations, and (5) mining common fix patterns in each class of fixed violations. We describe in details these steps as well as the techniques employed.

\subsection{Collecting violations}
\label{sec:collectvio}
To collect violations from a program, we apply a static analysis tool to every revision of the associated project's source code. Given the resource-intensive nature of this process, we focus in this study on the \findbugs~\cite{shen_efindbugs:_2011} tool, although our method is applicable to other static analysis tools such as Facebook Infer\footnote{http://fbinfer.com/}, Google ErrorProne\footnote{https://errorprone.info/}, etc.
We use the most sensitive option to detect all types of violations defined in \findbugs violation descriptions~\cite{findbugs:description}. For each individual violation instance, we record, as a six-tuple value, all information on the violation type, the enclosing program entity (e.g., project, class or method), the 
commit id, the file path, and the location (i.e., start and end line numbers) where the violation is detected.
Figure~\ref{fig:violationTupleExample} shows an example of a violation record in the collected dataset.% taken from {\em Report\_AnalysisPanel.java} file within Commit {\tt b0ed41} in \texttt{GWASpi}~\footnote{\url{https://github.com/GWASpi/GWASpi}} project.

\begin{figure}[ht]
    \centering
\begin{lstlisting}[linewidth={\linewidth},frame=tb,basicstyle=\footnotesize\ttfamily]
(@{\bf \textless ViolationInstance\textgreater}@)
    (@{\bf \textless ViolationType\textgreater}@)NP_NULL_ON_SOME_PATH(@{\bf \textless /ViolationType\textgreater}@)
    (@{\bf \textless ProjectName\textgreater}@)GWASpi-GWASpi(@{\bf \textless /ProjectName\textgreater}@)
    (@{\bf \textless CommitVersionID\textgreater}@)b0ed41(@{\bf \textless /CommitVersionID\textgreater}@)
    (@{\bf \textless FilePath\textgreater}@)src/main/java/org/gwaspi/gui/reports/
                Report_AnalysisPanel.java(@{\bf \textless /FilePath\textgreater}@)
    (@{\bf \textless StartLineNumber\textgreater}@)89(@{\bf \textless /StartLineNumber\textgreater}@)
    (@{\bf \textless EndLineNumber\textgreater}@)89(@{\bf \textless /EndLineNumber\textgreater}@)    
(@{\bf \textless /ViolationInstance\textgreater}@)
\end{lstlisting}
    \caption{Example record of a single-line violation of type \texttt{NP\_NULL\_ON\_SOME\_PATH} found in {\em ReportAnalysisPanel.java} file within Commit {\tt b0ed41} in {\tt GWASpi}\protect\footnotemark{} project. }
    \label{fig:violationTupleExample}
\end{figure}
\footnotetext{https://github.com/GWASpi/GWASpi}

% 1st round filtering.
Since \findbugs requires Java bytecode rather than source code, and given that violations must be tracked across all revisions in a project, 
it is necessary to automate the compilation process. In this study, we accept projects that 
support the {\em Apache Maven}~\cite{maven} build automation management tool. We apply maven build command (i.e., `{\em mvn package install}') to compiling each revision in \nallproject projects 
that we have collected. Eventually, we were able to successfully build \nmavenproject automatically.

\subsection{Tracking violations}
\label{sec:tracking}

Violation tracking consists in identifying identical violation instances between consecutive revisions: after applying a static analysis tool to a specific revision of a project, one can obtain a set of violations. In the next version, another set of violations can be produced by the tool. If there is any change in the next revision, new violations can be introduced and existing ones may disappear. In many cases however, code changes can move violation positions, making this process a non-trivial task.

Static analysis tools often report violations with line numbers in source code files. Even when a commit modifies other lines in different source file than the location of a violation, it might be unable to use line numbers for matching identical violation pairs between two consecutive revisions. Yet, if the tracking is not precise, the identification of fixed violations may suffer from many false positives and negatives (i.e., identifying unfixed ones as fixed ones or vice versa). Thus, to match potential identical violations between revisions, our study follows the method proposed by Avgustinov et al.~\cite{avgustinov_tracking_2015}. 
This method has three different violation matching heuristics when a file containing violations is changed.
The first heuristic is (1) {\em location-based matching}: if a (potential) matching pair of violations is in code change diffs\footnote{A ``code change diff'' consists of two code snapshots. One snapshot represents the code fragment that will be affected by a code change, and another one represents the code fragment after it has been affected by the code change.}, 
% i.e., code churn~\cite{munson1998code}, a pair of deleted and added code lines in a code change fragment of a file from the buggy version to the fixed one, such as the deleted and added lines in Figure~\ref{fig:expfixchange}.
% sets of files. One set represents the files (i.e., a version of their contents) that will be affected by a code change, and the other set represents the files (i.e., the updated version) after they have been affected by the code change.
it compares the offset of the corresponding violations in the code change diffs. If the difference of the offset is equal to or lower than 3, we regard the matching pair as an identical violation. When a matching pair is located in two different code snapshots, we use (2) {\em snippet-based matching}: if two text strings of the code snapshots (corresponding to the same type of violations in two revisions) are identical, we can match those violations. When the two previous heuristics are not successful, our study applies (3) {\em hash-based matching}, which is useful when a file containing a violation is moved or renamed. This matching heuristic first computes the hash value of adjacent tokens of a violation. It then compares the hash values between two revisions. We refer the reader to more details on the heuristics in~\cite{avgustinov_tracking_2015}.

There have been several other techniques developed to do this task. For example, Spacco et al.~\cite{spacco_tracking_2006} proposed a fuzzy matcher. It can match violations in different source locations between revisions even when a source code file has been moved by package renaming. Other studies~\cite{heckman_establishing_2008,hanam_finding_2014} also provide violation matching heuristics based on software change histories. However, these are not precise enough to be automatically applied to a large number of violations in a long history of revisions~\cite{avgustinov_tracking_2015}.

\subsection{Identifying fixed violations}
\label{sec:vioidentify}

Once violation tracking is completed, we can figure out the resolution of an individual violation. Violation resolution can result in three different outcomes. (1) A violation can disappear due to deleting a file or a method enclosing the violation. (2) A violation exists at the latest revision after tracking (even some code is changed), which indicates that the violation has not been fixed so far. (3) A violation can be resolved by changing specific lines (including code line deletion) of source code. The literature refer to the first and second outcomes as {\em unactionable violations}~\cite{hanam_finding_2014,heckman_systematic_2011,heckman_establishing_2008} or {\em false positives}~\cite{spacco_tracking_2006,yoon_reducing_2014,yi_empirical_2007} while the third one is called {\em actionable violations} or {\em true positives}. In this study we inspect violation tracking results, focusing on the second outcome (which yields the set of {\bf unfixed~violations}) and the third outcome (which yield the set of {\bf fixed~violations}).
% \textcolor{red}{Some violations might be removed because of code fragment deletion. 
% %It is challenge to identify the exact code causing a violation from the deleted code fragments without other changes. 
% Removing the violation code might be an effective resolution of fixing violations. 
% Removing the code fragment containing violations(s) can also be code refactory, such as the first outcome, which is treated as \textit{unfixed violations} in this study. \kui{why?}
% }
%code changes resolve violations
%In this study, we thus represent the first and second outcomes as \textit{unfixed violations}, and the third one as \textit{fixed violations}. %since this study mainly focuses on how violations are induced and resolved. 

Starting from the earliest revision where a violation is seen, we follow subsequent revisions until a later revision has no matching violation (i.e., the violation is resolved by removal of the file/method or the code has been changed). If the violation location in the source code is in a diff pair, we classify it as a \textit{fixed violation}. Otherwise, it is an \textit{unfixed violation}.
\subsection{Mining common code patterns}
\label{sec:miningCP}
Our goal in this step is to understand how a violation is induced. To achieve this goal, we mine code fragments where violations are localized and identify common patterns, not only in fixed violations but also in unfixed violations. 
Before describing our approach of mining common code patterns,
we formalize the definition of a code pattern, and provide justifications for the techniques selected in the approach (namely CNNs~\cite{masakazu_subject_2003, peng_building_2015, mou_convolutional_2016} and {\em X-means} clustering algorithm~\cite{pelleg_xmeans_2000}).%~\kui{The reason.}.

%\subsubsection{Pattern mining process}
%\label{sec:miningMethodology}

\subsubsection{Preliminaries}
\label{sec:definitions}

\paragraph*{\em Definition of code patterns}
%A pattern is a discernible regularity in the world or in manmade designs\footnote{\url{https://en.wikipedia.org/wiki/Pattern}}.\dongsun{unnecessary; too verbose.}
In this study, a \textit{code pattern} refers to a generic representation of similar source code fragments. Its definition is related to the definition of a \textit{source code entity} and of a {\em code context}.
% which are the elements of a {\em code pattern}.
% A code pattern is defined as below:

\begin{definition}{{\bf Source Code Entity (Sce):}}
\label{def:sce}
A \textit{source code entity} (hereafter entity) is a pair of type and identifier, which denotes a node in an Abstract Syntax Tree (AST) representation, i.e., 
\begin{equation}
Sce = (Type, Identifier)
\end{equation}
\end{definition}

\noindent 
where \textit{Type} is an AST node type and \textit{Identifier} is a textual representation (i.e., raw token) of an AST node, respectively.

\noindent 
\begin{definition}{{\bf Code Context (Ctx):}}
\label{definitionCCon}
A \textit{code context} is a three-element tuple, which is extracted from a fined-grained AST subtree (see Section~\ref{sec:refinedast}) associated to a code block, i.e.,
\begin{equation}
Ctx = (Sce, Sce_p, cctx)
\end{equation}
\end{definition}

\noindent
where \textit{Sce} is an entity and % of the current node. 
% \textit{$id_e$} denotes the identifier of \textit{et}.
$Sce_p$ is the parent entity of \textit{Sce} (with $Sce_p = \emptyset$ when \textit{Sce} is a root entity).
% \textit{$id_p$} represents the identifier of \textit{pt}.
\textit{cctx} is a list of code contexts that are the children of \textit{Ctx}. When \textit{Sce} is a leaf node entity, $cctx=\emptyset$.

\noindent 
\begin{definition}{{\bf Code Pattern (CP):}}
    A \textit{code pattern} is 
    a three-value tuple as following:
    \begin{equation}
        CP = (Sce_{a}, Sce_{c}, cctx)
    \end{equation}
\end{definition}

\noindent
where $Sce_{a}$ is a set of abstract entities of which {\em identifiers} are abstracted from concrete representations of specific {\em identifiers} that will not affect the common semantic characteristics of the code pattern. $Sce_{c}$ is a set of concrete entities, of which {\em identifiers} are concrete, that can represent the common semantic  characteristics of the code pattern. Abstract entities represent that the entities of a code pattern can be specified in actual instances while concrete entities indicate characteristics of a code pattern
%These concrete entities are the intersecting entity set of similar code fragments 
and cannot be abstracted. Otherwise, the code pattern will be changed. 
$cctx$ is a set of code contexts (See Definition~\ref{definitionCCon}) that are used to explain the relationships among all entities in this code pattern.
%\dongsun{I will come back to this definition and revise later; still not clear.}

\begin{figure}[!h]
    \centering
    \includegraphics[width=\columnwidth]{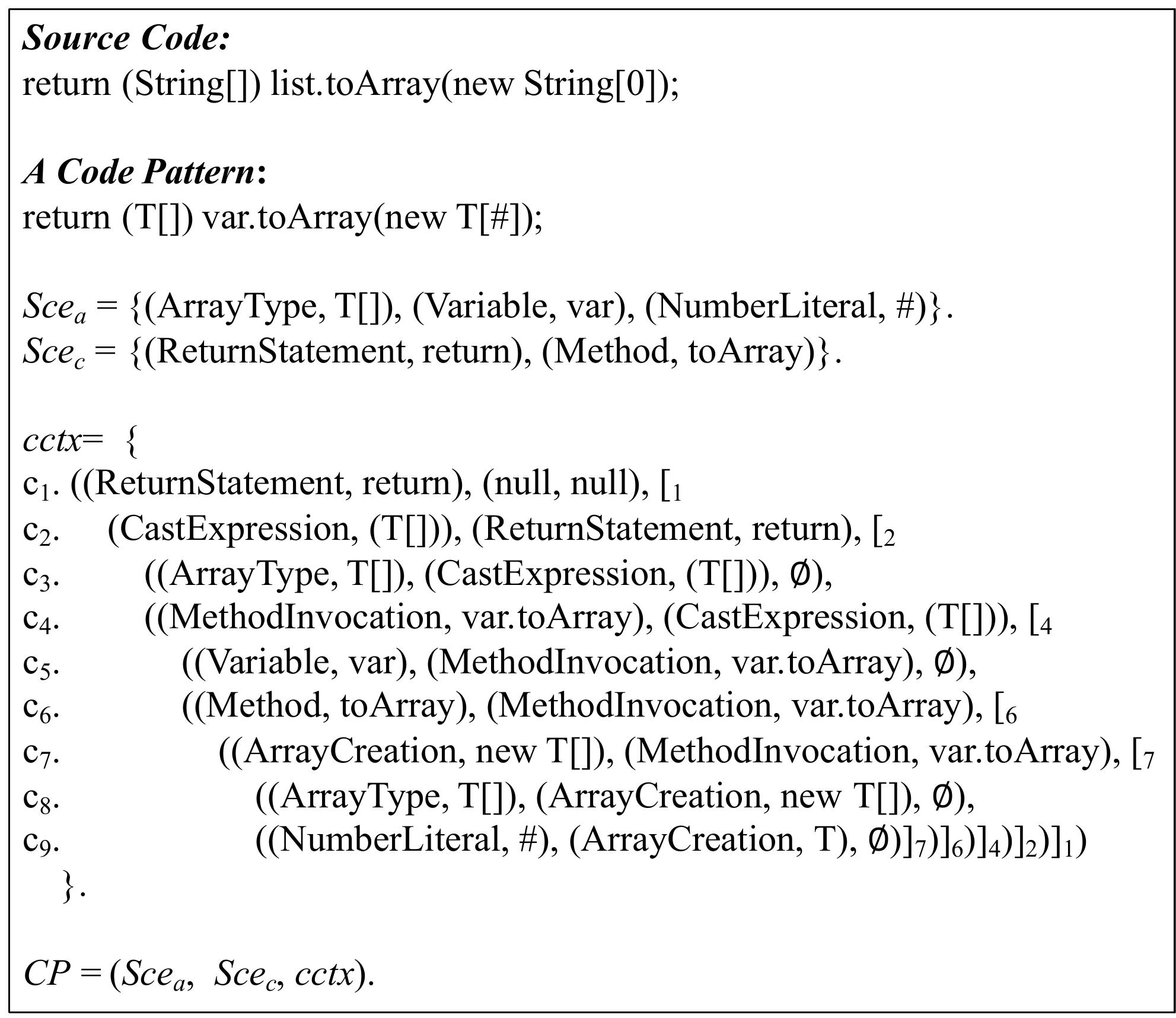}
    \caption{Example representation of a code pattern.}
    \label{fig:egCP}
\end{figure}

Figure~\ref{fig:egCP} shows an example of a code pattern extracted from the source code. 
$Sce_{a}$ contains an array type entity ({\tt ArrayType}, {\tt T[]}), a variable name entity ({\tt Variable}, {\tt var}), and a number literal entity ({\tt NumberLiteral}, {\tt \#}), where {\tt T[]} is abstracted from the identifier {\tt String[]} of ({\tt ArrayType}, {\tt String[]}), {\tt var} is abstracted from the identifier {\tt list} in ({\tt Variable}, {\tt list}), and identifier {\tt \#} is abstracted from the number literal {\tt 0}.
%The array type {\tt T[]} can be {\tt String[]}, {\tt Integer[]} and {\tt Double[]} etc. The {\tt List} variable name {\tt var} can be named {\tt strList}, {\tt intList} and {\tt douList} etc. The value of the number literal {\tt \#} can be changed as well. The anonymizations of the three entities will not affect the type of this code pattern.
The three identifiers of the three entities can also be abstracted from other related similar entities, which will not change the attributes of this pattern.
$Sce_{c}$ consists of a ({\tt ReturnStatement}, {\tt return}) entity and a method invocation entity ({\tt Method}, {\tt toArray}). The identifiers of the two entities cannot be abstracted, otherwise, the attributes of this pattern will be changed. If extracting code pattern from the code at the level of violated source code expression (i.e., the code pattern is {\tt (T[]) var.toArray(new T[\#])}), the ({\tt ReturnStatement}, {\tt return}) node entity can be abstracted as a null entity because this node entity will not affect this code pattern.

$cctx$ contains a code context that explains the relationships among these entities, of which code block is a \texttt{ReturnStatement}.
$c_1$ is the code context of the root source code entity \texttt{ReturnStatement} and consists of three values. The first one is the current $Sce$ that contains a \textit{Type} and an \textit{Identifier}. The second one is the $Sce_p$ of the current $Sce$ which is null as $Sce$ is a root entity. The last one is a list of code contexts which are $c_1$'s children. It is the same as others.
$c_2$ is the direct child of $c_1$. $c_3$ and $c_4$ are the direct children of $c_2$. The source code entity of $c_3$ is a leaf node entity, as a result, its child set is null. It is the same for others.

\paragraph*{\em Suitability of Convolutional Neural Networks}
Grouping code requires the use of discriminating code features to compute reliable metrics of similarity.
While the majority of feature extraction strategies perform well on fixed-length samples, it should be noted that code fragments often consist of multiple code entities with variable lengths. A single code entity such as a method call may embody some local features in a given code fragment, while
several such features must be combined to reflect the overall features of the whole code fragment. It is thus necessary to adopt a technique which can enable the extraction of both local features and the synthesis of global features that will best characterize code fragments so that similar code fragments can be regrouped together by a classical clustering algorithm. 
Note that the objective is not to train a classifier whose output will be some classification label given a code fragment or the code change of a patch.
Instead, we adopt the idea of unsupervised learning~\cite{hastie2009unsupervised} and lazy learning~\cite{david1997lazy} to extract discriminating features of code fragments and patch code changes. 

Recently, a number of studies~\cite{peng2015building, allamanis2015bimodal,mou2016convolutional,gu2016deep,jiang2017automatically,nguyen2017exploring, gu2018deep} 
have provided empirical evidence to support the {\em naturalness of software}~\cite{hindle2012naturalness, allamanis2018survey}. 
A recent work by Bui et al.~\cite{bui2017cross} has provided preliminary results showing that some variants of Convolutional Neural Networks (CNNs) are even effective to 
capture code semantics so as to allow the accurate classification of code implementations across programming languages.

Inspired by the {\em naturalness} hypothesis, we treat source code of violations 
as documents written in natural language and to which we apply CNNs to addressing the objective of feature learning. CNNs are biologically-inspired variants of multi-layer artificial neural networks~\cite{masakazu_subject_2003}. We leveraged the LeNet5~\cite{yann_gradient_1998} model, which involves lower- and upper-layers. Lower-layers are composed of alternating convolutional and subsampling layers which are {\em local-connected} to capture the local features of input data, while upper-layers are {\em fully-connected} and correspond to traditional multi-layer perceptrons (a hidden layer and a logistic regression classifier), which can synthesize all local features captured by previous lower-layers. 

\paragraph*{\em Choice of X-means clustering algorithm}
While K-Means is a classical algorithm that is widely used, it poses the challenge of a try-and-err protocol for specifying the number K of clusters. Given that we lack prior knowledge on the approximate number of clusters which can be inferred, we rely on X-Means~\cite{pelleg_xmeans_2000}, an extended version of K-Means, which effectively and automatically estimate the value of K based on Bayesian Information Criterion.
% \begin{figure}[ht]
%     \centering
%     \includegraphics[width=0.95\linewidth]{fig/CodeContext}
%     \caption{The code context of the buggy code in Figure~\ref{fig:patchExample}.}
%     \label{fig:codeContext}
% \end{figure}

%\noindent 

\subsubsection{Refining the Abstract Syntax Tree}
\label{sec:refinedast}

\begin{figure*}[!t]
    \centering
    \includegraphics[width=\linewidth]{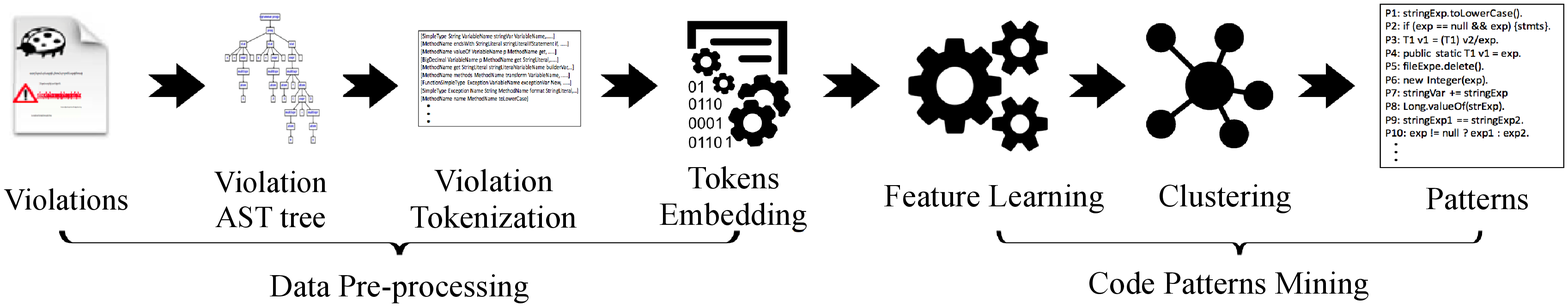}
    \caption{Overview of our code patterns mining method.}
    \label{fig:miningApproach}
\end{figure*}

\begin{figure}[!ht]
    \centering
    \includegraphics[width=\columnwidth]{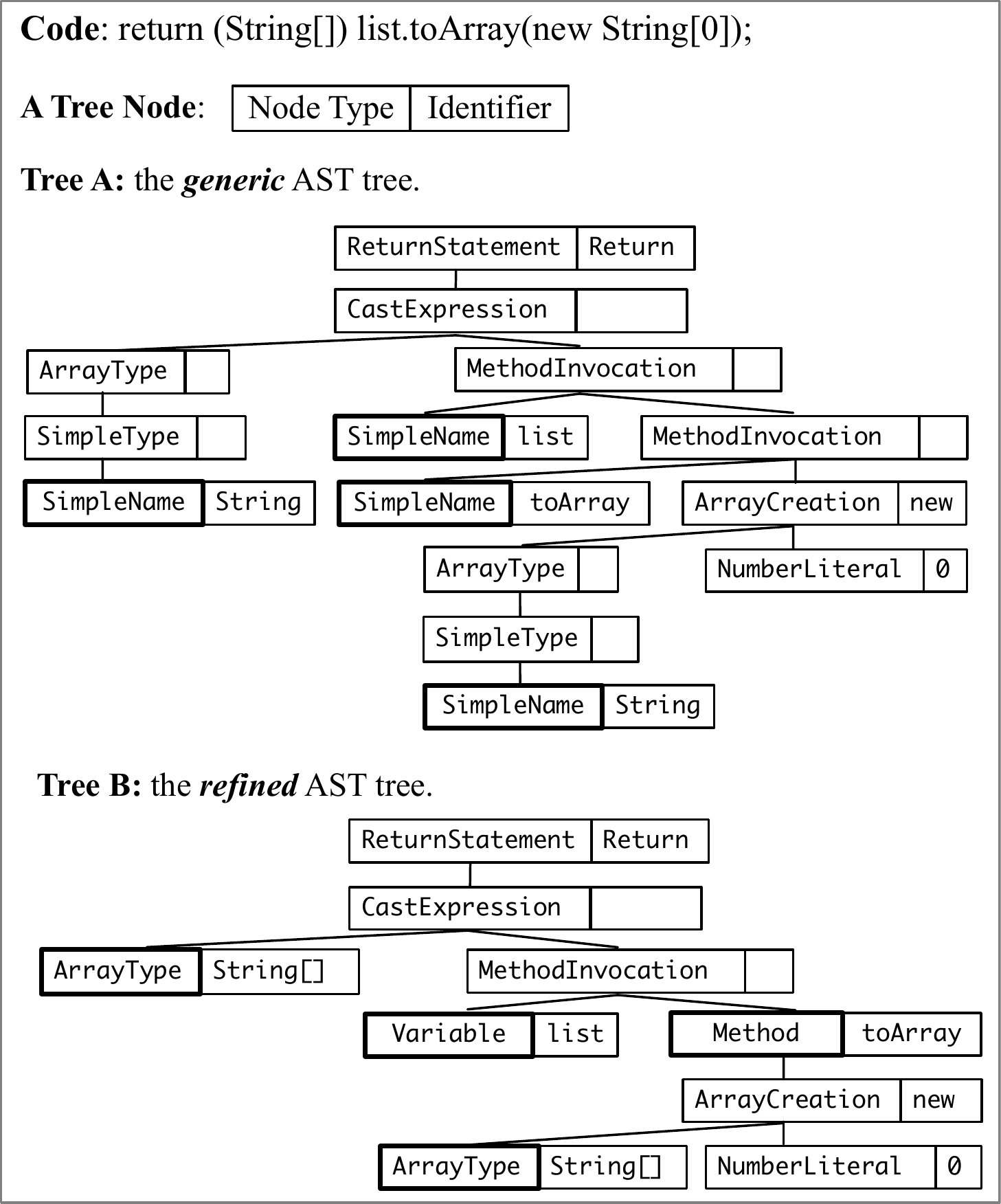}
    \caption{Generic and Refined AST of an example code fragment.}
    \label{fig:trees}
\end{figure}

In our study, code patterns are inferred based on the tokens that are extracted from the AST of code fragments, i.e., the node types and identifiers. Preliminary observations reveal that
some tokens generically tagged {\tt SimpleName} in leaf nodes can interfere feature learning of code fragments.
For example, in Figure~\ref{fig:trees}, the variable node \textit{list} is presented as (\texttt{SimpleName}, \texttt{list}), and the method node \texttt{toArray} is also presented as (\texttt{SimpleName}, \texttt{toArray}) at the leaf node in the generic AST tree. 
As a result, it may be challenging to distinguish the two nodes from each other. Hence, a method of refining the generic AST tree is necessary to reduce such confusions.

Algorithm~\ref{alg:refiningAST} illustrates the algorithm of refining a generic AST tree.
The refined AST tree keeps the basic construct of the generic AST tree. If the label of a current node can be specified as a {\tt SimpleName} leaf node in generic AST tree, the node will be simplified as a single-node construct by combining its discriminating grammar type and its label (i.e., identifier), and its label-related children will be removed in the refined AST tree.

\begin{algorithm}[!h]
%    \vspace{0.2cm}
    \caption{Refining a generic AST tree.}
    \label{alg:refiningAST}
    \LinesNumbered
    \footnotesize
    \KwIn{A generic AST tree $T$.}
    \KwOut{A refined AST tree $T_{rf}$.}
    \SetKwProg{Fn}{Function}{}{}
    \Fn{refineAST(T)}{
        r $\leftarrow$ $T.currentNode$\;
        $T_{rf}.currentNode$ $\leftarrow$ r\;
        \If{r's label can be a SimpleName node}{
            {\footnotesize// r's label can be specified as a SimpleName leaf node\;}
            Remove SimpleName-related children from r\;
            Update r to (r.Type, r.Label.identifier) in $T_{rf}$\;
        }
        \ForEach{child $\in$ r.children}{
            $children_{rf}$.add( $refineAST(child)$ )\;
            $T_{rf}.currentNode.children$ $\leftarrow$ $children_{rf}$\;
        }
        \Return $T_{rf}$\;
    }
\end{algorithm}

Figure~\ref{fig:trees} shows the models respectively of the {\em generic} AST tree and of the {\em refined} AST tree of a code fragment containing a return statement. First, the refined tree presents a simplified architecture.
Second, it becomes easier to distinguish some different nodes with the refined AST tree than the generic AST tree nodes. The node of array type {\tt String[]} is simplified as ({\tt ArrayType}, {\tt String[]}), the variable ({\tt SimpleName}, {\tt list}) is simplified as ({\tt Variable}, {\tt a}), and the method invocation of {\tt toArray} is simplified as ({\tt Method}, {\tt toArray}).
Although the method node \texttt{toArray} can be identified by visiting its parent node (i.e., {\tt MethodInvocation}), it requires more steps to obtain this information.
In the refined AST tree, the two nodes are presented as (\texttt{Variable}, \texttt{list}) and (\texttt{Method}, \texttt{toArray}) respectively. Consequently, it becomes easier to distinguish the two nodes with the refined AST tree than the generic AST tree nodes.

\begin{figure*}[!t]
    \centering
    \includegraphics[width=1\linewidth]{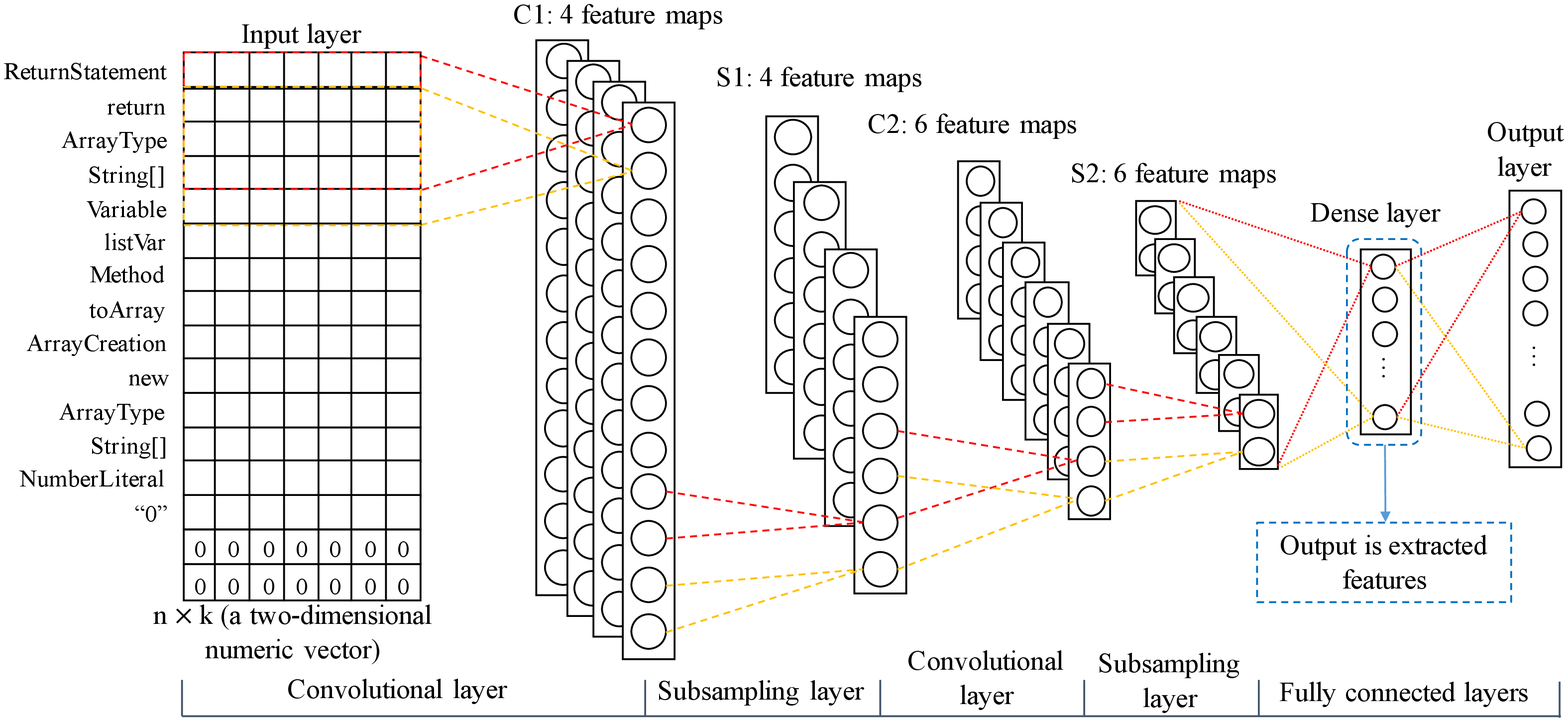}
    \caption{CNN architecture for extracting clustering features. {\em C1} is the first convolutional layer, and {\em C2} is the second one. {\em S1} is the first subsampling layer, and {\em S2} is the second one. The output of dense layer is considered as extracted features of code fragments and will be used to do clustering.}
    \label{fig:cnnarchitecture}
\end{figure*}

\vspace{5mm}
To understand which implementations induce static analysis violations, 
we design an approach for mining common code patterns of detected violations. The patterns
are expected to summarize the main ingredients of code violating a given static analysis rules.
This approach involves two phases: data preprocessing and violation patterns mining,
as illustrated in Figure~\ref{fig:miningApproach}.

\subsubsection{Data preprocessing}
\label{sec:dataPrepro}

\findbugs, reports violations by specifying the start and end lines of a code hunk which is relevant to the reported violation: this is considered as the location of the violation. It is challenging to mine common code patterns from these code hunks directly as they are just textual expression. A given violation code is therefore parsed into a refined AST tree and converted into a token vector.
Token vectors are further embedded with Word2Vec~\cite{tomas_efficient_2013} and converted into numeric vectors which can be fed to CNNs to learn discriminating features of violation code.

\subsubsection*{Violation tokenization}
In order to represent violations with numeric vectors, in this study, violations are tokenized into textual vectors in the first step.
All code hunks of violations are parsed with the refined AST tree and are tokenized into textual vectors by traversing their refined AST trees with the depth-first search algorithm to obtain two kinds of tokens: one is the AST node type and another is the identifier (i.e., raw token) of this node. 
For example, the code ``{\tt int a}'' is tokenized as a vector of four tokens ({\tt\footnotesize PrimitiveType, int, Variable, a}). A given violation is thus represented as a vector of such tokens. Noisy information of nodes (e.g., meaningless variable names such as `a', `b', etc.) can interfere with identifying similar violations. Thus, all variable names are renamed as the combination of their data type and string `{\em Var}'. For example, variable {\tt a} in ``{\tt int a}'' is renamed as {\tt intVar}.

\subsubsection*{Token embedding with Word2Vec}
Widely adopted deep learning techniques 
require numeric vectors with the same size as the format of input data. 
%Thus,  token vectors of violations are further embedded and converted into numeric vectors with the same size.
Tokens embedding is performed with Word2Vec~\cite{tomas_efficient_2013} which can yield a numeric vector for each unique token. Eventually,
a violation is then embedded as a two-dimensional numeric vector (i.e., a vector of the vectors embedding the tokens).
Since token vectors may have different sizes throughout violations, the
corresponding numeric vectors must be padded to comply with deep learning algorithms requirements. We follow the workaround tested
by Wang et al.~\cite{wang_automatically_2016} and append 0 to all vectors
to make all vector sizes consistent with the size of the longest vector.

Word2Vec\footnote{\url{https://code.google.com/archive/p/word2vec/}}~\cite{mikolov_efficient_2013} is a two-layer neural network,
whose main purpose is to embed words, i.e., convert each word into a numeric vector.
 
Numerical representations of tokens can be fed to deep learning neural networks or simply queried to identify relationships among words. For example, relationships among words can be computed by measuring cosine similarity of vectors, given that Word2Vec strives to regroup
similar words together in the vector space. Lack of similarity is expressed as a 90-degree angle, while complete similarity of 1 is 
expressed as a 0-degree angle. For example, in our experiment, `true' and `false' are \texttt{boolean} literal in Java. There is a cosine similarity of 0.9433768 between `true' and `false', the highest similarity between `true' and any other token. 

The left side of Figure~\ref{fig:cnnarchitecture} shows how a violation is vectorized. The $n \times k$ represents a two-dimensional numeric vector of an embedded and vectorized violation, where {\em n} is the number of rows and denotes the size of the token vector of a violation. A row represents a numeric vector of an embedded token. {\em k} is the number of columns and denotes the size of a one-dimensional numeric vector of an embedded token. The last two rows represent the appended 0 to make all numeric vector sizes consistent.

\subsubsection{Code Patterns Mining} 
Although violations can be parsed and converted into two-dimensional numeric vectors, it is still challenging to mine code patterns given that noisy information (e.g., specific meaningless identifiers) can interfere with identifying similar violations.
Deep learning has recently been shown promising in various software engineering tasks~\cite{mou_convolutional_2016, wang_automatically_2016, gu_deep_2016}.
In particular, it offers a major advantage of requiring less prior knowledge and human effort in feature design
for machine learning applications.
Consequently, our method is designed to deeply learn discriminating features for mining code patterns of violations.
We leverage CNNs to perform deep learning of violation features with embedded violations, and also use {\em X-means} clustering algorithm to cluster violations with learned features.

\subsubsection*{Feature learning with CNNs}

Figure~\ref{fig:cnnarchitecture} shows the CNNs architecture for learning violation features. The input is two-dimensional numeric vectors of preprocessed violations. The alternating local-connected convolutional and subsampling layers are used to capture the local features of violations. The dense layer compresses all local features captured by former layers. We select the output of the dense layer as the learned violation features to cluster violations. 
Note that our approach uses CNNs to extract features of violation code fragments, in contrast to normal supervised learning applications that classify labels with training process to show patterns clearly.

\subsubsection*{Violations Clustering and Patterns Labelling}
With learned features of violations, cluster violations with {\em X-means} clustering algorithm.
% which is an extended {\em K-means} algorithm and can estimate the K value of {\em K-means} based on Bayesian Information Criterion effectively and automatically.
In this study, we use Weka's implementation~\cite{witten_data_2016} of {\em X-means} to cluster violations.
Finally, we manually label each cluster with identified code patterns of violations from clustered similar code fragments of violations to show patterns clearly. Note that, the whole process of mining patterns is automated.

\subsection{Mining Common Fix Patterns}
\label{sec:miningApproach}
Our goal in this step is to summarize how a violation is resolved by developers. To achieve this goal, we collect violation fixing changes and proceed to identify their common fix patterns.
The approach of mining common fix patterns is similar to that of mining common code patterns. The differences lie in the data collection and tokenization process.
Before describing our approach of mining common fix patterns,
we formalize the definitions of patch and fix pattern.

\subsubsection{Preliminaries}

\begin{figure*}[!t]
    \centering
    \includegraphics[width=\linewidth]{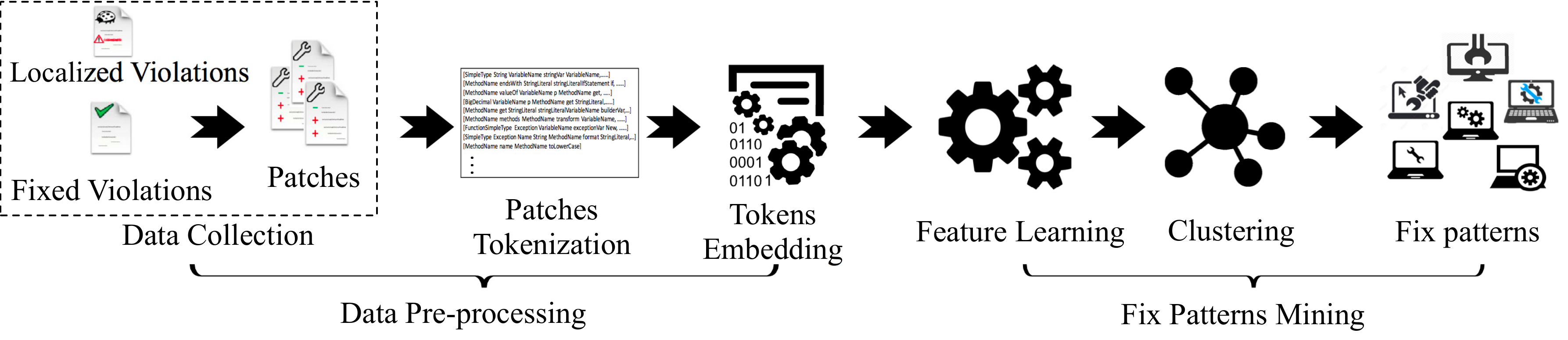}
    \caption{Overview of our fix patterns mining method.}
    \label{fig:miningProcess}
\end{figure*}
%\footnotetext{\url{https://commons.apache.org/proper/commons-io/}}

A patch represents a modification carried on a program source code 
to repair the program which was brought to an erroneous state at runtime.
A patch thus captures some knowledge on modification behavior, and similar
patches may be associated with similar behavioral changes. 
%To understand how violations are resolved, the patch and the fix pattern are formalized in following paragraphs.

\noindent 
\begin{definition}{{\bf Patch (P)}:}
A \textit{patch} is a pair of source code fragments, one representing a buggy version and another as its updated (i.e., bug-fixing) version. In the traditional GNU diff representation of patches, the buggy version is represented by lines starting with {\tt -}, while the fixed version is represented by lines starting with {\tt +}. A patch is formalized as:
\begin{equation}
\label{def:patch}
    P = (Frag_b, Frag_f)
\end{equation}
\end{definition}

\noindent
where $Frag_b$ and $Frag_f$ are fragments of buggy/fixing code, respectively; both are a set of text lines. Either of the two sets can be an empty set but cannot be empty simultaneously. If $Frag_b = \emptyset$, the patch purely adds a new line(s) to fix a bug. On the other hand, the patch only removes a line(s) if $Frag_f = \emptyset$. Otherwise (i.e., both sets are not empty), the patch replaces at least one line.

Figure~\ref{fig:patchExample} shows an example of a patch %, taken from {\em FilenameUtils.java} file within Commit \texttt{09a6cb} in the \texttt{commons-io}\footnote{\url{https://commons.apache.org/proper/commons-io/}} project, 
which fixes a bug of converting a \texttt{String List} into a \texttt{String Array}. $Frag_b$ is the line that starts with {\tt -} while $Frag_f$ is the lines that start with {\tt +}.

% \dongsun{this is completely out of scope in this section. This is more like approach. At least later than here.}
% \kui{I want to use these two paragraphs to introduce why we need fix patterns.}
By analyzing the differences between the buggy code and the fixing code of the patch in Figure~\ref{fig:patchExample}, the patch can be manually summarized as an abstract representation shown in Figure~\ref{fig:absexp1} which could be used to address similar bugs. Abstract representation indicates that specific identifiers and types are abstracted from concrete representation. 

Abstract patch representations can be formally defined as {\em fix patterns}.
Coccinelle~\cite{coccinelle} and its semantic patches provide a metavariable example of how fix patterns can be leveraged to systematically apply common patches, e.g., to address collateral evolution due to API changes~\cite{padioleau_documenting_2008}.
Manually summarizing fix patterns from patches is however time-consuming. Thus, we are investigating an automated approach of mining fix patterns. To that end, we first provide a formal definition of a fix pattern.

\noindent 
\begin{definition}{{\bf Fix Pattern (FP):}}
\label{definitionFP}
A \textit{fix pattern} is a pair of a \textit{code context} extracted from a buggy code block and a set of \textit{change operations}, which can be applied to a given buggy code block to generate fixing code. This can be formalized as:
 \begin{equation}
    FP = (Ctx, CO)
\end{equation}
\end{definition}

\noindent 
where \textit{Ctx} represents the code context that is an abstract representation of the buggy code block. $CO$ is a set of change operations (See Definition~\ref{defitionO}) to be applied to modifying the buggy code block.

% \noindent
% \begin{definition}{{\bf Change Operations (CO):}}
% \label{defitionCO}
% \dongsun{we don't need this definition. Need to define change operation only.}
% \textit{Change operations} is defined as an ordered and hierarchical set of \textit{Operations} (See Definition~\ref{defitionO}) at the refined AST tree level of buggy code acted on modifying the buggy code. This can be formalized as:
% \begin{equation}
%     CO = O^+
% \end{equation}
% \end{definition}

% \noindent
% where $O$ is a specific change operation worker on a specific source code entity.

\noindent
\begin{definition}{{\bf Change Operation (O):}}
\label{defitionO}
A \textit{change operation} is a three-value tuple which contains a change action, a source code entity and a set of sub change operations. This can be formalized as:
\begin{equation}
O = (Action, Sce, CO)
\end{equation}
\end{definition}

\noindent
where \textit{Action} is an element of an action set (i.e., \{{\tt UPD, DEL, INS, MOV}\}) working on the entity (\textit{Sce}). 
\texttt{UPD} is an \textit{update} action which means updating the target entity,
\texttt{DEL} is a \textit{delete} action which denotes deleting the target entity,
\texttt{INS} is an \textit{insert} action which represents inserting a new entity,
and \texttt{MOV} is a \textit{move} action which indicates moving the target entity.
{\em CO} is a set of sub change operations working on the sub entities of the current action's entity. When an operation acts on a leaf node entity, $CO=\emptyset$.

\begin{figure}[!tp]
    \centering
    \includegraphics[width=0.6\columnwidth]{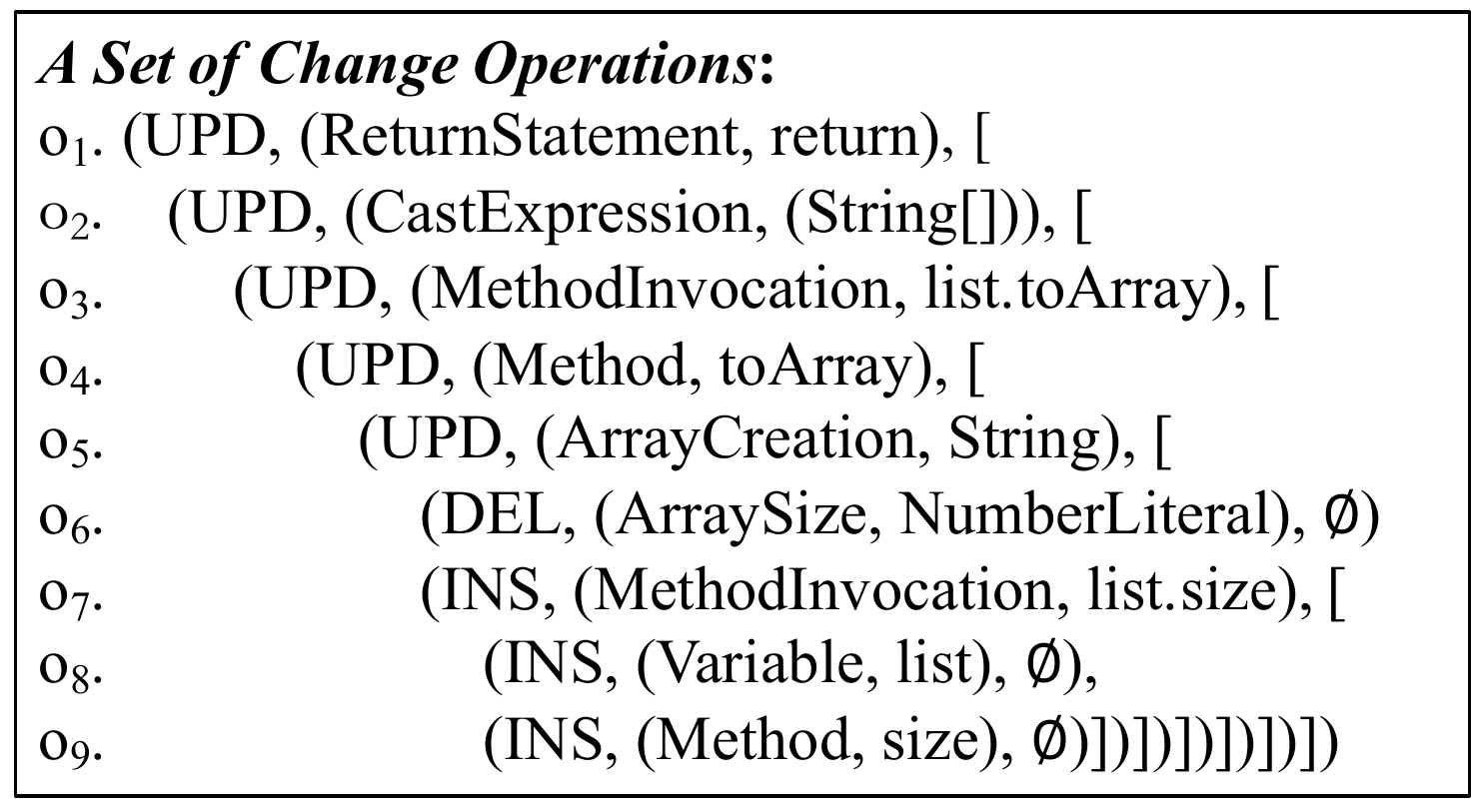}
    \caption{A set of change operations of the patch in Figure~\ref{fig:patchExample}.}
    \label{fig:changeOperations}
\end{figure}

For example, Figure~\ref{fig:changeOperations} shows the set of change operations of the patch in Figure~\ref{fig:patchExample}. 
$o_1$ is the change operation working on the root entity {\tt ReturnStatement}. {\tt UPD} is the {\em Action}, ({\tt ReturnStatement}, {\tt return}) is the root entity being acted, and $o_2$ is the sub change operation acting on the sub entity {\tt CastExpression} of the root entity.
It is the same as others. $o_6$, $o_8$, and $o_9$ are the change operations working on leaf node entities. So that, the sets of their sub change operations are null.

A fix pattern is used as a guide to fix a bug. The fixing process is defined as a \textit{bug fix process} presented in Appendix~\ref{appendix:fixProcess} for interested readers.
% Appendix A for interested readers.
% The process of applying fix patterns to fixing bugs is named {\em bug fix process}, which definition is presented in Appendix~\ref{appendix:fixProcess} for interested readers.

\subsubsection{Pattern mining process}
\label{sec:fixPatternsMining}

Figure~\ref{fig:patchExample} shows a concrete patch that can only be used to fix related specific bugs as it limits the syntax and semantic structure of the buggy code. The statement is limited to be a \texttt{Return Statement} and the parameterized type of the \texttt{List} and the \texttt{Array} is also limited to \texttt{String}. Additionally, the variable name \texttt{list} can also interfere with the matching between this patch and similar bugs.
However, the abstract patch in Figure~\ref{fig:absexp1} abstracts the aforementioned interferon, which can be matched with various mutations of the bug converting a \texttt{List} into an \texttt{Array}. Hence, it is necessary to mine common patch patterns from massive and various patches for specific bugs. 

\begin{figure}[!h]
    \centering
    %\vspace{0.2cm}
\begin{lstlisting}[linewidth={\linewidth},frame=tb,basicstyle=\footnotesize\ttfamily]
(@{\bf DiffEntry of a patch:}@)
@@ -1246,1 +1246,1 @@
(@{\color{red}-        return (String[]) list.toArray(new String[0]);}@)
(@{\color{codegreen}+        return (String[]) list.toArray(}@)
(@{\color{codegreen}+~~~~~~~~~~new String[list.size()]);}@)
\end{lstlisting}
    \caption{Example of a patch taken from {\em FilenameUtils.java} file within Commit \texttt{09a6cb} in project \texttt{commons-io}\protect\footnotemark{}.}
    \label{fig:patchExample}
\end{figure}
\footnotetext{\url{https://commons.apache.org/proper/commons-io/}}

\begin{figure}[!h]
    \centering
\begin{lstlisting}[linewidth={\linewidth},frame=tb,basicstyle=\footnotesize\ttfamily]
(@{\bf Abstract representation of a patch:}@)
@
var: a list variable.
T: the parameterized type of a list.
@
- (T[]) var.toArray(new T[0]);
+ (T[]) var.toArray(new T[var.size()])
\end{lstlisting}
    \caption{Example of an abstract representation of the patch in Figure~\ref{fig:patchExample}.}
    \label{fig:absexp1}
\end{figure}

Our conjecture is that {\em common fix patterns} can be mined from large change sets. 
Exposed bugs are indeed generally not new and common fix patterns may be an immediate and appropriate way to address them automatically. 
For example, when discussing the deluge
of buggy mobile software, Andy Chou, a co-designer of the Coverity bug finding tool, reported that, based on his experience, the found bugs are nothing new and are ``actually well-known and well-understood in the development community - the same {\em use after free} and {\em buffer overflow} defects we have seen for decades''~\cite{bessey_few_2010}. In this vein, we design an approach to mine common fix patterns for static analysis violations by extracting changes that represent developers' manual corrections. 
Figure~\ref{fig:miningProcess} illustrates our process for mining common fix patterns.

% that consists of two phases: data preprocessing and fix patterns mining.

%Overall, for a fixed violation, comparing the identified buggy code file (i.e., the localized violation file) against the fixed code file (i.e., the fixed violation file) can extract the associated patches that are further parsed with \gumtree\footnote{\url{https://github.com/GumTreeDiff/gumtree}} and converted into token vectors. Token vectors are then embedded using Word2Vec and represented with numeric vectors that are fed to CNNs to learn discriminating features of patches. Similar patches are gathered into a cluster by applying {\em X-means} clustering algorithm to learned features of patches. Finally, fix patterns of violations are manually identified from clustered patches.

% Thus, we must preprocess the patches that are collected via source code differencing. 

\subsubsection*{Data Preprocessing.}
\label{sec:dataPreprocess}
As defined in Definition~\ref{definitionFP}, a fix pattern contains a set of change operations, which can be inferred by comparing the buggy and fixed versions of source code files. In our study, code changes of a patch are described as a set of change operations in the form of Abstract Syntax Tree (AST) differences (i.e., AST diffs). In contrast with GNU diffs, which represent code changes as a pure text-based line-by-line edit script, AST diffs provide a hierarchical representation of the changes applied to the different code entities at different levels (statements, expressions, and elements). We leverage the open source GumTree~\cite{falleri_fine_2014} tool to extract and describe change operations implemented in patches. GumTree, and its associated source code, is publicly available,
allowing for replication and improvement, and is built on top of the Eclipse Java model\footnote{\url{http://www.vogella.com/tutorials/EclipseJDT/article.html}}. 
%It can extract fine-grained and accurate source code change operations of patches by comparing the buggy and fixed versions of source code files~\cite{falleri_fine_2014}.  Figure~\ref{fig:changeOperations} shows an example of the change operation set that is the extracted results of the patch in Figure~\ref{fig:patchExample}.

All patches are tokenized into textual vectors by traversing their AST-level diff tree 
with the deep-first search algorithm and extracting the action string (e.g., {\tt UPD}), the entity type (e.g., {\em
ReturnStatement}) and the entity identifier (e.g., {\em
return}) as tokens of a change action (e.g., {\em UPD ReturnStatement return}). A given patch is thus represented as a list of such tokens, further embedded and vectorized as a numeric vector using the same method described in Section~\ref{sec:dataPrepro}.

\subsubsection*{Fix Patterns Mining.}
Patches can be considered as a special kind of natural language text, 
which programmers leverage daily to request and communicate changes in their community. Currently available patch tools only perform directly
the specified operations (e.g., remove and add lines for GNU diff) so far without the interpretation of what the changes are about.
Although all patches can be parsed and converted into two-dimensional numeric vectors, it is still challenging to mine fix patterns given that noisy change information (e.g., specific changes)
can interfere with identifying similar patches.
Thus, our method is designed to effectively learn discriminating features of patches for mining fix patterns.

Similarly to the case of violation code pattern mining, we leverage CNNs to perform deep learning of patch features with preprocessed patches, and {\em X-means} clustering algorithm to automatically cluster similar patches together with learned features. 
%\tb{Here is where we should say how we automate patterns? Intersection of change operations? With what threshold of recurrence?}
Finally, we manually label each cluster with fix patterns of violations abstracted from clustered patches to show fix patterns clearly.
%\tb{Manual task should only be when we generate patches	}

\section{Empirical Study}
\label{sec:evaluation}

\subsection{Datasets}
\label{sec:data}
We consider project subjects based on a curated database of \github provided through {\tt GHTorrent}~\cite{ghtorrent}. We select projects satisfying three constraining criteria: (1) a project has, at least, \mincommits\footnote{A minimum number of commits is necessary to collect a sufficient number of violations, which will be used for violation tracking.} commits, (2) its main language is Java, and (3) it is unique, i.e., not a fork of another project. 
As a result, \nallproject projects are initially collected. We then filter out projects which are not automatically built with {\em Apache Maven}. Subsequently, for each project, we execute \findbugs on the compiled\footnote{\findbugs runs on compiled bytecode (cf. Section~\ref{sec:collectvio}).} code of its revisions (i.e., committed version). 
If a project has at least two revisions in which \findbugs can successfully identify violations, we apply the tracking procedure described in Section~\ref{sec:tracking} to collecting data.

Table~\ref{tab:subjects} shows the number of projects and violations used in this study. There are \nmavenproject projects with 291,615 commits where 250,387,734 violations are detected; these violations are associated with \nviolationtype types defined by \findbugs.
After applying our violation tracking method presented in Section~\ref{sec:tracking} to these violations, as a result, 16,918,530 distinct violations are identified.

\begin{table}[!ht]
    \centering
    \setlength\tabcolsep{2pt}
    \caption{Subjects used in this study.}
    \label{tab:subjects}
    \begin{threeparttable}
    	\begin{tabular}{l|r}
    	    \toprule
    	    \# Projects & \nmavenproject \\
    	    \# Commits &  291,615 \\
    	    \# Violations (detected) & 250,387,734 \\
    	    \# Distinct violations & \nalarms \\
    	    \# Violations types & \nviolationtype \\
    		\bottomrule
    	\end{tabular}
    \end{threeparttable}
\end{table}

% \begin{figure*}[!ht]
%     \centering
%     \includegraphics[width=\textwidth]{fig/empirical_study/Distribution-of-all-violation-types.pdf}
%     \caption[l]{Quantity distributions of violation types sorted by their occurrences. The x-axis represents arbitrary id numbers assigned to violation types. The y-axis represents the percentages of their occurrences in all violations.}
%     \label{fig:qpvt}
% \end{figure*}

\subsection{Statistics on detected violations}
\label{sec:empiricalStudy}

We start our study by quickly investigating {\bf RQ1}: ``{\em to what extent do violations recur in projects?}''.
We focus on three aspects of violations: number of occurrences, spread in projects and category distributions.
Given that such statistics are merely confirming natural distributions of the phenomenon of defects, we provide all the details in the Appendix~\ref{appendix:statsV} of this paper. Interested readers can also directly refer to the replication package (including code and data) at :
\begin{center}
\url{https://github.com/FixPattern/findbugs-violations}.
\end{center}

\find{Overall, we have found that around 10\% of violation types are related to about 80\% of violation occurrences. However, only 200 violation types are spread over more than 100 projects (i.e., 14\% of the subjects), and some violation types which are the most widespread (i.e., top-50) actually have less occurrences than lesser widespread ones. Finally, although most violation types defined by \findbugs are related to {\em Correctness}, the clear majority (66\%) of violation occurrences are associated with {\em Dodgy Code} and {\em Bad Practice}. {\em Security}-related violations account only for 0.5\% of violation occurrences, although they are widespread across 30\% of projects.
 }
 
\subsection{What types of violations are fixed?}
\label{sec:fixedViolations}
Although overall statistics of violation detections show that there is variety in recurrence of violations, we must investigate {\em what types of violations are fixed by developers?} ({\bf RQ2}).
We provide in Appendix~\ref{appendix:statsFV} more details on the following three sub-questions that are considered to thoroughly answer this question. 
\begin{itemize}
    \item RQ2-1: Which types of violations are developers most concerned about?
    \item RQ2-2: Are fixed violations per type proportional to all detected violation?
    \item RQ2-3: What is the distribution of fixed violations per category? 
\end{itemize}

We refer the interested reader to this part for more statistics and detailed insights.

Overall, we have identified 88,927 violation instances which have been fixed by developer code changes. We note that we could not identify fixes for some 69 (i.e., 17\%) types of violations, nor in 183 (i.e., 25\%) projects. Given the significantly low proportion of violations that eventually get fixed,
we postulate that some violation types must represent programming issues that are neglected by the large majority of developers. Another
plausible explanation is the limited use of violation checkers such as \findbugs in the first place since 36\% (273) of the projects associated with \findbugs include at least one commit referring to the \findbugs tool, and 1,944 (2\% of 88,927) cases where the associated commit messages refer to \findbugs. 

\find{
Only a small fraction of violations are fixed by developers. This suggests these violations are related to a potentially high false positive ratio in the static analysis tool, or lack developer interest due to their minor severity. There is thus a necessity to implement a practical prioritization of violations.
}

With respect to RQ2-1, we find that only 50 violation types, i.e., 15\% of the fixed violation types, are associated with 80\% of the fixed violations, and only 63 (19\%) fixed violation types are appearing in at least 10\% of the projects.

\find{Developers appear to be concerned about only a few number of violation types. The top-2 fixed violation types ({\tt \protect{SIC\_INNER\_SHOULD\_BE\_STATIC\_ANON}\protect\footnotemark{}} and {\tt \protect{DLS\_DEAD\_LOCAL\_STORE}\protect\footnotemark{}}) are respectively performance and Dodgy code issues.} 
\addtocounter{footnote}{-2}
\stepcounter{footnote}\footnotetext{Inner class could be refactored into a named static inner class.}
\stepcounter{footnote}\footnotetext{Dead store to local variable.}

With respect to RQ2-2, we compute a fluctuation ratio metric which, for a given violation type, assesses the differences of ranking in terms of detection and in terms of fixes. Indeed a given violation type may account for a very high x\% of all violation detections, but account for only a low y\% (i.e., $y\ll x$). Or vice versa. This metric allows to better perceive how violations can be prioritized: for example, we identified 4 violation types, including \texttt{\footnotesize NM\_CLASS\_NAMING\_CONVENTION}\footnote{Class names should start with an upper case letter.}, have fluctuation ratio values higher than 10, suggesting that, although they have high occurrence rates, they have lower fix rates by developers.  On the other hand, violation type \texttt{\footnotesize NP\_NONNULL\_RETURN\_VIOLATION}\footnote{Method may return null, but is declared @Nonnull.} has an inversed fluctuation ratio of over 20, suggesting that although it has low occurrences in detection, it has a high priority to be fixed by developers. 

\find{Our detailed study of the differences between detection and fix ratios provides data and insights to build detection report and fix prioritization strategies of violations.} 

Finally, with respect RQ2-3, our investigations revealed that the top-50 fixed violation types are largely dominated by {\em Dodgy code}, {\em Performance} and {\em Bad Practice} categories. Although {\em Correctness} overall regroups the largest number (33\%) of fixed violation types, its types have, each, a low number of fix occurrences. Interestingly, {\em Internationalization} is also a common fixed category, with 6,719 fixed instances across 347 (63.3\%) projects, with only two types (\texttt{\footnotesize DM\_CONVERT\_CASE\footnote{Consider using Locale parameterized version of invoked method.}} and \texttt{\footnotesize DM\_DEFAULT\_ENCODING}\footnote{Reliance on default encoding.}) which are among top-5 most occurring violation types and among top-10 most widespread throughout projects).

\find{Overall, {\em Dodgy code}, {\em Performance}, and {\em Bad Practice} issues are the most addressed by developers. {\em Correctness} issues, however, although they are with to the majority of fixed types, developers fail to address a large portion of them. Compared to {\em Internationalization}, which are straightforward and resolved uniformly, the statistics suggest that developers could accept to fix {\em Correctness} issues if there were tool support.}

\subsection{Comparison against other empirical studies on FindBugs violations}
The literature includes a number of studies related to \findbugs violations. While our work includes
such a study, it is substantially more comprehensive and is based on more representative subjects.
As presented in Table~\ref{tab:compEmpiricalStudy}, our study collects data from 730 real-world projects (i.e., in the wild) 
where 400 violation types (of 9 categories) can be found. Other studies have only considered overall only 3 real-world projects. Vetro et al.~\cite{vetro2011empirical} collect data from 301 projects, but they are in-the-lab projects which may not be representative of real-world development. Ayewah et al.~\cite{ayewah_evaluating_2007} only investigated some ($<100$) {\em Correctness}-related violations. Fixit~\cite{ayewah2010google} studied violations at the category level and limited violations into six categories. Vetro et al.~\cite{vetro2011empirical} studied 77 violation types but ignored violation categories. 
% Additionally, the three studies investigated violations limiting to violation occurrences with small data. Therefore, our study can provide all-round knowledge on violation distributions.
% Only Fixit~\cite{ayewah2010google} studied both detected and fixed violations.

\begin{table}[!h]
    \centering
    \caption{Comparison of empirical studies on FindBugs violations.}
    \setlength\tabcolsep{2pt}
    \resizebox{1.0\columnwidth}{!}{
        \begin{threeparttable}
            \begin{tabular}{l|c|c|c|c}
                \toprule
                 & Our study  & Ayewah et al.~\cite{ayewah_evaluating_2007} & Fixit~\cite{ayewah2010google} & Vetro et al.~\cite{vetro2011empirical}\\
                 \hline
                Projects & \makecell[c]{730 projects \\in the wild} & \makecell[c]{Two projects \\in the wild} & \makecell[c]{One student project, \\ One project in the wild} & \makecell[c]{301 projects \\ in the lab}\\\midrule
                \# types & 400 & $<$ 100 & - & 77 \\\midrule
                \# categories & 9 (all of them) & 1 (Correctness) & 6 & - \\\midrule
                % \# cases & \makecell[c]{detected: 16,918,530\\fixed: 88,927 } & detected: 1,506 & \makecell[c]{detected: 10,479\\fixed: 640}& detected: 1,692\\%\midrule
                \# detected cases & 16,918,530 & 1,506 & 10,479 & 1,692\\\midrule
                \# fixed cases & 88,927 & 518 & 640 & -\\\midrule
                Objective & Fix pattern mining & \makecell[c]{Evaluating static\\analysis warnings} & \makecell[c]{Look into the value\\of static analysis} & \makecell[c]{Assess percentage and\\type of violations}\\
                \bottomrule
            \end{tabular}
        \end{threeparttable}
    }
    \label{tab:compEmpiricalStudy}
\end{table}

Additionally, our study investigates detected violation distributions from three aspects: occurrences, spread, and categories, which provides three different metrics to prioritize violations. 
Nevertheless, it should be noted that the false positives of \findbugs could threaten the reliability of violation prioritization based on the statistics of detected violations.
Previous studies~\cite{ayewah_evaluating_2007, ayewah2010google, vetro2011empirical} do not discuss this aspect. 
To reduce this threat, we further investigate distributions of fixed violations, which represent violations that attract developer attention for resolution, thus suggesting higher probabilities for true positives.
Our results provide more reliable prioritization metrics for violations reporting. 
%Additionally, we compare the distributions of fixed violations with detected violations to further evaluate which violation types are really concerned by developers, of which results can provide other metrics of ranking violations for developers. 

We further note that these studies focused on objectives that are different from ours. 
Ayewah et al.~\cite{ayewah_evaluating_2007} focused on evaluating the importance of static analysis warnings in production software. 
In Fixit~\cite{ayewah2010google}, the authors looked into the value of FindBugs on finding program issues. 
Vetro et al.~\cite{vetro2011empirical} aimed at assessing the percentage and type of issues of FindBugs that are actual defects. 
After going through their research tracks, our work could be applied to their research questions, but our eventual goal is to mine fix patterns for FindBugs violations. 
\subsection{Code Patterns Mining}%{Code patterns of Fixed Violations VS. Unfixed Violations}
\label{sec:unfixedViolations}
Empirical findings on violation tracking across the projects showed that only a small fraction of violations are fixed by developers. Thus, overall, the distribution of unfixed violations follow that of detection violations. We now investigate the research question {\em what kinds of patterns do unfixed and fixed violations have respectively?} ({\bf RQ3}), focusing on the following sub-questions:
\begin{itemize}
    \item RQ3-1: What are the common code patterns for unfixed violations and fixed ones respectively?
    \item RQ3-2: What is the relationship or difference between the common source code patterns of unfixed violations and fixed ones?
    \item RQ3-3: What are possible reasons for some violations to remain unfixed?
\end{itemize}

To avoid noise in the dataset due to varying distributions, we focus on instances the instances of violations where the violation types are among the
top-50 types that developers are concerned about (i.e., the most fixed ones). Then, we apply the approach of mining code patterns presented in Section~\ref{sec:miningCP} to identify common code patterns of unfixed violations and fixed ones respectively.

\paragraph*{\em Disclaimer} Note that \findbugs produces a large number of false positives in two ways: 1) locations of detected violations can be incorrectly reported by \findbugs, or 2) the detected violations are correctly located, but developers may still treat it as a false positive warning since it could not be considered as a serious enough concern to fix. While the second kind of false positives does not threaten patterns mining, but the first kind does. To reduce the threat to validity due to false positives related to incorrect localization, we focus on the pattern mining process on the recurrent fixed violations: their locations are most likely correct given that developers manually checked and addressed the issue.
% since they are really concerned and resolved by developers. They are highly possible to be true positives, and their corresponding fix patterns are highly possible to be effective resolutions of fixing violations.

\subsubsection{Experiment Setup}
\label{sec:parametersSetting}
\begin{figure}[!h]
    \centering
    \includegraphics[width=\columnwidth]{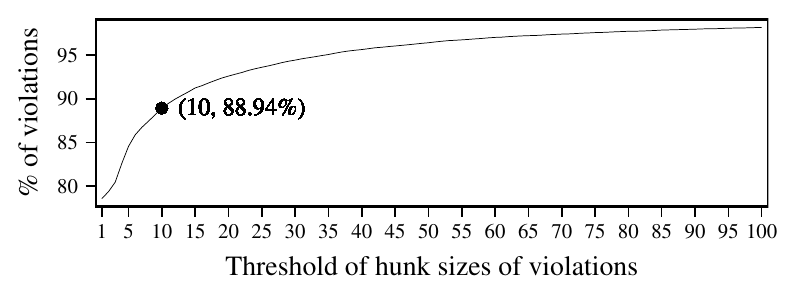}
    \caption{Hunk sizes' distribution of all violations.}
    \label{fig:hunkSizes}
\end{figure}

\findbugs reports violations by specifying the start line and the end line of the code hunk that is relevant to the violation.
Since it is challenging (and error-prone) to mine code patterns by considering big code hunks, we limit our
experiments on small hunks. Figure~\ref{fig:hunkSizes} illustrates the distribution of sizes (i.e., the code line numbers of hunks) of the code hunks associated with
all violations. 

\begin{figure}[!h]
    \centering
    \includegraphics[width=0.6\columnwidth]{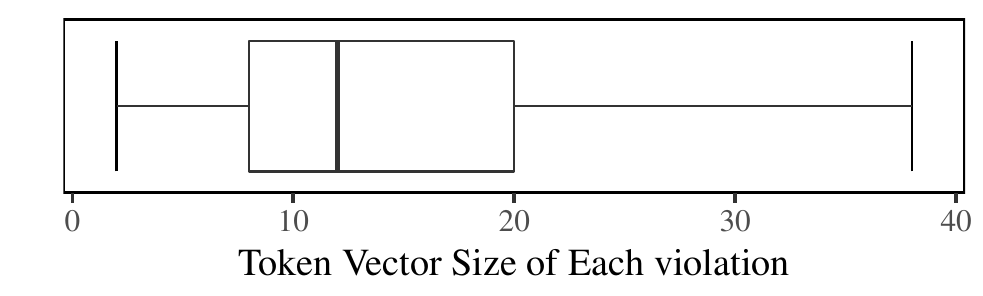}
    \caption{Sizes' distribution of all violation token vectors.}
    \label{fig:tokens}
\end{figure}

For 89\% of the violations, the relevant code hunk is limited to 10 code lines or less. 
We have further manually observed that a line-based calculation of hunk size is not reliable due to the presence of noise caused by comments, annotations and unnecessary blank lines, so we select violations by their tokens.
Figure~\ref{fig:tokens} provides the distribution of numbers of code tokens by violations. We discard outliers and thus focus on violations where the code includes at most 40 tokens extracted based on their refined AST trees (cf. tree B in Figure~\ref{fig:trees}).

Following the methodology described in Section~\ref{sec:miningCP}, 
violations are represented with numeric vectors using Word2Vec with the following parameters (Size of vector~=~300; Window size~=~4; Min word frequency~=~1)
%Table~\ref{tab:word2VecParameter} enumerates the key settings of Word2Vec. 

%\begin{table}[!h]
%    \centering
%    \vspace{0.15cm}
%    \setlength\tabcolsep{2pt}
%    \caption{Parameters Setting of Word2Vec.}
%    \label{tab:word2VecParameter}
%    \begin{threeparttable}
%        \begin{tabular}{l|c}
%        \toprule
%         Parameters & Values \\
%        \hline
%        Size of vector   & 300 \\
%        Window size & 4 \\
%        Min word frequency & 1 \\
%        \bottomrule
%        \end{tabular}
%    \end{threeparttable}
%\end{table}

Feature extraction is then implemented based on CNNs whose parameters are listed in Table~\ref{tab:cnnParameter}. The literature has consistently reported that effective models for Word2Vec and deep learning applications require
well-tuned parameters~\cite{yoav_word2vec_2014, matthew2012ADADELTA, tomas_distributed_2013, krizhevsky_imagenet_2012, ciregan_multi_2012}.
In this study, 
all parameters of the two models are tuned through a visualizing network training UI\footnote{\url{https://deeplearning4j.org/visualization}} provided by DeepLearning4J.

\begin{table}[!h]
    \centering
    \setlength\tabcolsep{2pt}
    \caption{Parameters setting of CNNs.}
    \label{tab:cnnParameter}
    \begin{threeparttable}
        \begin{tabular}{l|c}
        \toprule
        Parameters & Values \\
        \hline
         \# nodes in hidden layers & 1000 \\
         learning rate & 1e-3  \\
         Optimization algorithm & stochastic gradient descent \\
         pooling type & max pool \\
         activation (output layer ) & softmax \\
         activation (other layers) & leakrelu \\
         loss function & mean squared logarithmic error \\
        \bottomrule
        \end{tabular}
    \end{threeparttable}
\end{table}

Finally, Weka's~\cite{witten_data_2016} implementation of {\em X-means} clustering algorithm uses the extracted features to find
similar code for each violation type. Parameter settings for the clustering are enumerated in Table~\ref{tab:xmeans}.

%, and based on a supervised learning on the source code of ten violation types.

\begin{table}[!h]
    \centering
    \setlength\tabcolsep{2pt}
    \caption{Parameters setting of X-means.}
    \label{tab:xmeans}
    \begin{threeparttable}
        \begin{tabular}{l|c}
        \toprule
         Parameters & Values \\
        \hline
        Distance Function  & Euclidean Distance \\
        KD Tree & true \\
        \# max iterations & 1000 \\
        \# max K-means & 500\\
        \# max K-means of children & 500 \\
        \# seed & 500 \\
        \# max clusters & 500\\
        \# min clusters & 1\\
        \bottomrule
        \end{tabular}
    \end{threeparttable}
\end{table}

\begin{table*}[ht]
    \centering
    % \tiny
    \setlength\tabcolsep{2pt}
    \caption{Common code pattern examples of violations.}
    \label{tab:unfixedPatterns}
    \begin{threeparttable}
        \begin{tabular}{l|l}
        \toprule
        Violation Type     &  Common Source Code Pattern(s) \\
        \hline
% SIC\_INNER\_SHOULD\_BE\_STATIC\_ANON  & \makecell[l]{new Type() \{ override some methods(...)\{...\}\}.}\\
% DLS\_DEAD\_LOCAL\_STORE  & v1 = exp1, variable assignment statement.\\
DM\_CONVERT\_CASE  & \circled{1} stringExp.toLowerCase(),  \circled{2} stringExp.toUpperCase().\\ 
\hline
% DM\_DEFAULT\_ENCODING  & \makecell[l]{\circled{1} new BufferedReader(new InputStreamReader(input)), \\2) new StreamWriter(output); 3) stringExp.getBytes(),\\ 4) new String(byteExp), 5) byteExp.toString().}\\
% UWF\_FIELD\_NOT\_INITIALIZED\_IN\_CONSTRUCTOR  & null. \\ 
RCN\_REDUNDANT\_NULLCHECK\_OF\_NONNULL\_VALUE  & \makecell[l]{\circled{1} if (exp == null ...) \{...\}, \circled{2} if (exp != null ...) \{...\}, \\\circled{3} exp == null ? exp1 : exp2, \circled{4} exp != null ? exp1 : exp2.} \\
% \textbf{\color{blue}NM\_METHOD\_NAMING\_CONVENTION}  & null.\\% (method body) \\
% URF\_UNREAD\_FIELD  & \circled{1} this.v1 = v2, \circled{2} this.v1 = new T(), \circled{3} private Type v = exp.\\ %\textbf{only pattern 3 is consistent.} 
\hline
BC\_UNCONFIRMED\_CAST  & \circled{1} T1 v1 = (T1) v2/exp, \circled{2} v1 = (T1) v2/exp, \circled{3} ((T1) v2).exp.\\
% \hline
% REC\_CATCH\_EXCEPTION  & try\{...\} catch (Exception e) \{...\}, Exception is not any specific exception.\\
% BC\_UNCONFIRMED\_CAST\_OF\_RETURN\_VALUE  & \makecell[l]{\circled{1} T1 v1 = (T1) method\_invocation(),  \circled{2} v1 = (T1) method\_invocation(),\\ \circled{3} ((T1) method\_invocation()).exp.}\\ 
% SE\_NO\_SERIALVERSIONID  & null. \\ 
% UPM\_UNCALLED\_PRIVATE\_METHOD  & null.\\ %(method body) \\ 
% \textbf{\color{red}VA\_FORMAT\_STRING\_USES\_NEWLINE}  & "$\backslash$n", this new line symbol is in a StringLiteral of a format method.\\ 
\hline
MS\_SHOULD\_BE\_FINAL  & public/protected static T1 v1 = exp.\\ %e.g., public static int maxStringLength = 65535. 
\hline
RV\_RETURN\_VALUE\_IGNORED\_BAD\_PRACTICE  & \makecell[l]{\circled{1} fileExpe.mkdirs(), \circled{2} fileExpe.mkdir(), \circled{3} fileExpe.delete(), \\ \circled{4} fileExpe.createNewFile(), \circled{5} other exp.method\_invoation() returns a value.}\\ 
\hline
% ST\_WRITE\_TO\_STATIC\_FROM\_INSTANCE\_METHOD  & 1) static\_v1 = instance\_exp1.method\_invocation(), 2) v1 = instance\_exp1.method\_invocation(), 3) v1 = exp1, other variable assignment statement.\\ 
% EI\_EXPOSE\_REP2  & 1) this.v1 = v2, 2) v1 = v2. \textbf{inconsistent.}\\
% URF\_UNREAD\_PUBLIC\_OR\_PROTECTED\_FIELD  & \circled{1} this.v1 = v2, \circled{2} public/protected (static) T1 v1 = exp; \circled{3} v1 = exp1.\\ %\textbf{only 2 is consistent.}
% WMI\_WRONG\_MAP\_ITERATOR  & MapVariable.get(keyExp).\\ 
% OBL\_UNSATISFIED\_OBLIGATION  & \makecell[l]{1) StreamType v = new StreamType(...), 2) streamVariable = new StreamType(new StreamType2(...)), 3) sqlStatementExp.executeMethod(). \\e.g., 1. FileOutputStream fos = new FileOutputStream(outFile); 2. ResultSet rs = sqlStmt.executeQuery().\textbf{not very consistent.}} \\
% EI\_EXPOSE\_REP  & 1) return v; 2) return this.v;\\
% NP\_LOAD\_OF\_KNOWN\_NULL\_VALUE  & null.\\ 
DM\_NUMBER\_CTOR  & \makecell[l]{\circled{1} new Long(...), \circled{2} new Integer(...), \circled{3} new Short(...), \\\circled{4} new Byte(...), \circled{5} new Char(...).}\\
\hline
% SIC\_INNER\_SHOULD\_BE\_STATIC  & 1) this.v1 = v2; 2) v1 = v2; 3) super(...) \textbf{inaccurate localization of violation instances}\\ 
SBSC\_USE\_STRINGBUFFER\_CONCATENATION  & \circled{1} stringVariable += stringExp, \circled{2} stringVariable = stringExp1 + stringExp2.\\ 
\hline
% OS\_OPEN\_STREAM\_EXCEPTION\_PATH  & 1) StreamType v = new StreamType(...), 2) streamVariable = new StreamType(...).\textbf{not very consistent.}\\ 
% \textbf{\color{red}NP\_NONNULL\_RETURN\_VIOLATION}  & return null. \textbf{Inconsistent.}\\ 
% SF\_SWITCH\_NO\_DEFAULT  & switch(exp) \{case1:...case2:...\}, without default statement.\\ 
% \hline
% \textbf{\color{green}UWF\_UNWRITTEN\_FIELD}  & null.\\ 
% DE\_MIGHT\_IGNORE  & try\{...\} catch (Exception e) \{...\}, Exception is not any specified exception.\\%\textbf{To be discussed.} 
% IS2\_INCONSISTENT\_SYNC  & 1) return v1, 2) if(v1 != null) \{ v1.method\_invocation(); v1 = null;\}, 3)sqlStmt = stringLiteral1 + stringExp + stringLiteral2.\textbf{not very consistent.}\\ 
DM\_BOXED\_PRIMITIVE\_FOR\_PARSING  & \circled{1} Integer.valueOf(str), \circled{2} Long.valueOf(str). \\ 
% \textbf{\color{green}RV\_RETURN\_VALUE\_IGNORED\_NO\_SIDE\_EFFECT}  & method\_invocation\_exp, Method invocation statement returns a value, but does not assign the value to any variable. (High false positives.) \\ 
% \textbf{\color{red}ODR\_OPEN\_DATABASE\_RESOURCE}  & \makecell[l]{1) conn.createStatement(), 2) conn.createStatement.execute(...), \\3) PreparedStatement st = conn.prepareStatement(), 4)DataBaseConnection conn = exp.getConnection(...).}\\ 
\hline
PZLA\_PREFER\_ZERO\_LENGTH\_ARRAYS  & return null. \\
% RI\_REDUNDANT\_INTERFACES  & super(...). \textbf{not very consistent. Locate on the position of super classes.}\\ 
% \textbf{\color{green}NP\_NULL\_ON\_SOME\_PATH\_FROM\_RETURN\_VALUE}  & File file = filesArray[index]/filesList.get(index). \\
% UCF\_USELESS\_CONTROL\_FLOW  & if (conditional expression) \{empty statements.\}.\\ 
% UC\_USELESS\_CONDITION  & if (conditional expression) \{...\}.\\ 
% \textbf{\color{green}NP\_NULL\_ON\_SOME\_PATH}  & null.\\ 
% \textbf{\color{green}UC\_USELESS\_OBJECT}  & T1 v1 = new T1().\\ 
% DM\_FP\_NUMBER\_CTOR  & \circled{1} new Double(exp), \circled{2} new Float(exp).\\ 
% MS\_PKGPROTECT  & public static (final) T1 v1 = initializationExp.\\ 
% \textbf{\color{red}SQL\_PREPARED\_STATEMENT\_GENERATED\_FROM\_NONCONSTANT\_STRING}  & sqlExecutionStatement = StringLiteral1 + stringV1 + StringLiteral2 + stringV2.\\ 
% \textbf{\color{green}OBL\_UNSATISFIED\_OBLIGATION\_EXCEPTION\_EDGE}  & \makecell[l]{1) StreamType v = new StreamType(...), 2) streamVariable = new StreamType(new StreamType2(...)), 3) sqlStatementExp.executeMethod(). \\e.g., 1. FileOutputStream fos = new FileOutputStream(outFile); 2. ResultSet rs = sqlStmt.executeQuery(). \textbf{not very consistent.}} \\ 
\hline
ES\_COMPARING\_STRINGS\_WITH\_EQ  & \circled{1} stringExp1 == stringExp2, \circled{2} stringExp1 != stringExp2. \\ 
% \textbf{\color{green}OS\_OPEN\_STREAM}  & \makecell[l]{1) StreamType v = new StreamType(...), 2) streamVariable = new StreamType(new StreamType2(...)), 3) sqlStatementExp.executeMethod(). \\e.g., 1. FileOutputStream fos = new FileOutputStream(outFile); 2. ResultSet rs = sqlStmt.executeQuery(). \textbf{the 3rd one is inconsistent.}} \\
% \textbf{\color{green}RCN\_REDUNDANT\_NULLCHECK\_WOULD\_HAVE\_BEEN\_A\_NPE}  & null.\\ 
% \textbf{\color{blue}NP\_PARAMETER\_MUST\_BE\_NONNULL\_BUT\_MARKED\_AS\_NULLABLE} & 1) method\_invocation(...), 2) class\_instance\_creation(...). \textbf{(not very consistent.)}\\
        \bottomrule
        \end{tabular}
    \end{threeparttable}
\end{table*}

\subsubsection{Code Patterns}
\label{sec:codePatterns}
Given that violation code fragments are represented in the generic form of an AST, we can automatically mine patterns by simply considering the most recurring fragment in a cluster yielded by our approach as the pattern. We then manually assess each pattern to assign a label to it.  
We investigate code patterns on fixed violations and unfixed ones respectively.
Overall, while unfixed violations yield a few more patterns than fixed violations, we find that most patterns are shared by both unfixed and fixed sets. Table~\ref{tab:unfixedPatterns} shows some examples of identified common code patterns of 10 violation types.

We manually checked the patterns yielded for the top-50 violation types and assessed these patterns with respect to \findbugs' documentation. For example, \texttt{\footnotesize DM\_NUMBER\_CTOR} violation refers to the use of a number constructor to create a number object, which is inefficient~\cite{findbugs:description}.
For instance, using \texttt{\footnotesize new Integer(...)} is guaranteed to always result in a new \texttt{\footnotesize Integer} object whereas \texttt{\footnotesize Integer.valueOf(...)} allows caching of values to be done by the compiler, class library, or JVM. Using cached values can avoid object allocation and the code will be faster. Our mined patterns are the five types of number creations with number constructors. \texttt{\footnotesize DM\_FP\_NUMBER\_CTRO} has the similar patterns with it. This example shows how violation code patterns mined with our approach are consistent with the static analysis tool documentation. We have carefully checked the patterns for the top-50 violation types, and found that for 76\%, the patterns are adequate with respect to the documentation. Appendix~\ref{appendix:VCP} provides details on 10 example violation types.

\find{Our code pattern mining approach yields patterns that are consistent with the violation descriptions in documentation of the static analysis tool.}

We focused our investigations on some of the patterns that are yielded only from unfixed violation code, and found that in some cases, there are inconsistencies between the pattern and the bug description provided by \findbugs.

First, we consider a case where the number of patterns discovered for a given violation type exceeds the number of cases enumerated by \findbugs in its documentation.
\texttt{\footnotesize MS\_SHOULD\_BE\_FINAL} is a violation type raised when the analyzer encounters a static field that is public but not final: such a field could be changed by malicious code or accidentally from another package~\cite{findbugs:description}. Besides public static field declarations, the identified patterns on violation code of this type include protected static field declarations, which is inconsistent with the description by \findbugs. Figure~\ref{fig:case_1} shows an example of such inconsistent detection by \findbugs in project {\tt BroadleafCommerce}. When developers confront \findbugs' warning message against their code, they may decide not to address such an undocumented bug.
%This may be caused by the incorrect detection of violations, which could further enlarge the number of false positives of \findbugs.

\begin{figure}[!ht]
    \centering
\begin{lstlisting}[linewidth={\linewidth},frame=tb,basicstyle=\footnotesize\ttfamily]
(@{\bf Violation Type:}@)
MS_SHOULD_BE_FINAL

(@{\bf Violation Code:}@)
    protected static String FACETS_ATTRIBUTE_NAME = 
        "facets";
\end{lstlisting}
    \caption{Example of a detected {\tt MS\_SHOULD\_BE\_FINAL} violation, taken from project {\tt BroadleafCommerce}\protect\footnotemark{}.} %datanucleus-datanucleus-core_ecd8df_ea876bsrc#main#java#org#datanucleus#store#types#wrappers#Map.java
    \label{fig:case_1}
\end{figure}% 6273+2 26107+136 36077
~\footnotetext{\url{https://github.com/BroadleafCommerce/BroadleafCommerce}}

Second, we consider a case where the mined pattern is inconsistent with the documentation of the violation.
\texttt{\footnotesize RI\_REDUNDANT\_INTERFACES} is a warning on a class which implements an interface that has already been implemented by one of the class' super classes~\cite{findbugs:description}. Its mined common code pattern is associated to a super constructor invocation. However, the violation location is positioned on the class declaration line. After manually checking some \texttt{\footnotesize RI\_REDUNDANT\_INTERFACES} cases, we find that the Java classes with \texttt{\footnotesize RI\_REDUNDANT\_INTERFACES} violations indeed have a redundant interface(s) in their class declaration code part. However, some detected \texttt{\footnotesize RI\_REDUNDANT\_INTERFACES} violations locate on the super constructor invocations but not the class declaration code, which could confuse developers and increase the perception of high false positives rates. For example, in Figure~\ref{fig:case_2}, the exact position of the {\tt\footnotesize RI\_REDUNDANT\_INTERFACES} violation should be the ``{\sf\fontsize{8.5}{8.5}\selectfont implements Serializable}'' part (L-33). \findbugs however reports the position at L-49 (highlighted with red background) which is not precise and can even confuse developers on {\em why the code is a violation and how to resolve it}.

%public abstract class AbstractFormat extends NumberFormat implements Serializable {
\begin{figure}[!ht]
    \centering
    \vspace{0.3cm}
\begin{lstlisting}[linewidth={\linewidth},frame=tb,basicstyle=\footnotesize\ttfamily]
(@{\bf Violation Type:}@)
RI_REDUNDANT_INTERFACES

(@{\bf Violation Code:}@)
L-33(@{\hspace{0.5cm}public abstract class AbstractFormat extends}@)
(@{\hspace{2cm}NumberFormat \textcolor{red}{implements Serializable} \{}@)
                ......
L-48(@{\hspace{1.35cm}protected AbstractFormat() \{}@)
L-49(@{\hspace{2cm}\colorbox{red}{this(getDefaultNumberFormat());}}@)
(@{\hspace{2cm}\}}@)
\end{lstlisting}
    \caption{Example of a miss-located {\tt\footnotesize RI\_REDUNDANT\_ INTERFACES} violation, taken from {\tt commit 84a642} in project {\tt commons-math}.} 
    \label{fig:case_2}
\end{figure}
% ~\footnotetext{\url{https://github.com/datanucleus/datanucleus-core}}

\find{Some violations remain unfixed as a result of their imprecise detection. False positives in \findbugs can be improved by addressing some issues with accurate reporting of violation locations, as well as updating the documentation.%\kui{R1.2}
}

Finally, we note that
it is challenging to identify common code patterns for some violation types for two main reasons.

First, some clusters are too small, indicating that the violation instances, despite the abstraction with AST, are too specific. For example, \texttt{\footnotesize DLS\_DEAD\_LOCAL\_STORE} violations are about variable assignments which are specific operators in source code. It is challenging to identify any common code pattern except for the pattern, {\em variable assignment statement}, identified at the level of AST node types. With this information alone, it is practically impossible to figure out why a code fragment is related to a \texttt{\footnotesize DLS\_DEAD\_LOCAL\_STORE} violation. This is a potential reason why some \texttt{\footnotesize DLS\_DEAD\_LOCAL\_STORE} violations remain unfixed.

Second, again, \findbugs cannot locate some violations accurately. We enumerate three scenarios:
\begin{itemize}[leftmargin=*]
    \item The detected violation code is the method body but not the method name. For example, \texttt{\footnotesize NM\_METHOD\_NAMING\_CONVENTION} violations violate the method naming convention but not method bodies, however the source code of these violations tracked with their position provided by \findbugs is the method bodies. Similar source code can be clustered into the same cluster to identify some patterns which cannot explain how the violation is induced, but could help interpret the behavior of these methods. Actually, the method name is the abstract description of method body, so we think that it is inefficient to identify the violation of method names by their naming convention without considering the behavior of method bodies.
    \item The second case is that the source code of violations is irrelevant source code. For instance, \texttt{\footnotesize UWF\_FIELD\_NOT\_INITIALIZED\_IN\_CONSTRUCTOR} indicates that a field is never initialized within any constructor, loaded and referenced without a null check~\cite{findbugs:description}. 
According to observing the instances of this violation type, the source code of these violations is the statements of one method body in these violated Java class, which is irrelevant to the violation type. Some similar source code can be clustered together to obtain some patterns which still cannot explain the violation type. Therefore, it is inconsistent with the bug description of this type.
    \item The third case is that the violation locates on class body rather the declaration of class name. \texttt{\footnotesize SE\_NO\_SERIALVERSIONID} means the current violated Java class implements the {\tt Serializable} interface, but does not define a {\tt serialVersionUID} field~\cite{findbugs:description}. The positions of this kind of violations provided by \findbugs are located in the class body. It is impossible to identify the common code patterns of this violation type which can interpret why the source code makes the violations.
\end{itemize}

These inaccurate localized violations could mislead or confuse developers, which may cause that developers do not prefer to fix these kinds of violations.
In this study, we re-locate the violations of {\tt serialVersionUID} and \texttt{\footnotesize RI\_REDUNDANT\_INTERFACES} to class declarations. Combining the results with source code changes of type-related fixed violations, it is easy to follow why the source code fragment is a violation. Figure~\ref{fig:RI_Example} shows an example of fixing a \texttt{\footnotesize RI\_REDUNDANT\_INTERFACES} violation. Interface {\tt java.util.Map} has been implemented in the super class {\tt AbstractMap} of the current class {\tt Map}. Thus, it is fixed by removing the redundent {\tt java.util.Map} interface.

\begin{figure}[ht]
    \centering
\begin{lstlisting}[linewidth={\linewidth},frame=tb,basicstyle=\footnotesize\ttfamily]
(@{\bf Violation Type:}@) RI_REDUNDANT_INTERFACES

(@{\bf Fixing Patch:}@)
    public class Map extends AbstractMap implements 
(@{\color{red}-~~~~~java.util.Map, SCOMap<java.util.Map>,  }@)
(@{\color{codegreen}+~~~~~SCOMap<java.util.Map>, }@)
        Cloneable, java.io.Serializable {
\end{lstlisting}
    \caption{Example of a fixed {\tt\footnotesize RI\_REDUNDANT\_INTERFACES} violation, taken from {\tt commit ea876b} in {\tt datanucleus-core}\protect\footnotemark{} project.} %datanucleus-datanucleus-core_ecd8df_ea876bsrc#main#java#org#datanucleus#store#types#wrappers#Map.java
    \label{fig:RI_Example}
\end{figure}
~\footnotetext{\url{https://github.com/datanucleus/datanucleus-core}}

\find{Many violation types are associated with code from which patterns can be inferred. Such patterns are relevant for immediately understanding how violations are induced. For some other violations code however it is difficult to mine patterns, partly due to the limitation of \findbugs and the fact that the code fragment is too specific.}

\subsection{Fix Patterns Mining}%{How are the violations resolved if fixed?}
\label{sec:fixPatterns}
We now investigate our ultimate research question on {\em how are the violations resolved if fixed?} ({\bf RQ4}).  To 
that end, we first dissect the violation fixing changes and propose to 
cluster relevant fixes to infer common fix patterns following the CNN-based approach described in Section~\ref{sec:miningApproach}.
% to mining common fix patterns from fixed violations.

%\begin{table}[!ht]
%    \centering
%    \vspace{0.15cm}
%    \setlength\tabcolsep{2pt}
%    \caption{Selected fixed violations for fix patterns mining.}
%    \label{table:FixedALarms}
%    \begin{threeparttable}
%        \begin{tabular}{l|r}
%        \toprule
%        Fixed violations & Quantity \\
%        \hline
%        Collected fixed violations     &  88,927 \\
%        \hline
%        Testing files     & 4,682\\
%        Without changing source code    &  7,010 \\
%        Non-local fixed violations & 7,121 \\
%        Failed to match GumTree results & 25,464 \\
%        Large Hunk &  9,060\\ \hline
%        Final selected instances   & 35,590 \\
%        \bottomrule
%        \end{tabular}
%    \end{threeparttable}
%\end{table}

We curate our dataset of 88,927 violation fixing changes by filtering out changes related to:

\begin{itemize}[leftmargin=*]
    \item 4,682 violations localized in test files. Indeed, we focus on mining patterns related to developer changes on actual program code.
   
    \item 7,010 violations whose fix do not involve a modification in the violation location file. This constraint, which excludes cases where long data flow may require a fixing change in other files, is dictated by our automation strategy for computing the AST edit script, which is simplified by focusing on the violation location file.

    \item 7,121 violations where the associated fix changes are not local to the method body of the violation.  

    \item 25,464 violations where the fixing changes are applied relatively far away from the violation location. We consider that the corresponding AST edit script matches if the change actions are performed within $\pm 3$ lines of the violation location. This constraint conservatively helps to further remove false positive cases of violations which are actually not fixed but are identified as fixed violations due to limitations in violation tracking.

    \item 9,060 violations whose code or whose fix code contain a large number of tokens.  In previous works,  
    Herzig et al.~\cite{herzig_impact_2013} and Kawrykow et al.~\cite{kawrykow_non_2011} have found that large source code change hunks generally address feature additions, refactoring needs, etc., rather than  bug fixes.
    Pan et al.~\cite{pan_toward_2008} also showed that large bug fix hunk pairs do not contain meaningful bug fix patterns, and most bug fix hunk pairs (91-96\%) are small ones. Ignoring large hunk pairs has minimal impact on analysis. Consequently, we use the same threshold (i.e., 40, presented in Section~\ref{sec:unfixedViolations}) of tokens to select fixed violations. 
\end{itemize}

\begin{table*}[ht]
    \centering
    % \tiny
    \setlength\tabcolsep{2pt}
    \caption{Common fix pattern examples of fixed violations.}
    \label{tab:fixedPatterns}
    \begin{threeparttable}
        \begin{tabular}{l|l}
        \toprule
        Violation Type     &  Fix Pattern(s) \\
        \hline
%SIC\_INNER\_SHOULD\_BE\_STATIC\_ANON  & \makecell[l]{1) DEL a parameter of a ClassInstanceCreation in an AnonymousClassDeclaration, \\2) DEL the statement contains an AnonymousClassDeclaration, \\3) Replace the AnonymousClassDeclaration with an defined Static Object of this class.}\\
%DLS\_DEAD\_LOCAL\_STORE  & 1) DEL the buggy statement. \\
DM\_CONVERT\_CASE  & ADD a rule of Locale.ENGLISH into toLowerCase()/toUpperCase().\\ 
\hline
%DM\_DEFAULT\_ENCODING  & \makecell[l]{1) ADD a parameter (i.e., "UTF-8") into the ClassInstanceCreation of a Stream object. \\2) Replace ClassInstanceCreation with a MethodInvocation. 3) DEL the buggy statement.}\\
%UWF\_FIELD\_NOT\_INITIALIZED\_IN\_CONSTRUCTOR  & null. \\ 
RCN\_REDUNDANT\_NULLCHECK\_OF\_NONNULL\_VALUE  & \circled{1} Delete the null check expression. \circled{2} Delete the null check IfStatement. \\
\hline
%NM\_METHOD\_NAMING\_CONVENTION  &  Update Method Name of MethodDeclaration.\\
%URF\_UNREAD\_FIELD  & DEL the buggy statement.\\ 
BC\_UNCONFIRMED\_CAST  & \makecell[l]{\circled{1} Delete the violated statement, \circled{2} Delete the cast type, \\\circled{3} Replace CastExpression with a null value.}\\
\hline
% REC\_CATCH\_EXCEPTION  & Change Exception to a specific exception class.\\
% \hline
%BC\_UNCONFIRMED\_CAST\_OF\_RETURN\_VALUE  & \makecell[l]{1) UPD T1 v1 = (T1) method\_invocation() with T2 v1 = method\_invocation(), 2) DEL the buggy statement, \\3) Replace the buggy CastExpresion with a MethodInvocation.}\\ 
%SE\_NO\_SERIALVERSIONID  &  Insert a private static final field serialVersionUID. \\ 
%UPM\_UNCALLED\_PRIVATE\_METHOD  & DEL the buggy statements.\\ 
%VA\_FORMAT\_STRING\_USES\_NEWLINE  & 1) DEL "$\backslash$n", 2) Replace "$\backslash$n" with "\%n"\\ 
MS\_SHOULD\_BE\_FINAL  & Add a ``final'' modifier.\\ 
\hline
RV\_RETURN\_VALUE\_IGNORED\_BAD\_PRACTICE  & \makecell[l]{\circled{1} Add an IfStatement to check the return value of violated source code.\\ \circled{2} Replace violated expression with a new method invocation.}\\
\hline
%ST\_WRITE\_TO\_STATIC\_FROM\_INSTANCE\_METHOD  & DEL the buggy statement. \\ 
%EI\_EXPOSE\_REP2  & \makecell[l]{1) DEL the buggy statement, 2) Replace the being assigned Expression with a ConditionalExpression.\\ 3) Replace the being assigned Expression with other Expression} \\
%URF\_UNREAD\_PUBLIC\_OR\_PROTECTED\_FIELD  & 1) ADD a ``final'' modifier, 2) DEL the buggy statement.\\ 
%WMI\_WRONG\_MAP\_ITERATOR  & Replace the KeySet iterating ForStatement with an EntrySet iterating ForStatement.\\ 
%OBL\_UNSATISFIED\_OBLIGATION  & null.\\
%EI\_EXPOSE\_REP  & \makecell[l]{1) DEL the buggy statement, 2) Replace the being assigned Expression with a ConditionalExpression.\\3) Replace the Expression being assigned with other Expression} \\
%NP\_LOAD\_OF\_KNOWN\_NULL\_VALUE  & 1) DEL Null-check expression or statement, 2) Replace the null-known variable with NULL.\\ 
DM\_NUMBER\_CTOR  & Replace the number constructor with a static number.valueOf() method.\\ 
\hline
%SIC\_INNER\_SHOULD\_BE\_STATIC  & 1) ADD a ``static'' modifier, 2) DEL the buggy statements.\\ 
SBSC\_USE\_STRINGBUFFER\_CONCATENATION  & \makecell[l]{Replace the String type with the StringBuilder, and replace plus operator of \\StringVarialbe with the append method of StringBuilder.}\\ 
\hline
%OS\_OPEN\_STREAM\_EXCEPTION\_PATH  & null.\\ 
%NP\_NONNULL\_RETURN\_VIOLATION  & 1) DEL the null-returned ReturnStatement. 2) Replace NullLiteral with an Expression which returns an empty list.\\ 
% SF\_SWITCH\_NO\_DEFAULT  & null. \\ 
% \hline
%UWF\_UNWRITTEN\_FIELD  & null. \\ 
%DE\_MIGHT\_IGNORE  & Change Exception to a specific exception class.\\ 
%IS2\_INCONSISTENT\_SYNC  & null. \\ 
DM\_BOXED\_PRIMITIVE\_FOR\_PARSING  & Replace Number.valueOf() with Number.parseXXX() method. \\ 
\hline
%ODR\_OPEN\_DATABASE\_RESOURCE  & null. \\ 
PZLA\_PREFER\_ZERO\_LENGTH\_ARRAYS  & \circled{1} Delete the buggy statement, \circled{2} Replace the null value with an empty array. \\
\hline
%RI\_REDUNDANT\_INTERFACES  & DEL the implemented Interface type. \\ 
%NP\_NULL\_ON\_SOME\_PATH\_FROM\_RETURN\_VALUE  & 1) ADD a Null-check statement or expression, 2) DEL the buggy statement.\\ 
%UCF\_USELESS\_CONTROL\_FLOW  & 1) DEL the buggy statement, 2) DEL the useless ConditionalExpression.\\ 
%UC\_USELESS\_CONDITION  & 1) UPD the operator of a ConditionalExpersion, 2) UPD an sub-expression of a ConditionExpression.\\ 
%NP\_NULL\_ON\_SOME\_PATH  & ADD a Null-check statement or expression. \\ 
%UC\_USELESS\_OBJECT  & DEL the buggy statement. \\ 
%DM\_FP\_NUMBER\_CTOR  & \makecell[l]{1) DEL the buggy statement, 2) Replace Double or Float ClassInstanceCreation with (Double or Float).valueOf(), \\3) Remove the Double or Float ClassInstanceCreation from the buggy expression. }\\ 
%MS\_PKGPROTECT  & \makecell[l]{1) DEL the buggy statement, 2) DEL the ``public'' or ``protected'' modifier \\3) Replace the ``public'' or ``protected'' modifier with the ``private'' modifier.\\ 4) DEL the initialization expression.}\\ 
%SQL\_PREPARED\_STATEMENT\_GENERATED\_FROM\_NONCONSTANT\_STRING  & null.\\ 
%OBL\_UNSATISFIED\_OBLIGATION\_EXCEPTION\_EDGE  & null. \\ 
ES\_COMPARING\_STRINGS\_WITH\_EQ  & Replace the ``=='' or ``!='' InfixExpression with a equals() method invocation.\\ 
%OS\_OPEN\_STREAM  & null. \\
%RCN\_REDUNDANT\_NULLCHECK\_WOULD\_HAVE\_BEEN\_A\_NPE  & ADD a Null-check statement or expression.\\ 
%NP\_PARAMETER\_MUST\_BE\_NONNULL\_BUT\_MARKED\_AS\_NULLABLE & null. \\
        \bottomrule
        \end{tabular}
    \end{threeparttable}
\end{table*}

Overall, our fix pattern mining approach is applied to 35,590 violation fixing changes, which are associated with 288 violation types. Parameter values of Word2Vec, CNNs and {\em X-means} are identical to those used for common code patterns mining (cf. Section~\ref{sec:unfixedViolations}).
In this study, once a cluster of similar changes, for a given violation type, are found, we can automatically mine the patterns based on the AST diffs. Although  approaches such as the computation of longest common subsequence of repair actions could be used to mine fix patterns, we observe that they do not always produce semantically meaningful patterns. Thus, we consider a naive but empirically effective approach of inferring fix patterns by considering the most recurring AST edit script in a given cluster, i.e., the code change that occurs identically the most.
Finally, labels to each change pattern are assigned manually after a careful assessment of the pattern relevance. 

For the experiments, we focus on the top-50 fixed violation types for the mining of fix patterns. Table~\ref{tab:fixedPatterns} summarizes 10 example cases of violation types with details, in natural language, on the fix patterns.

%\subsubsection{Fix Patterns}
%In this study, we

Figure~\ref{fig:FP_Example} presents an inferred pattern in terms of AST edit script for violation type \texttt{\footnotesize RCN\_REDUNDANT\_NULLCHECK \_OF\_NONNULL\_VALUE} described in Table~\ref{tab:fixedPatterns}. For AST-level representation of patterns of other violations, we refer the reader to the replication package.
%

%
%The abstract fix patterns can be used to change the violated source code in a heuristic way or should be specified by actual context. For example, in Table~\ref{tab:fixedPatterns}, the fix patterns of \texttt{\footnotesize RCN\_REDUNDANT\_NULLCHECK\_OF\_NONNULL\_VALUE}, \texttt{\footnotesize BC\_UNCONFIRMED\_CAST}, \texttt{\footnotesize RV\_RETURN\_VALUE\_IGNORED\_BAD\_PRAC}-\texttt{\footnotesize TICE}, \texttt{\footnotesize SBSC\_USE\_STRINGBUFFER\_CONCATENATION} and \texttt{\footnotesize PZLA\_PR}-\texttt{\footnotesize EFER\_ZERO\_LENGTH\_ARRAYS} types are abstract fix patterns.
%
Overall, the pattern presented in AST edit script format, which should be translated into fix changes to 
``{\em delete the null check expression}'' requires some code context to be concretized. When the \texttt{\footnotesize var\footnote{{\tt var} represents any variable being checked.} != null} expression is the {\em null-checking} conditional expression of an \texttt{\footnotesize IfStatement}, the concrete patch must delete the violated expression. Similarly, when the \texttt{\footnotesize exp == null} expression is the condition expression of an \texttt{\footnotesize IfStatement}, the patch also removes the {\em null-checking} expression. When \texttt{\footnotesize exp == null} or \texttt{\footnotesize exp != null} expression is one of the condition expressions of an \texttt{\footnotesize IfStatement}, the patch is deleting the violated expression. This example shows the complexity of automatically generating patches from abstract fix patterns, an entire research direction which is left for future work.
%\tb{Is this correct? or can we have clusters for the case of ifstatement?} 
For now, we generate the patches manually based on the mined fix patterns.
%The fix pattern of \texttt{\footnotesize PZLA\_PREFER\_ZERO\_LENGTH\_ARRAYS} is replacing {\tt null} with {\tt new T[0]} in \texttt{\footnotesize ReturnStatement}, where {\tt T} represents a specified type of an array.

\begin{figure}[!h]
    \centering
\begin{lstlisting}[linewidth={\linewidth},frame=tb,basicstyle=\footnotesize\ttfamily]
(@{\bf DiffEntry of a patch:}@)
(@{\color{red}{\tt -}\hspace{0.5cm}if (dictionaryObject!=null \&\& !onlyEmptyFields)}@){
(@{\color{codegreen}{\tt +}\hspace{0.5cm}if (!onlyEmptyFields) }@){
(@{\hspace{1.5cm}signatures.put(new COSObjectKey(dict), }@)
(@{\hspace{2.5cm}(COSDictionary)dict.getObject());}@)
(@{\hspace{0.8cm}}@)}

(@{\bf Repair actions parsed by GumTree at AST level:}@)
(@{UPD IfStatement@@if(dictionaryObject!=null \&\& }@)
(@{\hspace{2.5cm}!onlyEmptyFields)}@)
(@{---UPD InfixExpression@@dictionaryObject!=null \&\& }@)
(@{\hspace{2.5cm}!onlyEmptyFields}@)
(@{------DEL InfixExpression@@dictionaryObject!=null}@)
(@{---------DEL VariableName@@dictionaryObject}@)
(@{---------DEL Operator@@!=}@)
(@{---------DEL NullLiteral@@null}@)
(@{------DEL Operator@@\&\&}@)

(@{\bf Inferred Fix Pattern:}@)
UPD IfStatement@@if(Null-Check-Expression
(@{\hspace{2.5cm}Operator Other-Expression)}@)
(@{---UPD InfixExpression@@Null-Check-Expression }@)
(@{\hspace{2.5cm}Operator Other-Expression}@)
(@{------DEL Null-Check-Expression}@)
(@{------DEL Operator}@)
\end{lstlisting}
    \caption{Example of a fix pattern for {\tt\footnotesize RCN\_REDUNDANT\_NULL CHECK\_OF\_NONNULL\_VALUE} violation inferred from a violation fix instance taken from {\tt commit a41eb9} in project {\tt apache-pdfbox}\protect\footnotemark{}.} 
%    apache-pdfbox_710c65_a41eb9pdfbox#src#main#java#org#apache#pdfbox#cos#COSDocument.java
    \label{fig:FP_Example}
\end{figure}
~\footnotetext{\url{https://github.com/apache/pdfbox}}

%DLS\_DEAD\_LOCAL\_STORE
%NM\_METHOD\_NAMING\_CONVENTION
%URF\_UNREAD\_FIELD
%UPM\_UNCALLED\_PRIVATE\_METHOD
%ST\_WRITE\_TO\_STATIC\_FROM\_INSTANCE\_METHOD

\find{Our proposed fix pattern mining approach can effectively cluster similar changes of fixing violations together. And the fix pattern mining protocol is applicable to derive meaningful patterns.
}

\begin{lstlisting}[caption=Violation types failed to be identified fix pattern, label=lst:nonPattern]
1. UWF_FIELD_NOT_INITIALIZED_IN_CONSTRUCTOR
2. SF_SWITCH_NO_DEFAULT
3. UWF_UNWRITTEN_FIELD
4. IS2_INCONSISTENT_SYNC 
5. VA_FORMAT_STRING_USES_NEWLINE
6. SQL_PREPARED_STATEMENT_GENERATED_FROM_NONCONSTANT_STRING
7. OBL_UNSATISFIED_OBLIGATION
8. OBL_UNSATISFIED_OBLIGATION_EXCEPTION_EDGE
9. OS_OPEN_STREAM
10.OS_OPEN_STREAM_EXCEPTION_PATH
11.ODR_OPEN_DATABASE_RESOURCE
12.NP_PARAMETER_MUST_BE_NONNULL_BUT_MARKED_AS_NULLABLE
\end{lstlisting}

Listing~\ref{lst:nonPattern} enumerates 12 violation types for which our mining approach could not yield patterns, given that the number of samples per cluster was small, or that within a cluster we could not find strictly redundant change actions sequences. Our observations of such cases revealed the following causes of failure in fix pattern mining:

\begin{itemize}[leftmargin=*]
	\item violations can be fixed by adding completely new node types. For example, one fix pattern of \texttt{\footnotesize RV\_RETURN\_VALUE\_ IGNORED\_BAD\_PRACTICE} violations is replacing the violated expression with a method invocation which encapsulates the detailed source code changes.
	\item violations can occur on specific source code fragments from which it is even difficult to mine patterns. Fixes for such violations generally do not share commonalities.
	\item violations can have fix changes applied in separate region than the violation code location. Since we did not consider such cases for the mining, we systematically miss bottom-7 violation types of Listing~\ref{lst:nonPattern} which are in this case.
	\item  violations can be associated to a \texttt{\footnotesize String} literal. For example, we observe that the fixing changes of \texttt{\footnotesize VA\_FORMAT\_STRING\_USES\_NEWLINE} violations are replacing ``$\backslash$n'' with ``\%n'' within strings. Unfortunately, our AST nodes are focused on compilable code tokens, and thus changes in \texttt{\footnotesize String} literal are ignored to guarantee sufficient abstraction from concrete patches.
	\end{itemize}

%
%\find{
%With our observation, it can infer four reasons why it is challenging to identify common fix patterns from some fixed violation instances.
%}

\subsection{Usage and effectiveness of fix patterns}
We finally investigate whether {\em fix patterns can actually help resolve  violations in practice?} ({\bf RQ5}). To that end, we consider the following sub-questions:

\begin{itemize}
    \item RQ5-1: Can fix patterns be applied to automate the management of some {\em unfixed} violations?
    \item RQ5-2: Can fix patterns be leveraged as ingredients for automated  repair of {\em buggy} programs?
    \item RQ5-3: Can fix patterns be effective in systematizing the resolution of \findbugs violations {\em in the wild}?
\end{itemize}

We recall that our work automates the generation of fix patterns. Patch generation is out of scope, and thus will be performed manually (based on the mined fix patterns), taking into account the code context. 

\subsubsection{Resolving unfixed violations}
\label{subsec:resolve-unfixed}
\begin{figure*}[!ht]
    \centering
    \includegraphics[width=\textwidth]{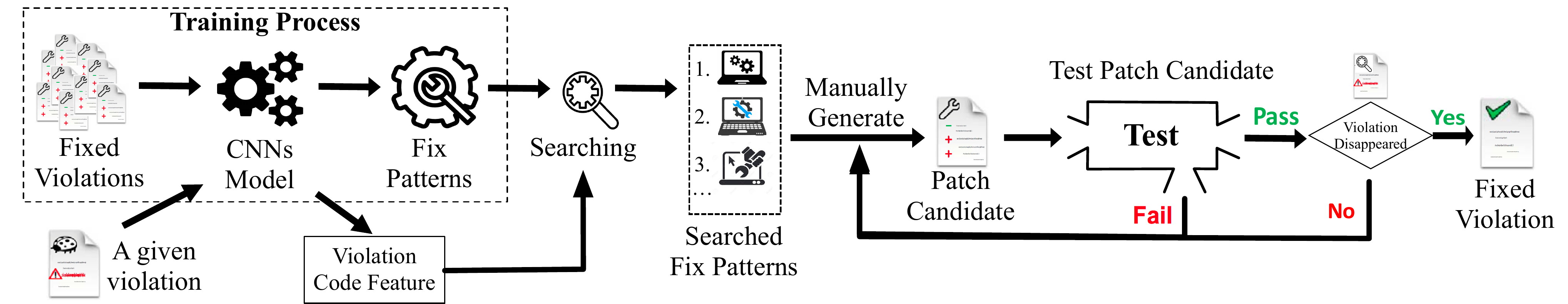}
    \caption{Overview of fixing similar violations with fix patterns.}
    \label{fig:PatternMatching}
\end{figure*}

We apply fix patterns to a subset of unfixed violations in our subjects following the process illustrated in 
Figure~\ref{fig:PatternMatching}. For a given unfix violation, we search for the top-k\footnote{$k=10$ in our experiments} most suitable fix patterns to generate patches. To that end, we consider cosine similarity between the violation code features vector (built with CNNs in Section~\ref{sec:dataPrepro}) and the features vector of the centroid fixed violation in the cluster associated to each fix pattern.

A fix pattern is regarded as a true positive fix pattern for an unfixed violation, if a patch candidate derived from this pattern is addressing the violation. We check this by ensuring that the resulting program after applying the patch candidates passes compilation and all tests, FindBugs no longer raises a warning at this location, and manual checking by the authors has not revealed any inappropriate change of semantics in program behaviour.

%\subsubsection{Can fix patterns be applied to similar unfixed violations?}
%\label{sec:evaluation1}
%To answer this question, we apply fix patterns to a subset of unfixed violations in our subjects.

\paragraph*{\bf Test data}

We collect a subset of unfixed violations in the top-50 fixed violation types (described in Section~\ref{sec:unfixedViolations}) as the testing data of this experiment to evaluate the effectiveness of fixed patterns. 
% \sout{Additionally, we use the method described in Section~\ref{sec:unfixedViolations} to extract source code features of fixed violations and unfixed ones, 
% % used to mine fix patterns in Section~\ref{sec:fixPatterns}, 
% and cluster unfixed violations.}
% \sout{With extracted source code features, we match fix patterns for unfixed violations. }
For each violation type, at most 10 unfixed violation instances, which are the most similar to the centroids of the corresponding fixed violations clusters, are selected as the evaluation subjects.

\paragraph*{\bf Results}

\begin{table}
    \centering
    \vspace{0.2cm}
    \setlength\tabcolsep{2pt}
    \caption{Unfixed-violations resolved by fix patterns.}
    \label{tab:fixedViolations}
    \resizebox{\columnwidth}{!}{
    \begin{threeparttable}
        \begin{tabular}{l|c|c|c|c}
        \toprule
        Violation types & Top 1 & Top 5 & Top 10 & Total\\
        \hline
        \tiny{RI\_REDUNDANT\_INTERFACES} & 10 & 10 & 10 & 10 \\ 
        \tiny{SE\_NO\_SERIALVERSIONID} & 10 & 10 & 10 & 10 \\ 
        \tiny{UPM\_UNCALLED\_PRIVATE\_METHOD} & 10 & 10 & 10 & 10 \\ 
        \tiny{DM\_NUMBER\_CTOR} & 9 & 10 & 10 & 10\\ 
        \tiny{DM\_FP\_NUMBER\_CTOR} & 9 & 10 & 10 &10 \\ 
        \tiny{DM\_BOXED\_PRIMITIVE\_FOR\_PARSING} & 8 & 9 & 10 & 10\\
        \tiny{DM\_CONVERT\_CASE} & 7  & 9 & 10 & 10\\
        \tiny{MS\_SHOULD\_BE\_FINAL} & 7 & 9 & 9 & 10\\
        \tiny{PZLA\_PREFER\_ZERO\_LENGTH\_ARRAYS} & 7 & 7 & 8 & 10 \\
        \makecell[l]{\tiny{RCN\_REDUNDANT\_NULLCHECK\_WOULD}\\\hspace{0.3cm}\tiny{\_HAVE\_BEEN\_A\_NPE}} & 6 &8 & 8& 10\\
        \tiny{RV\_RETURN\_VALUE\_IGNORED\_BAD\_PRACTICE} & 6 & 7 & 8 & 10 \\
        \tiny{SBSC\_USE\_STRINGBUFFER\_CONCATENATION} & 4 & 10 & 10 & 10 \\
        \tiny{MS\_PKGPROTECT} & 4 & 9 & 9 & 10 \\
        \tiny{EI\_EXPOSE\_REP2} & 4 & 4 & 5  & 10\\
        \tiny{DM\_DEFAULT\_ENCODING} & 4 & 5 & 5 & 10 \\
        \tiny{WMI\_WRONG\_MAP\_ITERATOR} & 3 & 7 & 9 & 10 \\
        \tiny{UC\_USELESS\_CONDITION} & 3 & 6 & 6 & 10 \\
        \tiny{ES\_COMPARING\_STRINGS\_WITH\_EQ} & 2 & 8 & 10 & 10\\
        \tiny{RCN\_REDUNDANT\_NULLCHECK\_OF\_NONNULL\_VALUE} & 3 & 4 & 4 & 10\\
        \tiny{SIC\_INNER\_SHOULD\_BE\_STATIC\_ANON}   &  3 & 3 & 3 & 10 \\
        \tiny{UCF\_USELESS\_CONTROL\_FLOW} & 2 & 9 & 10 & 10 \\
        \tiny{BC\_UNCONFIRMED\_CAST\_OF\_RETURN\_VALUE} & 2 & 4 & 4 & 10\\
        \tiny{DLS\_DEAD\_LOCAL\_STORE} & 2 & 3 & 4 & 10\\
        \tiny{NP\_NULL\_ON\_SOME\_PATH} & 1 & 5 & 7 & 10\\
        \tiny{BC\_UNCONFIRMED\_CAST} & 1 & 1 & 1& 10 \\
        \tiny{UC\_USELESS\_OBJECT} & 0 & 8 & 8 & 10 \\
        \tiny{NP\_NULL\_ON\_SOME\_PATH\_FROM\_RETURN\_VALUE} & 0 & 3 & 5 & 10 \\
        % VA\_FORMAT\_STRING\_USES\_NEWLINE & 1 & 8 & 9 &  10\\
        \tiny{VA\_FORMAT\_STRING\_USES\_NEWLINE} & 0 & 0 & 0 &  10\\
        \tiny{UWF\_FIELD\_NOT\_INITIALIZED\_IN\_CONSTRUCTOR} & 0 & 0 & 0 & 10 \\
        \tiny{DE\_MIGHT\_IGNORE} & 0 & 0 & 0 & 10 \\
        \tiny{EI\_EXPOSE\_REP} & 0 & 0 & 0 & 10 \\
        \tiny{IS2\_INCONSISTENT\_SYNC} & 0 & 0 & 0 & 10 \\
        \tiny{NM\_METHOD\_NAMING\_CONVENTION} & 0 & 0 & 0 & 10 \\
        \tiny{NP\_LOAD\_OF\_KNOWN\_NULL\_VALUE} & 0 & 0 & 0 & 10 \\
        \tiny{NP\_NONNULL\_RETURN\_VIOLATION} & 0 & 0 & 0 & 10 \\% False positive of Findings, or DEL the whole method declaration.
        \makecell[l]{\tiny{NP\_PARAMETER\_MUST\_BE\_NONNULL}\\\hspace{0.3cm}\tiny{\_BUT\_MARKED\_AS\_NULLABLE}} & 0 & 0 & 0 & 10 \\
        \tiny{OBL\_UNSATISFIED\_OBLIGATION} & 0 & 0 & 0 & 10 \\
        \tiny{OBL\_UNSATISFIED\_OBLIGATION\_EXCEPTION\_EDGE} & 0 & 0 & 0 & 10 \\
        \tiny{ODR\_OPEN\_DATABASE\_RESOURCE} & 0 & 0 & 0 & 10 \\
        \tiny{OS\_OPEN\_STREAM} & 0 & 0 & 0 & 10 \\
        \tiny{OS\_OPEN\_STREAM\_EXCEPTION\_PATH} & 0 & 0 & 0 & 10 \\
        \tiny{REC\_CATCH\_EXCEPTION} & 0 & 0 & 0 & 10 \\
        \tiny{RV\_RETURN\_VALUE\_IGNORED\_NO\_SIDE\_EFFECT} & 0 & 0 & 0 & 10 \\
        \tiny{SF\_SWITCH\_NO\_DEFAULT} & 0 & 0 & 0 & 10 \\% too many irrelevant tokens in the switch body, which interfere the matching resutls.
        \tiny{SIC\_INNER\_SHOULD\_BE\_STATIC} & 0 & 0 & 0 & 10 \\
        \makecell[l]{\tiny{SQL\_PREPARED\_STATEMENT\_GENERATED}\\
        \hspace{0.3cm}\tiny{\_FROM\_NONCONSTANT\_STRING}} & 0 & 0 & 0 & 10 \\
        \tiny{ST\_WRITE\_TO\_STATIC\_FROM\_INSTANCE\_METHOD} & 0 & 0 & 0 & 10 \\
        \tiny{URF\_UNREAD\_PUBLIC\_OR\_PROTECTED\_FIELD} & 0 & 0 & 0  & 10 \\
        \tiny{URF\_UNREAD\_FIELD}  & 0 & 0 & 0 &  10 \\
        \tiny{UWF\_UNWRITTEN\_FIELD} & 0 & 0 & 0 &  10\\
        \hline
        Total & \makecell[c]{127(25.4\%)} & \makecell[c]{188(37.6\%)} & \makecell[c]{203(40.6\%)} & 500 \\
        \bottomrule
        \end{tabular}
        Identified fix patterns are applied to fixing a subset of unfixed violations in our subjects.
    \end{threeparttable}
    }
\end{table}

Table~\ref{tab:fixedViolations} presents summary statistics on unfixed violations resolved by our mined fix patterns.
Overall, among the selected 500 unfixed violations in the test data, 127 (25.4\%) are fixed by the most similar matched fix patterns (i.e., top-1), 188 (37.6\%) are fixed by a pattern among the top-5, and 203 (40.6\%)  are fixed within the top-10. 
The matched positive fix patterns mainly cluster on top-5 fix pattern candidates, which are a few less than the top-10 range. This suggests that enlarging the search space of fix pattern candidates cannot effectively find positive fix patterns for more target violations.

Among the 203 fixed unfixed-violations, only 3 of them are fixed by matched fix patterns collected across violation types. We observe that \texttt{\footnotesize DM\_NUMBER\_CTOR} and \texttt{\footnotesize DM\_FP\_NUMBER\_CTOR} have similar fix patterns. We use the fix patterns of \texttt{\footnotesize DM\_FP\_NUMBER\_CTOR} to match fix pattern candidates for \texttt{\footnotesize DM\_NUMBER\_CTOR} violations. The fix patterns of \texttt{\footnotesize DM\_FP\_NUMBER\_CTOR} can fix the \texttt{\footnotesize DM\_NUMBER\_CTOR} violations, and vice versa.

\find{
Almost half of the unfixed violations in a sampled dataset can be systematically resolved with mined fix patterns from similar violations fixed by developers. 1 out of 4 of these unfixed violations are immediately and successfully fixed by the first selected fix pattern.
}

We note that fix patterns for 23 violation types are effective in resolving any of the related unfixed violations. There are various reasons for this situation, notably related to the specificity of some violation types and code, the imprecision in \findbugs violation report, or the lack of patterns. We provide detailed examples in Appendix~\ref{appendix:RFU}.

%The features of the syntax and semantic structure of these violations is differentiating, the localization of these violations is in a high accuracy, and other statements have little influence on them.
%Thus, the accuracy of matching a appropriate fix pattern and successfully removing violations with the guidance of fix patterns for these violations is high.

\subsubsection{Fixing real bugs}
We attempt to apply fix patterns to relevant faults documented in the Defects4J~\cite{just2014defects4j} collection of real-world defects in Java programs. This dataset is largely used in studies of program repair~\cite{martinez_automatic_2016, xiong_precise_2017, le_history_2016}. 

\paragraph*{\bf Test data}
We run \findbugs on the 395 buggy versions of the 6 Java projects used to establish Defects4J. As a result, it turns out that 14 bugs can be detected as static analysis violations detectable \findbugs. 
This is a reasonable number since most of the bugs in Defects4J are functional bugs which fail under specific test cases rather than programming rule violations. 
% \textcolor{red}{

For each relevant bug, we consider the fix patterns associated to their violation types, and manually generate the patches. When the generated patch candidate can (1) pass the failed test cases of the corresponding bug and (2) \findbugs cannot identify any violation at the same position, then the matched fix pattern is regarded as a positive fix pattern for this bug.

% % \vspace{\baselineskip}
% % \noindent
\paragraph*{\bf Results}

Table~\ref{tab:fixedBugs} shows the results of this experiment.
4 out of the 14 bugs are fixed with the mined fix patterns and the generated patches by fix patterns are \underline{semantically equivalent} to the patches provided by developers for these bugs.
The violations of 2 bugs are indeed eliminated by fix patterns, but the generated patches lead to new bugs (in terms of test suite pass).
There are 2 bugs that can be matched with fix patterns, but more context information was necessary to fix them. For example, bug \texttt{\footnotesize Lang23} is identified as a \texttt{\footnotesize EQ\_DOESNT\_OVERRIDE\_EQUALS} violation and matched with a fix pattern: overriding the \texttt{\footnotesize equals(Obj o)} method. It is difficult to generate a patch of the bug with this fix pattern without knowing the property values of the object being compared.
The remaining 6 (out of 14 bugs) occurred on specific code, which is challenging to match plausible fix patterns for them without any context.

\begin{table}[!h]
    \centering
    \setlength\tabcolsep{2pt}
    \caption{Fixed bugs in Defects4J with fix patterns.}
    \label{tab:fixedBugs}
    \begin{threeparttable}
        \begin{tabular}{l|c}
        \toprule
        Classification  & \# bugs\\
        \hline
        Fixed bugs   & 4 \\
        Violations are removed but generates new bugs & 2\\
        Need more contexts & 2\\
        Failed to match plausible fix patterns & 6\\
        \hline
        Total & 14\\
        \bottomrule
        \end{tabular}
    \end{threeparttable}
\end{table}

% \begin{table}[!h]
%     \centering
%     \setlength\tabcolsep{2pt}
%     \caption{Alarm Types of Fixed Bugs.}
%     \label{tab:fixedType}
%     \begin{threeparttable}
%         \begin{tabular}{|c|c|}
%         \toprule
%         Alarm type & quantity \\
%         \hline
%         NP\_ALWAYS\_NULL & 1 \\
%         NP\_NULL\_ON\_SOME\_PATH & 1 \\
%         FE\_FLOATING\_POINT\_EQUALITY & 1 \\
%         \bottomrule
%         \end{tabular}
%     \end{threeparttable}
% \end{table}

\find{
Static analysis violations can represent real bugs that make programs fail functional test cases. Our mined fix patterns can contribute to automating the fix of such bugs as experimented on the Defects4J dataset.
}

\subsubsection{Systematically fixing \findbugs violations in the wild}

\label{sec:liveStudy}
We conduct a live study to evaluate the effectiveness of fix patterns 
to systematize the resolutions of violations in open source projects.
We consider 10 open source Java projects collected from \github on 30th September 2017 and presented in Table~\ref{tab:javaProjects}.
\findbugs is then run on compiled versions of the associated programs to localize static analysis violations.

\begin{table}[!ht]
    \centering
    \setlength\tabcolsep{2pt}
    \caption{Ten open source Java projects.}
    \label{tab:javaProjects}
    \begin{threeparttable}
        \begin{tabular}{l|r|r}
        \toprule
        Project Name & \# files & \# lines of code \\
        \hline
        json-simple & 12 & 2,505 \\
        commons-io & 117 & 28,541 \\
        commons-lang & 148  & 77,577  \\
        commons-math & 841 & 186,425 \\
        ant & 859 & 219,506\\
        cassandra &1,625  & 216,192\\
        mahout & 1,145 & 222,345 \\
        aries & 1,570 & 216,646\\
        poi & 4,562 & 894,514\\
        camel & 8,119  & 1,079,671  \\
        \bottomrule
        \end{tabular}
    \end{threeparttable}
\end{table}

\paragraph*{\bf Test data}
We focus on violation instances in the top-50 fixed violation types (presented in Section~\ref{sec:fixedViolations}) are randomly selected as our evaluating data.
% \textcolor{red}{
For each violation, patches are generated manually in a similar process than the previous experiments: a patch must lead to a program that compiles, passes the test cases, and the previous violation location should not be flagged by \findbugs anymore. For each of such patch, we create a pull request and submit the patch to the project developers.

\paragraph*{\bf Results}
Overall, we managed to push 116 patches to the developers of the 10 projects (cf. Table~\ref{tab:liveStudy}). 30 patches have been ignored while 15 have been rejected. Nevertheless, 2 patches have been improved by developers and 67 have been immediately merged. 1 of our pull requests to the  {\tt json-simple} project  was not merged, but an identical patch has been applied later by the developers to fix the violation. Finally, the last patch (out the 116) has not been applied yet, but was attached to the issue tracking system, probably for later replacement.

%The last one of these 116 patches for {\tt mahout} project is attached to its bug report system.%\dongsun{fixed? or ignored?}\dongsun{In Table~\ref{tab:liveStudy}, at the line of ``Total'', 116 is not equal to 30+15+2+67. The note below of the table is missing two patches?}

\begin{table}[!h]
    \centering
    \vspace{0.2cm}
    \setlength\tabcolsep{2pt}
    \caption{Results of live study.}
    \label{tab:liveStudy}
    \begin{threeparttable}
        \begin{tabular}{l|ccccc}
        \toprule
        \multirow{2}{*}{Project Name} & \multicolumn{5}{c}{\# Patches} \\\cline{2-6}
        & pushed & ignored & rejected & improved & merged \\
        \hline
        json-simple & 2 & 1 & 0 & 0 & 0 \\
        commons-io & 2 & 0 & 2 & 0 & 0 \\
        commons-lang & 7 & 5 & 1 & 1 & 0  \\
        commons-math & 6 & 6 & 0 & 0 & 0\\
        ant & 16 & 2 & 4 & 1 & 9 \\
        cassandra & 9 & 9 & 0 & 0 & 0\\
        mahout & 3 & 2 & 0 & 0 & 0 \\
        aries & 5 & 5 & 0 & 0 & 0\\
        poi & 44 & 0 & 0 & 0 & 44 \\
        camel & 22 & 0 & 8 & 0 & 14  \\ % mvn compile successfully, but failed to mvn compile -P sourcecheck
        \hline
        Total$^\dagger$ & 116 & 30 & 15 & 2 & 67 \\
        \bottomrule
        \end{tabular}
        $^\dagger$One patch of {\tt json-simple} is the same as a patch of the same violation which has been fixed by its developer in another version. One patch of {\tt mahout} is attached to its bug report system but has not yet been merged.
    \end{threeparttable}
\end{table}

Table~\ref{tab:delay} presents the distribution of delays before acceptance for the 69 accepted (merged + improved) patches. 67\% of the patches are accepted within 1 day, while 97\% (67\% +30\%) are accepted within 2 days. Only 2 patches took a longer time to get accepted. We note that this acceptance delay is much shorter than the median distributions of the three kinds of patches submitted for the Linux kernel~\cite{koyuncu_impact_2017}.

\begin{table}[!h]
    \centering
    \setlength\tabcolsep{2pt}
    \caption{Delays until acceptance.}
    \label{tab:delay}
    \begin{threeparttable}
        \begin{tabular}{c|c|c|c}
        \toprule
        Delay &  less than 1 day & 1 to 2 days & 17 days\\
        \hline
        Number of Patches & 46 (67\%) & 21 (30\%) & 2 (3\%)\\ 
        \bottomrule
        \end{tabular}
        Acceptance indicates one of improved or merged patches.
    \end{threeparttable}
\end{table}

As summarized in Table~\ref{tab:verify}, we note that 21 accepted patches were verified by at least two developers. 
Although 48 accepted patches were verified by only one developer, we argue that this does not bias the results: first, the common source code patterns of these accepted fixed violation types are consistent with the descriptions documented by \findbugs; second, the matched fix patterns are likely acceptable by developers since the patterns are common in fixing violations as mined in the revision histories of real-world projects.

\begin{table}[!h]
    \centering
    \setlength\tabcolsep{2pt}
    \caption{Verification of accepted patches.}
    \label{tab:verify}
    \begin{threeparttable}
        \begin{tabular}{c|c|c|c}
        \toprule
        Verified by &  1 developer & 2 developers & 3 developers\\
        \hline
        Number of Patches & 48 & 19 & 2 \\ 
        \bottomrule
        \end{tabular}
    \end{threeparttable}
\end{table}

\find{
Our mined fix patterns are effective to fix violations in the wild. Furthermore, the generated patches are eventually quickly accepted by developers.
}

The live study further yields a number of insights related to static analysis violations.

\noindent
{\bf Insight 1.} Well-maintained projects are not prone to violating commonly-addressed violation types.
We note that 8 violation types (presented in Listing~\ref{lst:unhappenedTypes}) do not appear at the current revisions of the selected 10 projects.
Type \texttt{\footnotesize RI\_REDUNDANT\_INTERFACES} occurs only one time in {\tt json-simple} project.
%As a result, we \sout{infer the reason is} presume that these projects are well-maintained by developers for latest versions of these projects.
This finding suggests that violation recurrences may be time-varying, so that, there is a time-variant issue of violation recurrences in revision histories of software projects, which may help to prioritize violations. It is included in our future work.

%\vspace{10pt}
\begin{lstlisting}[caption={Violation types not seen in the selected 10 projects.}, label=lst:unhappenedTypes]
1. SIC_INNER_SHOULD_BE_STATIC
2. NM_METHOD_NAMING_CONVENTION
3. SIC_INNER_SHOULD_BE_STATIC_ANON
4. NP_PARAMETER_MUST_BE_NONNULL_BUT_MARKED_AS_NULLABLE
5. NP_NONNULL_RETURN_VIOLATION
6. UPM_UNCALLED_PRIVATE_METHOD
7. ODR_OPEN_DATABASE_RESOURCE
8. SE_NO_SERIALVERSIONID
\end{lstlisting}
%\vspace{10pt}

\noindent
{\bf Insight 2.} Developers can write positive patches to fix bugs existing in their projects based on the fix patterns inferred with our method.
For example, the developers of {\tt commons-lang}~\footnote{\url{https://github.com/apache/commons-lang}} project fixed a bug\footnote{\url{https://garygregory.wordpress.com/2015/11/03/java-lowercase-conversion-turkey/}} reported as a \texttt{\footnotesize DM\_CONVERT\_CASE} violation by \findbugs by improving the patch that was proposed using our method (cf. Figure~\ref{fig:finding1}). Our method cannot generate the patch they wanted because there is no fix pattern that is related to adding a rule of \texttt{\footnotesize Locale.ROOT} in our dataset, so that there might be a limitation of existing patches in revision histories. 

\begin{figure}[!ht]
    \centering
\begin{lstlisting}[linewidth={\linewidth},frame=tb,basicstyle=\footnotesize\ttfamily]
(@{\bf The patch generated by fix patterns:}@)
(@{\color{red}- final TimeZone tz = TimeZone.getTimeZone(}@)
(@{\color{red}-~~~~~~~~value.toUpperCase());}@)
(@{\color{codegreen}+ final TimeZone tz = TimeZone.getTimeZone(}@)
(@{\color{codegreen}+~~~~~~~~value.toUpperCase(Locale.ENGLISH));}@)

(@{\bf The patch improved by developers:}@)
(@{\color{red}- final TimeZone tz = TimeZone.getTimeZone(}@)
(@{\color{red}-~~~~~~~~value.toUpperCase());}@)
(@{\color{codegreen}+ final TimeZone tz = TimeZone.getTimeZone(}@)
(@{\color{codegreen}+~~~~~~~~value.toUpperCase(Locale.ROOT));}@)
\end{lstlisting}
    \caption{Example of an improved patch in real project.}
    \label{fig:finding1}
\end{figure}

\noindent
{\bf Insight 3.} Developers will not accept plausible patches that appear unnecessary even if those are likely to be useful.
For example, Figure~\ref{fig:finding2} shows a rejected patch that adds an {\tt instanceof} test to the implementation of {\tt equals(Object obj)}. 
The developers want to accept this patch at the first glimpse, but they reject to change the source code after reading the context of these violations since the implementation of {\tt equals(Object obj)} belongs to an inner static class which is only used in a generic type that will not compare against other Object types. %Although this kinds of patches are rejected, we argue that it should be better to add an {\tt instanceof} test in the implementation of {\tt equals(Object obj)} for a good practice and in case of any possible refactoring of these source code in the future.\dongsun{too strong argument without evidence.}

\begin{figure}[!ht]
    \centering
\begin{lstlisting}[linewidth={\linewidth},frame=tb,basicstyle=\footnotesize\ttfamily]
(@{\bf Rejected Patch:}@)
(@{\color{red}- return Arrays.equals(keys,((MultipartKey)obj).keys);}@)
(@{\color{codegreen}+ return obj instanceof MultipartKey \&\& }@)
(@{\color{codegreen}+~~~~~~Arrays.equals(keys, ((MultipartKey)obj).keys);}@)
\end{lstlisting}
    \caption{Example of a rejected patch in real projects.}
    \label{fig:finding2}
\end{figure}

\noindent
{\bf Insight 4.} Some violations fixed based on the mined fix patterns may break the backward compatibility of other applications, leading developers to reject patches for such violations. For example, Figure~\ref{fig:finding4} shows a rejected patch of a \texttt{\footnotesize MS\_SHOULD\_BE\_FINAL} violation in {\em Path.java} file of the \texttt{ant} project, which breaks the backward compatibility of \texttt{\footnotesize systemClasspath} in {\em InternalAntRunner} class~\footnote{\url{https://github.com/eclipse/eclipse.platform/blob/R4_6_maintenance/ant/org.eclipse.ant.core/src_ant/org/eclipse/ant/internal/core/ant/InternalAntRunner.java#L1484}} of Eclipse project.

\begin{figure}[!tp]
    \centering
    \vspace{0.3cm}
\begin{lstlisting}[linewidth={\linewidth},frame=tb,basicstyle=\footnotesize\ttfamily]
(@{\bf Rejected Patch breaks the backward compatibility:}@)
(@{\color{red}-public static Path systemClasspath =}@)
(@{\color{codegreen}+public static final Path systemClasspath = }@)
(@~new Path(null,System.getProperty("java.class.path"));@)

(@{\bf Code @Line 1484 in InternalAntRunner.java in Eclipse project:}@)
org.apache.tools.ant.types.Path.systemClasspath = 
    systemClasspath;
\end{lstlisting}
    \caption{Example of a rejected patch breaking the backward compatibility.}
    \label{fig:finding4}
\end{figure}

\noindent
{\bf Insight 5.} Some violation types have low impact. For example, \texttt{\footnotesize PZLA\_PREFER\_ZERO\_LENGTH\_ARRAYS} refers to the \findbugs' rule that an array-return method should consider returning a zero-length array rather than null. Its fix pattern is replacing the null reference with a corresponding zero-length array. Developers ignored or rejected patches for this type of violations because they do have null-check to prevent these violations. 
If there is no null-check for these violations, the invocations of these methods would be identified as \texttt{\footnotesize NP\_NULL\_ON\_SOME\_PATH} violations. Thus, \texttt{\footnotesize PZLA\_} \texttt{\footnotesize PREFER\_ZERO\_LENGTH\_ARRAYS} might not be useful in practice.

\noindent
{\bf Insight 6.} Some fix patterns make programs fail to compile.
For example, the common fix pattern of \texttt{\footnotesize RV\_RETURN\_} \texttt{\footnotesize VALUE\_IGNORED\_BAD\_PRACTICE} violations is adding an {\tt if} statement to check the return {\tt boolean} value of the violated source code.
We note that return values of some violated source code of this violation type is not {\tt boolean} type. Copying the change behavior of the fix pattern directly to this kind of violations will lead to compilation errors.

\noindent
{\bf Insight 7.} Some fix patterns make programs fail to checkstyle. Figure~\ref{fig:finding7} presents an example of a patch generated by our method for a \texttt{\footnotesize MS\_SHOULD\_BE\_FINAL} violation in {\em XmlConverter.java} file of {\tt camel}~\footnote{\url{https://github.com/apache/camel}} project, which makes the project fail to checkstyle.

\begin{figure}[!tp]
    \centering
\begin{lstlisting}[linewidth={\linewidth},frame=tb,basicstyle=\footnotesize\ttfamily]
(@{\bf A Patch makes program fail to checkstyle:}@)
(@{\color{red}-public static String defaultCharset =}@)
(@{\color{codegreen}+public static final String defaultCharset = }@)
(@~~~~~ObjectHelper.getSystemProperty(@)
(@~~~~~~~~Exchange.DEFAULT\_CHARSET\_PROPERTY, "UTF-8");@)
\end{lstlisting}
    \caption{Example of a patch making program fail to checkstyle.}
    \label{fig:finding7}
\end{figure}

\noindent
{\bf Insight 8.} Some fix patterns of some violations are controversial. For example, the fix patterns of \texttt{\footnotesize DM\_NUMBER\_CTOR} violations are replacing the Number constructor with static Number \texttt{valueOf} method. It has been found that changing \texttt{new Integer()} to \texttt{Integer.valueOf()} and changing \texttt{Integer.valueOf()} to \texttt{new Integer()} were reverted repeatedly. Some developers find that \texttt{new Integer()} outperforms \texttt{Integer.valueOf()}, and some other developers find that \texttt{Integer.valueOf()} outperforms \texttt{new Integer()}. Additionally, some developers report that \texttt{Double.doubleToLongBits()} could be more efficient than \texttt{new Double()} and \texttt{Double.valueOf()} when comparing two double values with \texttt{equals()} method. We infer that the \texttt{\footnotesize DM\_NUMBER\_CTOR} or \texttt{\footnotesize DM\_FP\_NUMBER\_CTOR} violations should be identified and revised based on the specific function, otherwise, developers may be prone to ignoring these kinds of violations.
\section{Discussion}
\label{sec:discussion}

\subsection{Threats to validity}
A major threat to external validity of our study is the focus on \findbugs 
as the static analysis tool, with specific violation types and names.
Fortunately, the code problems described by \findbugs violations are similar to the violations described by other static analysis tools. For example, {\tt NP\_NULL\_ON\_SOME\_PATH} violations in FindBugs, {\tt Null dereference} violations in Facebook Infer, and {\tt ThrowNull} violations in Google ErrorProne are about the same issue: A NULL pointer is dereferenced and will lead to a NullPointerException when the code is executed. 
With the fix pattern of  {\tt NP\_NULL\_ON\_SOME\_PATH} of FindBugs mined in this study, we fixed 9 out of 10 different cases (each is from a distinct project in our subjects) of {\tt Null dereference} violations detected by Facebook Infer and 8 out of 10 different cases of {\tt ThrowNull} violations detected by Google ErrorProne, respectively. It shows the potential generalizability of the inferred fix patterns. 
We acknowledge, however, that there are some differences between \findbugs violations and other static analysis violations.  
Another threat to external validity of our study is that the fix patterns of violations are mined from open-source projects. 
Our findings might not applicable to industry projects that could have specific policies related to code quality.

Threats to internal validity include the limitations of the underlying tools used in this study (i.e., \findbugs and \gumtree).
\gumtree may produce unfeasible edit scripts. To reduce this threat, we have added extra labels into \gumtree. \findbugs may produce some violations with inaccurate positions. To reduce this threat, we re-locate and re-visit the violated source code with the bug descriptions of some violation types by \findbugs. \findbugs may yield high false positives. In order to reduce this threat, we focus on the common fixed violations in this study since common fixed violations are really concerned by developers. If the common fixed violations were addressed by common fix patterns, the common fixed violations are highly possible to be true positives and the common fix patterns are highly possible to be effective resolutions. These threats could be further reduced by developing more advanced tools.

Threats to internal validity also involve limitations in our method.
Violation tracking may produce false positive fixed violations. We combine the commit {\tt DiffEntry} and diffs parsed by \gumtree to reduce this threat.
Irrelevant code contexts can interfere with patterns mining. 
For example, one statement contains complex expressions, which may lead to a high number of irrelevant tokens. If this kind of violations were not filtered out in this study, it would increase the interference of noise. To reduce this validity, our study should be replicated in future work by extracting and analyzing the key violated source code with relevant code contexts identified using system dependency graphs.
In this study, we also find that some violations are replaced by method invocations which encapsulate the detailed source code changes of fixing the corresponding violations. The method we proposed extracts source code changes from source code changing positions of violations. It is challenging to extract source code changes from these kinds of fixed violations. In order to reduce this validity, we are planning to integrate static analysis technique into our method to get more detailed source code changes.

\subsection{Insights on unfixed violations}
Given the high proportion of violations that were found to remain unfixed in software projects, we investigate the potential reasons for this situation.
By comparing, in Section~\ref{sec:fixedViolations}, the code patterns of unfixed violations against those of fixed patterns, we note that they are commonly shared, suggesting that the reasons are not mainly due to the violation code characteristics. Instead, we can enumerate other implicit reasons
based on the observation of statistical data as well as the comments received during our live study to fix violations in ten open source projects. 

\begin{itemize}[leftmargin=*]
    \item Actually, many developers do not use \findbugs as part of their development tool chain. For example, we found that only 36\% of projects in our study include a commit mentioning \findbugs. Also, interestingly, in the cases of projects where we found that only 2\% (1,944/88,927) of fixed \findbugs violations explicitly refer to the \findbugs tool in commit messages.
    
    \item As a static analysis tool, \findbugs yields a significant number of false positives: i.e., violations that developers do not consider as being true violations. We indeed highlighted some code patterns of detected violations that they are inconsistent with the descriptions provided by \findbugs (cf. Section~\ref{sec:unfixedViolations}).
    
    \item Our interaction with developers helped us confirm that developers do not consider most \findbugs violations as being severe enough to deserve attention in their development process.
    
    \item Some violations identified by \findbugs might be controversial because we find that some fix patterns of some violations are controversial (cf. Insight 8 in Section~\ref{sec:liveStudy}).
    
    \item Finally, with our live study, we note that some developers may be willing to fix violations if they had in hand some fix patterns. Unfortunately, \findbugs only reports the violations, and does not provide in many cases any hint on how to deal with them. Our work is towards filling this gap systematically based on harvested knowledge from developer fixes.
    
\end{itemize}

% \subsection{Fix Patterns} 
% Practitioners and researchers can have different definitions of fix patterns.

\section{Related work}
\label{sec:relatedwork}

\subsection{Static analysis}
\paragraph*{\em Classification of Actionable and Unactionable Violations} Static analysis violations are studied and investigated from different aspects. Several studies attempted to classify actionable (likely to be true positive) and unactionable (false positive) violations by using machine learning techniques~\cite{heckman_model_2009,hanam_finding_2014,yoon_reducing_2014}. Classifying new and recurring alarms is necessary to prune identical alarms between subsequent releases. Hash code matching~\cite{spacco_tracking_2006} and coding pattern analysis~\cite{venkatasubramanyam_automated_2014} can be used for identifying recurring violations. 
Model checking techniques~\cite{darke_precise_2012,muske_efficient_2015} and constraint solvers~\cite{kim_filtering_2010,junker_smt-based_2012} can also verify true violations and prune false positive. 
As discussed in Section~\ref{sec:unfixedViolations}, trivial violations reported by FindBugs can be treated as false positives by developers, but they cannot be identified by previous work since they are negligible issues and too trivial to be addressed by developers. Investigating the violations recurrently addressed by developers like this study could reduce this threat to identify true positive violations. 

\paragraph*{\em Violation Prioritization} Violation prioritization can provide a ranked list so that developers focus on important ones first. Z-ranking~\cite{kremenek_zranking:_2003} prioritizes violations based on observations of real error trends. Jung et al. leveraged Bayesian statistics to rank violations~\cite{jung_taming_2005}. History-based prioritization~\cite{williams_automatic_2005,kim_prioritizing_2007,kim_which_2007} utilizes history of program changes to prioritize violations. In addition, several studies attempted to leverage user feedback to rank violations~\cite{kremenek_correlation_2004,heckman_establishing_2008,shen_efindbugs:_2011}. However, these works did not investigate violations with the big number of violations as our work, from multiple aspects as we done. Thus, our work can provide more reliable insights for violation ranking than these works.

\subsection{Change pattern mining}
\paragraph*{\em Empirical Studies on Change Patterns} Common change patterns are useful for various purposes. Pan et al.~\cite{pan_toward_2008} explored common bug fix patterns in Java programs to understand how developers change programs to fix a bug. Their fix patterns are, however, in a high-level schema (e.g. ``If-related: Addition of Post-condition Check (IF-APTC)'').
Based on the insight, PAR~\cite{kim_automatic_2013} leveraged common pre-defined fix patterns for automated program repair, that only contain six fix patterns which can only be used to fix a small number of bugs. 
Martinez and Monperrus further investigated repair models that can be utilized in program fixing while Zhong and Su~\cite{zhong_empirical_2015} conducted a large-scale study on bug fixing changes in open source projects. Tan et al.~\cite{tan_antipatterns_2016} analyzed anti-patterns that may interfere with the process of automated program repair. However, all of them studied code changes at the statement level, which is not as fine-grained as our work that extracts fine-grained code changes with an extended version of GumTree~\cite{falleri_fine_2014}.

\paragraph*{\em Pattern Mining for Code Change} SYDIT~\cite{meng2011systematic} and Lase~\cite{meng2013lase} generate code changes to other code snippets with the extracted edit scripts from examples in the same application. 
RASE~\cite{meng2015does} focuses on refactoring code clones with Lase edit scripts~\cite{meng2013lase}.  
FixMeUp~\cite{son2011rolecast} extracts and applies access control templates to protect sensitive operations. Their objectives are not to address issues caused by faulty code in program, such as the static analysis bugs studied in this study. 
REFAZER implements an algorithm for learning syntactic program transformations for C\# programs from examples~\cite{rolim2017learning} to correct defects in student submissions, which however are mostly useless across assignments~\cite{long2017automatic} and are not really defects in the wild as the violations in our study. Genesis~\cite{long2017automatic} heuristically infers application-independent code transform patterns from multiple applications to fix bugs, but its code transform patterns are tightly coupled with the nature and syntax of three kinds of bugs (i.e., null pointer, out of bounds, and class cast defects). 
Koyuncu et al.~\cite{koyuncu2018fixminer} have generalized this approach with FixMiner to mining fix patterns for all types of bugs given a large dataset. 
Our work tries to mine the common fix patterns for general static analysis violations which are not application-independent. 
Closely related to our work is the concurrent work of Reudismam et al.~\cite{rolim2018learning} who try to learn quick fixes by mining code changes to fix PMD violations~\cite{pmd}. Their approach aims at learning code change templates to be systematically applied to refactor code. 
Our approach can be used for a similar scenario, and scales to a huge variety of violation types.

\subsection{Bug datasets}
Several datasets of real-world bugs have been proposed in the literature
to evaluate approaches in software testing, software repair, and defect prediction approaches.
Do et al.~\cite{do_supporting_2005} have thus contributed to testing
techniques with a controlled experimentation platform. The associated dataset was added to the SIR database,
which provides a widely-used test bed for debugging and test suite
optimization.
Lu et al.~\cite{lu_bugbench_2005} and Cifuentes et al.~\cite{cifuentes_begbunch2009} have respectively proposed BugBench and BegBunch
as benchmarks for bug detection tools. Similarly, Dallmeier et al.~\cite{dallmeier_extraction_2007} have
proposed iBugs, a benchmark for bug
localization. Similarly to our process, their benchmark was obtained by extracting historical bug data.
Bug data can also be found in the PROMISE repository~\cite{promiseRepo} which includes a large variety of datasets for software engineering research. Le Goues
et al.~\cite{le_manybugs_2015} have designed the GenProg benchmark with C bugs.
Just et al.~\cite{just2014defects4j} have proposed Defects4J to evaluate software testing and repair approaches. 
Their dataset was collected from the recent history of five widely-used Java bugs, for which
they could include the associated test suites. 
To ensure the reliability of our experiments, we also collect subjects to identify violations and corresponding patches from real-world projects. The existing bug datasets focus on the bugs that make programs fail to pass some test case(s), but our data is about static analysis violations which may not fail to pass test cases.

\subsection{Program repair}

% This paragraph can be omitted.
Recent studies of program repair have presented several achievements. There are mainly two lines of research: (1) fully automated repair and (2) patch hint suggestion. The former focuses on automatically generating patches that can be integrated into a program without human intervention. The patch generation process often includes patch verification to figure out whether the patch does not break the original functionality when it is applied to the program. The verification is often achieved by running a given test suite. Automatize violation repair is included in our future work. The latter techniques suggest code fragments that can help create a patch rather than generating a patch ready to integrate. Developers may use the suggestions to write patches and verify them manually, that is similar to the patch generation of our work.

\paragraph*{\em Fully Automated Repair} Automated program repair is pioneered by GenProg~\cite{weimer_automatically_2009,le_genprog_2012}. This approach leverages genetic programming to create a patch for a given buggy program. It is followed by an acceptability study~\cite{fry_human_2012} and systematic evaluation~\cite{le_systematic_2012}. Regarding the acceptability issue, Kim et al.~\cite{kim_automatic_2013} advocated GenProg may generate nonsensical patches and proposed PAR to deal with the issue. PAR leverages human-written patches to define fix templates and can generate more acceptable patches. 
HDRepair~\cite{le_history_2016} leverages bug fixing history of many projects to provide better patch candidates to the random search process.
Recently, LSRrepair~\cite{kui2018live} proposes a live search approach to the ingredients of automated repair using code search techniques. 
While GenProg relies on randomness, utilizing program synthesis techniques~\cite{nguyen_semfix_2013,mechtaev_angelix_2016,mechtaev_directfix_2015} can directly generate patches even though they are limited to a certain subset of bugs.
Other notable approaches include contract-based fixing~\cite{wei_automated_2010}, program repair based on behavior models~\cite{dallmeier_generating_2009}, and conditional statement repair~\cite{xuan_nopol_2017}. 
This study does not focus on the fully automated program repair but the automated fix pattern mining for violations.

\paragraph*{\em Patch Hint Suggestion} Patch suggestion studies explored diverse dimensions. MintHint~\cite{kaleeswaran_minthint_2014} generates repair hints based on statistical analysis. Tao et al.~\cite{tao_automatically_2014} investigated how automatically generated patches can be used as debugging aids. 
Bissyand{\'e} suggests patches for bug reports based on the history of patches~\cite{bissyande2015harvesting}. 
Caramel~\cite{nistor_caramel_2015} focuses on potential performance defects and suggests specific types of patches to fix those defects. Our study is closely related to patch hint suggestion since we can suggest top-10 most similar fix patterns for targeting violations. The difference is that fix patterns in this work are mined from developers' patch submissions of static analysis violations.

\paragraph*{\em Empirical Studies on Program Repair} Many studies have explored properties of program repair. Monperrus~\cite{monperrus_critical_2014} criticized issues of patch generation learned from human-written patches~\cite{kim_automatic_2013}. Barr et al. discussed the plastic surgery hypothesis~\cite{barr_plastic_2014} that theoretically illustrates graftibility of bugs from a given program.
Long and Rinard analyzed the search space issues for population-based patch generation~\cite{long_analysis_2016}. Smith et al. presented an argument of overfitting issues of program repair techniques~\cite{smith_is_2015}. Koyuncu et al.~\cite{koyuncu_impact_2017} compared the impact of different patch generation techniques in Linux kernel development.
Benchmarks for program repair are proposed for different programming languages~\cite{le_manybugs_2015, just2014defects4j}. Based on a benchmark, a large-scale replication study was conducted~\cite{martinez_automatic_2016}.
More recently, Liu et al.~\cite{kui2018closer} investigated the distribution of code entities impacted by bug fixes with fine-grained granularity, and found that some static analysis tools (e.g., FindBugs~\cite{findbugs:description} and PMD~\cite{pmd}) are involved in some bug fixes.

%Our study results may provide 

\section{Conclusion}
\label{sec:conclusion}

In this study, we investigate recurrences of violations as well as their fixing changes, collected from open source Java projects.
% in terms of quantity, spread, and category respectively. 
%We note that a small number of violation types are responsible for the majority of recurrent violations and fixed violations in software projects.
%Additionally, there is no correlation between the spread of a violation type and its number of occurrences, and there is no correlation between the recurrences of all detected violation types and the recurrences of fixed violation types.
The yielded findings provide a number of insights into prioritization of violations for developers, as well as for researchers to improve violation reporting.

In this paper, we propose an approach to mine code patterns and fix patterns of static analysis violations by leveraging CNNs and {\em X-means}. The identified fix patterns are evaluated through three experiments. They are first applied to fixing many unfixed violations in our subjects. Second, we manage to get 67 of 116 generated patches accepted by the developer community and eventually merged into 10 open source Java projects.
Third, interestingly, the mined fix patterns were effective for addressing 4 real bugs in the Defects4J benchmark. 
%Finally, findings of our experiments provide useful insights for improving the state-of-the-art static analysis and automated program repair techniques.

As further work, we plan to combine fix pattern mining with automated program repair techniques to generate violation fixes more automatically. In the live study, we find that some common violations never occurred in latest versions of those projects. We postulate that violation recurrences may be time-varying. Our future work also includes studies on the time-variant issue of violation recurrences to further figure out the historic changes of fixed violations and the latest trend of violations, which may help new directions of violation prioritization.

{ %\balance
\bibliographystyle{IEEEtran}   %reference style
\bibliography{bib/violations,bib/repair,bib/change,bib/static,bib/tools}
}
% \vfill
\newpage

~
\newpage
\begin{appendices}

\section{Bug Fix Process}
\label{appendix:fixProcess}
A fix pattern is used as a guide to fix a bug, of which fixing process is defined as a \textit{bug fix process}.

\noindent
\begin{definition}{{\bf Bug Fix Process (FIX):}}
\label{definitionFIX}
A \textit{bug fix process} is a function of fixing a bug with a set of fix patterns.
\begin{equation}
FIX:(bc, FP^+) \rightarrow P^*
\end{equation}
\end{definition}

\noindent 
where \textit{bc} is the code block of a bug.
\textit{$FP^+$} means a set of fix patterns, and some of them could be applied to \textit{bc}. 
\textit{$P^*$} is a set of patches for \textit{bc}, which is generated by the bug fix process.
A \textit{bug fix process} is specified by a \textit{bug fix function} in this study.

\noindent
\begin{definition}{{\bf Bug Fix Function (FixF):}}
A \textit{bug fix function} consists of two domains %(i.e., a code block of a bug, a set of fix patterns)
and three sub functions. They can be formalized as:
% which are a function of converting buggy code into a code context, a function of matching the code context of the given buggy code block with fix patterns, and a function of repairing the bug with matched fix patterns which can generate a set of patch candidates for this bug.
\begin{equation}
FixF:(bc, FP^+) = CtxF + M + R \rightarrow P^*
\end{equation}
\begin{equation}
CtxF:bc \rightarrow Ctx_{bc}
\end{equation}
\begin{equation}
M: (Ctx_{bc}, Ctx_{fp} \in FP^+) \rightarrow FP^* % , \exists\,\, ccon_{bc}\leadsto ccon_{fp}
\end{equation}
\begin{equation}
R:(bc, Ctx_{bc}, CO \in FP^*) \rightarrow P^* %= CO\leadsto Ccon_{bc}\leadsto bc
\end{equation}
\end{definition}

\noindent
where \textit{bc} is the code block of a given bug and $FP^+$ is a set of fix patterns.
\textit{CtxF} denotes the function of converting buggy code into a code context (i.e., \textit{$Ctx_{bc}$}).
\textit{M} means the matching function of matching the code context of the given buggy code block with fix patterns to find appropriate fix patterns (i.e., \textit{$FP^*$, $FP^*$ $\subseteq$ $FP^+$}) for the bug, where $Ctx_{fp}$ is the code context of a fix pattern. If $FP^*=\emptyset$, it indicates that there is no fix pattern matched for the bug in the whole set of fix patterns.
\textit{R} represents the function of repairing the bug with change operations (i.e., \textit{$CO$}) in matched fix patterns.
%\textit{$\{O^+\}\leadsto ccon_{bc}\leadsto bc$} is a hierarchical modifying function of generating patch candidates ($P^*$).
If $P^*=\emptyset$, it indicates that there is no any patch which could be generated by the provided fix patterns and pass test cases of the bug.

% Figure~\ref{fig:fixFunction} shows an example of a bug fix process. The cases of $FP^*=\emptyset$ and $P^*=\emptyset$ are ignore in this graph.

% \begin{figure}[!tp]
%     \centering
%     \includegraphics[width=\columnwidth]{fig/FixProcess}
%     \caption{An example of a bug fix process. Solid arrow lines represent the control flow and data flow, and dotted arrow lines only represent the data flow.}
%     \label{fig:fixFunction}
% \end{figure}

\section{Statistics on Violations in the Wild}
\label{appendix:statsV}
In this section, we present the distributions of violations from three aspects of violations: number of occurrences, spread in projects and category. 
There are 16,918,530 distinct violations distributed throughout \nviolationtype types in our dataset. 
%Our conjecture is that {\bf only some violation types are recurrent in software projects}. 
We investigate which violation types are common by checking their recurrences in terms of quantity (i.e., how many times they occur overall) and  in terms of spread (i.e., in how many projects they occur). 
%The code and data are available at:% \url{https://github.com/FixPattern/static-analysis-violations}. 
%\begin{center}
%\url{https://github.com/FixPattern/static-analysis-violations}.
%\end{center}

\subsection*{Common types by number of occurrences.}
\begin{figure*}[!tp]
    \centering
    \includegraphics[width=\textwidth]{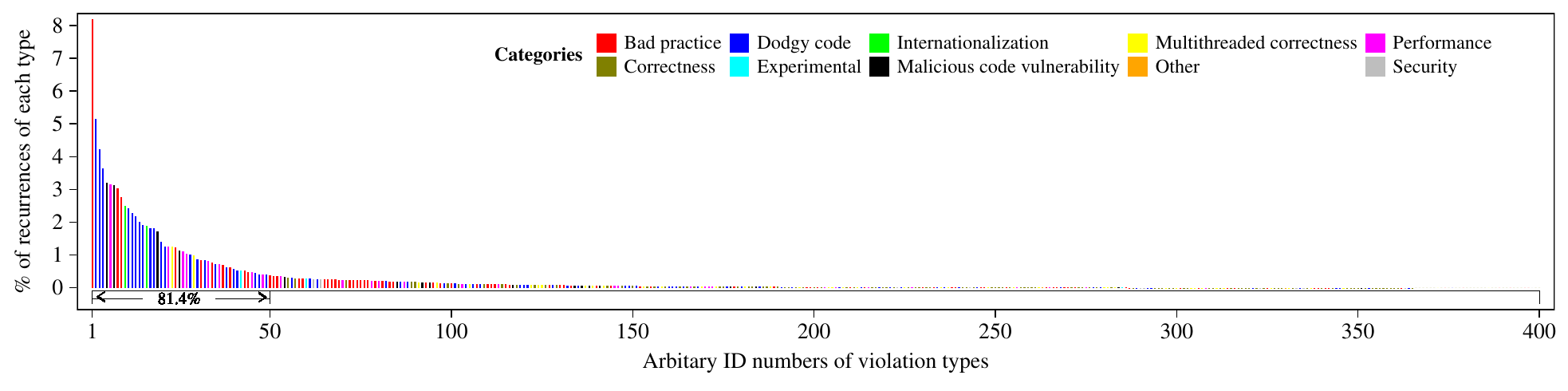}
    \caption[l]{Quantity distributions of violation types sorted by their occurrences. The x-axis represents arbitrary id numbers assigned to violation types. The y-axis represents the percentages of their occurrences in all violations.}
    \label{fig:qpvt}
\end{figure*}

Figure~\ref{fig:qpvt} shows the quantity distributions of all detected violation types. The x-axis represents arbitrary id numbers assigned to violation types following the number of times that occur in our dataset. 
The id mapping considered in this figure by sorting occurrences (i.e., id=1 corresponds to the most occurring violation type) will be used in the rest of this paper unless otherwise indicated. The {\em Order\_1} of Table~\ref{tab:comparisonVTvsFVT} presents the mapping of top 50 types. The whole mapping is available at the aforementioned website for interested readers.

It is noted from the obtained distribution that violation occurrences for the top 50 violation types account for 81.4\% of all violation occurrences. These types correspond only to about 12\% of \findbugs violation types. These statistics corroborate our conjecture that most violation instances are associated with a limited subset of violation types.

Figure~\ref{fig:qpvt} further highlights the category of each violation type according to the categorization by \findbugs. We note that all categories are represented among most and least occurring violations alike.

%\begin{tcolorbox}
%Only a small number of violation types are recurrent throughout real-world projects.
%\end{tcolorbox}

\subsection*{Common types by spread in projects}
\begin{figure*}[!tp]
    \centering
    \includegraphics[width=\textwidth]{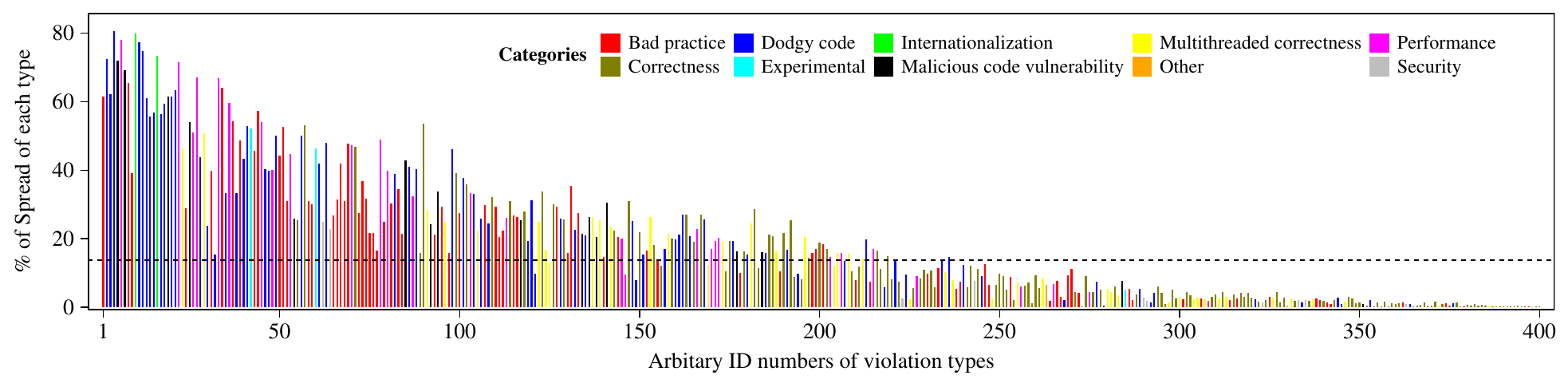}
    \caption[l]{Spread distributions of all violation types. The order of id numbers on x-axis is consistent with the order in Figure~\ref{fig:qpvt}. The y-axis represents the percentages of their spread in all projects.}
    \label{fig:wpvt}
\end{figure*}

Figure~\ref{fig:wpvt} illustrates to what extent the various violation types appear in projects. The id numbers for violation types are from the mapping produced before (i.e., as in Figure~\ref{fig:qpvt}). Almost 200 (50\%) violation types have been associated with over 100 (about 14\%) projects. It is further noted that there is no correlation between the spread of a violation type and its number of occurrences: some violation types among the most widespread types (e.g., top-50) actually occur less than some lesser widespread ones. Nevertheless, the data indicate that, together, the top-50 most widespread violations account also for the majority of violation instances.

%\begin{tcolorbox}
%Only a small number of violation types widely spread in a large number of real-world projects.
%There is no correlation between the recurrences in terms of occurred times of violations and numbers of their occurred projects.
%\end{tcolorbox}
%\tb{Maybe we skip Figure-\ref{fig:wpvt2}... text is enough}Figure~\ref{fig:wpvt2} illustrates the widespread distribution of all violation types which are sorted by their widespread decreasingly.
%The violations of the top 50 types occupy about 74.8\% of all violations. All of them occur in at least 323 (43.6\%) projects. It indicates that the violation types with high recurrences in terms of spread have high recurrences in quantity as well.

% Table~\ref{tab:topCommonVT} further details the differences between the top-50 violation types in terms of overall occurrence and the top-50 ones in terms of project spread. There are 12 most occurring violation types that are not among the top-50 most widespread violation types, and 2 of them are not even found in the top-100 most widespread, suggesting that a few projects may disregard specific programming rules. Similarly, we can identify 12 widespread violation types which are not among the top occurring, suggesting that
% some programming rules are transgressed in many projects but in limited code locations.

\subsection*{Category distributions of violations}

Table~\ref{tab:categories} provides the statistics on the categories
of violation types regrouped in the \findbugs documentation. The ranking of violation types is based on overall occurrences as in Figure~\ref{fig:qpvt}.
Category \texttt{\footnotesize Other} contains \texttt{\footnotesize SKIPPED\_CLASS\_TOO\_BIG} and \texttt{\footnotesize TESTING} that actually are not violation types defined in \findbugs. In the remainder of our experiments, instances of the two types are ignored.

\begin{figure*}[!tp]
    \centering
    \includegraphics[width=\textwidth]{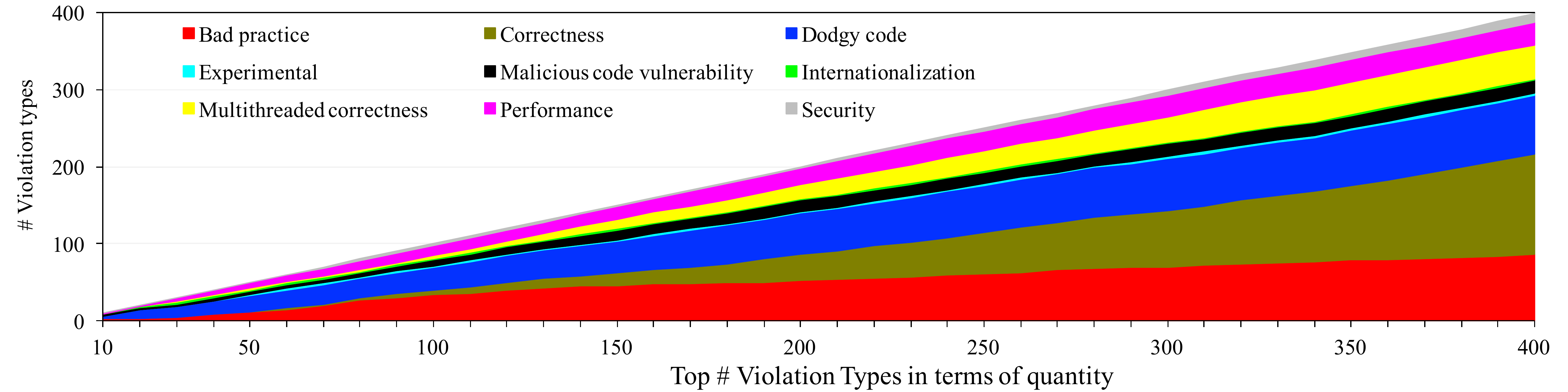}
    \caption[l]{Category distributions of violation types. The values on x-axis represent the threshold of violation types ranking used to carry out the category distributions, of which the order is consistent with the order in Figure~\ref{fig:qpvt}. For example, 50 means that top 50 violation types are used to carry out the category distributions.} %The value on {\em Y-axis} is the number of violation types of per-category.}
    \label{fig:categoriesDistribution}
\end{figure*}

\begin{table}[!tp]
    \centering
    \vspace{0.15cm}
    \setlength\tabcolsep{2pt}
    \caption{Category Distributions of Violations}
    \label{tab:categories}
    \resizebox{1.00\columnwidth}{!}{
        \begin{threeparttable}
            \begin{tabular}{l|r|ccc|c}
            \toprule
            \multirow{2}{*}{Category} & \multirow{2}{*}{\makecell[c]{\# Violation \\instances}} & \multicolumn{3}{c|}{\# Violation types}  & \multirow{2}{*}{\# Projects}
            \\\cline{3-5}
             & & top-50 & top-100 &  All & \\
            \hline
            Dodgy code	&	\textbf{6,736,692} & 22 & 29  & \textbf{75} & \textbf{703} \\ \midrule
            Bad practice	&	\textbf{4,467,817} & 11 & 34  & \textbf{86} & \textbf{696}\\ \midrule
            Performance	&	1,822,063 & 8 & 13 & 29 & \textbf{685} \\ \midrule
            \makecell[l]{Malicious code vulnerability} &	1,774,747 & 4 & 8  & 17 & \textbf{634}\\ \midrule
            Internationalization	&	740,392 & 2 & 2  & 2 & \textbf{632} \\ \midrule
            \makecell[l]{Multithreaded correctness} &	602,233 & 2 & 4 & 44 & 517 \\ \midrule
            Correctness	&	542,687 & 0 & 6 &  \textbf{131} & \textbf{636} \\ \midrule
            Experimental	&	135,559 & 1 & 2 & 3 & 446 \\ \midrule
            Security	&	95,258 & 0 & 2  & 11 & 219 \\ \midrule
            Other	&	1,082 & 0 & 0 &  2 & 51\\ 
            \bottomrule
            \end{tabular}
        \end{threeparttable}
    }
\end{table}

{\tt Dodgy code} and {\tt Bad practice} appear as the top two most common categories in terms of occurrence and spread. 
{\tt Security} violations are the least common, although they 
could be found in 30\% of the projects.

In terms of violation types, {\tt Correctness} regroups the largest number of types, but its types are not among the top occurring. 
Figure~\ref{fig:categoriesDistribution} illustrates the detailed distributions of categories. The number of violation types of {\tt Correctness} increases sharply from the ranking 100 to 400, while there is no correctness-related violation type in top 50 types. The violation types out of top 100 have much lower number of occurrences compared against top 100 types. Thus, {\tt Correctness} has a low number of overall occurrences, although it contains a large number of violation types and is seen in many projects. These findings suggest that developers commit few violations of these types.

Overall, {\tt Dodgy code} and {\tt Bad practice} are the top two most common categories.
{\tt Internationalization} is also found to be common since it contains only two violation types (i.e., \texttt{\footnotesize DM\_CONVERT\_CASE} and \texttt{\footnotesize DM\_DEFAULT\_ENCODING}) which are among top-20 most occurring violation types and among top-10 most widespread ones throughout projects. %For detailed definitions of categories and descriptions of violation types, please reference paper~\cite{root2009evaluation} and \findbugs Bug Descriptions~\cite{findbugs:description}.

%\begin{tcolorbox}
%{\tt Dodgy code}, {\tt Bad practice} and {\tt Internationalization} are found to be the common violation categories.
%\end{tcolorbox}

{\tt Dodgy code} represents either confusing, anomalous, or error-prone source code~\cite{root2009evaluation}.
Figure~\ref{fig:dodgyCodeExample} shows an example of a fixed {\tt Dodgy code} violation, 
%taken from {\em BytecodeDisassembler.java} file within Commit {\tt e2713c} in the {\tt ANTLR Stringtemplate4} project\footnote{\url{https://github.com/antlr/stringtemplate4}}. This
which is a fixed violation of \texttt{\footnotesize BC\_VACUOUS\_INSTANCEOF} type that denotes that the \texttt{instanceof} test would always return {\tt true}, unless the value being checked was \texttt{null}~\cite{root2009evaluation}. Although this is safe, make sure it is not an indication of some misunderstandings or some other logic errors. If the programmer really wants to check the value for being \texttt{null}, it would be clearer and better to do a \texttt{null} check rather than an \texttt{instanceof} test. Consequently, this violated instance is fixed by replacing the \texttt{instanceof} test with a \texttt{null} check.

\begin{figure}[!tp]
    \centering
    \vspace{0.3cm}
\begin{lstlisting}[linewidth={\linewidth},frame=tb,basicstyle=\footnotesize\ttfamily]
(@{\bf Violation Type:}@) BC_VACUOUS_INSTANCEOF.

(@{\bf Fixing DiffEntry:}@)
(@{\color{red}-  if (code.strings[poolIndex] instanceof String) \{}@)
(@{\color{codegreen}+  if (code.strings[poolIndex] != null) \{}@)
\end{lstlisting}
    \caption{Example of a fixed {\tt Dodgy code} violation taken from {\em BytecodeDisassembler.java} file within Commit {\tt e2713c} in project {\tt ANTLR Stringtemplate4}\protect\footnotemark{}.}
    \label{fig:dodgyCodeExample}
\end{figure}
\footnotetext{\url{https://github.com/antlr/stringtemplate4}}

\begin{figure}[!tp]
    \centering
\begin{lstlisting}[linewidth={\linewidth},frame=tb,basicstyle=\footnotesize\ttfamily]
(@{\bf Violation Type:}@) DM_CONVERT_CASE.

(@{\bf Fixing DiffEntry:}@)
(@{\color{red}-                cookieDomain = domain.toLowerCase();}@)
(@{\color{codegreen}+               cookieDomain = domain.toLowerCase(Locale.ENGLISH);}@)

\end{lstlisting}
    \caption{Example of a fixed {\tt Internationalization} violation taken from {\em BytecodeDisassembler.java} file within Commit {\tt 17bacf} in project {\tt Apache httpclient}\protect\footnotemark{}.} %apache-httpclient_f761a1_17bacfmodule-client#src#main#java#org#apache#http#impl#cookie#BasicClientCookie.java
    \label{fig:internationalizationEg}
\end{figure}
\footnotetext{\url{https://github.com/apache/httpclient}}

{\tt Bad practice} means that source code violates recommended coding practices~\cite{root2009evaluation}. The fixed violation in Figure~\ref{fig:expfixchange} is an example of a corrected {\tt Bad practice} violation.
It is not recommended to ellipsis an \texttt{instanceof} test when implementing an  \texttt{equals(Object o)} method, so that this violation is fixed by adding an \texttt{instanceof} test.

\texttt{Internationalization} denotes that source code uses non-localized method invocations~\cite{root2009evaluation}.
Figure~\ref{fig:internationalizationEg} presents an example of a fixed \texttt{Internationalization} violation, 
%taken from {\em BasicClientCookie.java} file within Commit \texttt{\footnotesize 17bacf} in \texttt{ Apache httpclient} project\footnote{\url{https://github.com/apache/httpclient}}.
which is a fixed \texttt{\footnotesize DM\_CONVERT\_} \texttt{\footnotesize CASE} violation that means that a String is being converted to upper or lower case by using the default encoding of the platform~\cite{findbugs:description}. This may result in improper conversions when used with international characters, therefore, this violation is fixed by adding a rule of \texttt{\footnotesize Locale.ENGLISH}.
For more definitions of categories and descriptions of violation types, please reference paper~\cite{root2009evaluation} and \findbugs Bug Descriptions~\cite{findbugs:description}.

Static analysis techniques are widely used in modern software projects\footnote{\url{http://findbugs.sourceforge.net/users.html}}. However, developers and researchers have no clear knowledge on the distributions of violations in the real world, especially for the fixed violations (See Section~\ref{sec:fixedViolations}). The empirical analysis can provide an overview of this knowledge from three different aspects: occurrences, spread and categories of violations, that can be used to rank violations for developers. The high false positives of \findbugs and the common non-severe violations could threaten the validity of the violation ranking. To reduce this threat, we further investigate the distributions of fixed violations in the next section. Fixed violations are resolved by developers, which means that they are detected with correct positions and are treated as issues being addressed, Thus they are likely to be true violations.

% \begin{figure*}[ht]
%     \centering
%     \includegraphics[width=\textwidth]{fig/empirical_study/fixed/Distribution-of-all-fixed-violation-types2.pdf}
%     \caption[l]{Quantity distributions of all fixed violation types sorted by their occurrences. The values on x-axis are the id numbers assigned to fixed violation types, which are different from the id numbers in Figure~\ref{fig:qpvt}. The values of y-axis are the percentages of their occurrences in all fixed violations.}
%     \label{fig:fixedQD}
% \end{figure*}
% \begin{figure*}[ht]
%     \centering
%     \includegraphics[width=\textwidth]{fig/empirical_study/fixed/Widespread-of-all-violation-types.pdf}
%     \caption[l]{Spread distributions of all fixed violation types. The x-axis is the same order as in Figure~\ref{fig:fixedQD}. The values of y-axis are percentages of all fixed types in projects.}
%     \label{fig:fixedWD}
% \end{figure*}

\section{Statistics on Fixed violations}
\label{appendix:statsFV}

This section presents the distributions of fixed violations with their recurrences in terms of quantity and in terms of spread. We further compare the distributions of fixed violations and detected ones.

\subsection*{Common types of fixed violations}

\begin{figure*}[ht]
    \centering
    \includegraphics[width=\textwidth]{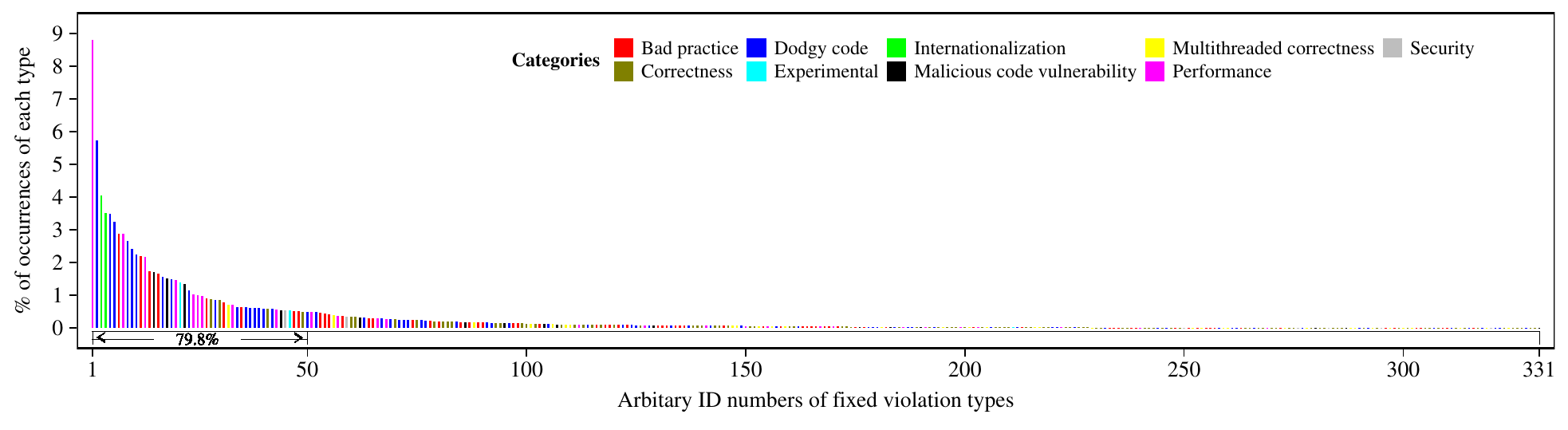}
    \caption[l]{Quantity distributions of all fixed violation types sorted by their occurrences. The values on x-axis are the id numbers assigned to fixed violation types, which are different from the id numbers in Figure~\ref{fig:qpvt}. The values of y-axis are the percentages of their occurrences in all fixed violations.}
    \label{fig:fixedQD}
\end{figure*}

Figure~\ref{fig:fixedQD} presents the distributions (in terms of quantity) of fixed violation types sorted by the number of their instances. 
Fixed violation instances of the top 50 (15\%) fixed types (presented by {\em Order\_2} in Table~\ref{tab:comparisonVTvsFVT}) account for about 80\% of all fixed violations. Additionally, 122 (about 37\%) types are represented in less than 20 instances, 91 (about 27\%) types are represented in less than 10 instances, and 20 (6\%) types are associated with a single fixed violation instance.
These data further suggest that only a few violation types are concerned by developers.

\begin{figure*}
\begin{center}
    \centering
    \includegraphics[width=\textwidth]{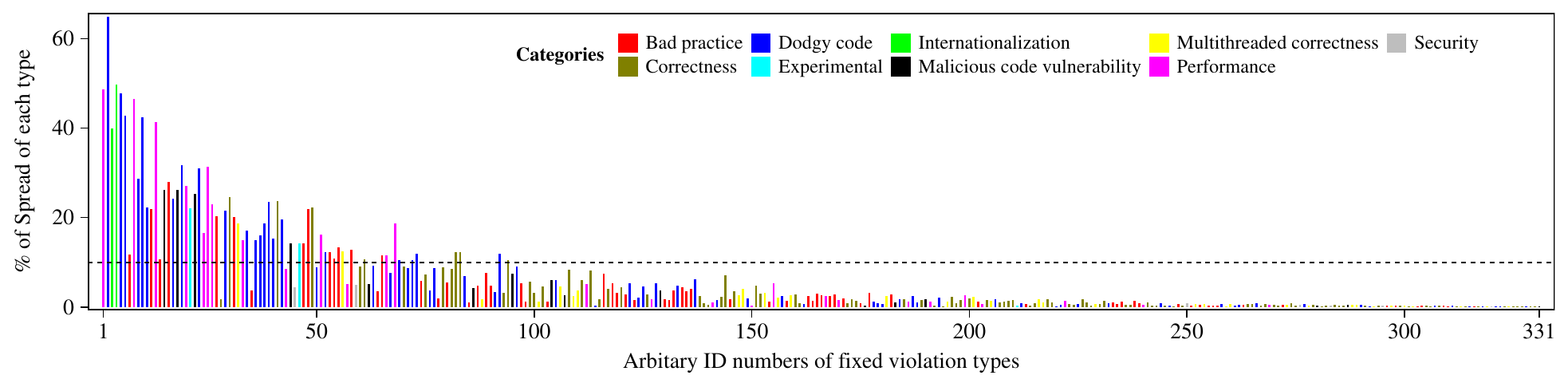}
    \caption[l]{Spread distributions of all fixed violation types. The x-axis is the same order as in Figure~\ref{fig:fixedQD}. The values of y-axis are percentages of all fixed types in projects.}
    \label{fig:fixedWD}
\end{center}
\end{figure*}

\begin{figure*}[!ht]
    \centering
    \includegraphics[width=\textwidth]{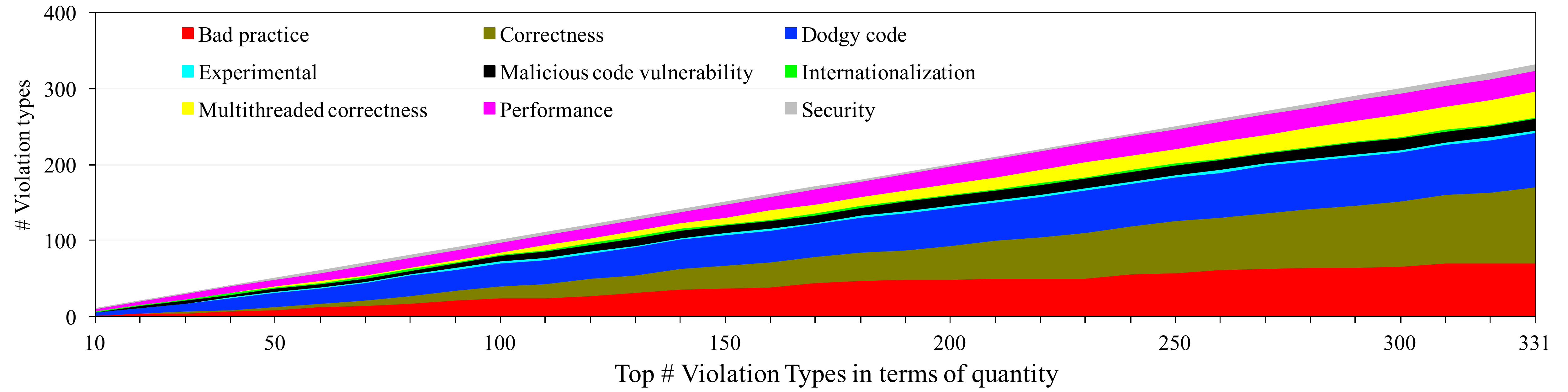}
    \caption[l]{Category distributions of fixed violation types. 
    The values on x-axis represent the threshold of fixed types ranking used to carry out the category distributions, of which the order is consistent with the order in Figure~\ref{fig:fixedQD}. For example, 50 means that top 50 violation types are used to carry out the category distributions.}
    \label{fig:fixedCategory}
\end{figure*}

Figure~\ref{fig:fixedWD} illustrates the appearance of violation types throughout software projects.
There is no correlation between the spread of a fixed violation type and its number of instances: some fixed violation types among the most spread actually occur less than some lesser spread ones. Nevertheless, the top-50 most spread violations account for the majority of fixed violation instances. Additionally, we note that 63 (19\%) fixed violation types occur in at least 10\% (55/547) projects, which further suggests that only a few violation types are concerned by developers.

%\begin{tcolorbox}
%A few violation types are really concerned by developers.
%\end{tcolorbox}
% \input{table/comparison2.tex}

\subsection*{Recurrences of types: fixed types VS. all detected ones}

 \begin{table*}[ht]
    \centering
    \setlength\tabcolsep{2pt}
    \caption{Comparison of distributions of fixed violation types against all detected violation types. {\em Order\_1} refers to the sorting order of violation types by the quantities of all detected violations (cf. the order in Figure~\ref{fig:qpvt}).  {\em Order\_2} refers
    to the sorting of violation types by the quantities of all fixed violations (cf., order in Figure~\ref{fig:fixedQD}). {\em Ratio\_1} represents the occurred ratio of a given violation type in the all detected violations. {\em Ratio\_2} represents the occurred ratio of a given fixed violation type in all fixed violations. {\em R1/R2} is used to measure the fluctuation ratio of a violation type from all detected violations to fixed violations, the same as R2/R1.}
    \label{tab:comparisonVTvsFVT}
    \resizebox{1.00\textwidth}{!}{
    
    \begin{threeparttable}
        \begin{tabular}{c|lccc|lccc}
        \toprule id &
Order\_1 (Top 50 all detected types) & \makecell[c]{Ratio\_1(\%)} & \makecell[c]{Ratio\_2(\%)} & R1/R2 & Order\_2 (Top 50 all fixed types)  & \makecell[c]{Ratio\_1(\%)} & \makecell[c]{Ratio\_2(\%)} & R2/R1 \\
\hline
1 & \tiny{SE\_NO\_SERIALVERSIONID} & \textbf{\color{red} 8.19} & \textbf{\color{red} 2.19} & \textbf{\color{red} 3.73} & \tiny{SIC\_INNER\_SHOULD\_BE\_STATIC\_ANON} & {\bf 3.15} & {\bf 8.81} & {\bf 2.80} \\
2 & \tiny{RCN\_REDUNDANT\_NULLCHECK\_OF\_NONNULL\_VALUE} & {\bf 5.15} & {\bf 3.24} & {\bf 1.59} & \tiny{DLS\_DEAD\_LOCAL\_STORE} & {\bf 3.64} & {\bf 5.74} & {\bf 1.58} \\
3 & \tiny{BC\_UNCONFIRMED\_CAST} & {\bf 4.24} & {\bf 2.65} & {\bf 1.60} & \tiny{DM\_CONVERT\_CASE} & {\bf 1.89} & {\bf 4.04} & {\bf 2.14} \\
4 & \tiny{DLS\_DEAD\_LOCAL\_STORE} & 3.64 & 5.74 & 0.63 & \tiny{DM\_DEFAULT\_ENCODING} & 2.49 & 3.52 & 1.41 \\
5 & \tiny{EI\_EXPOSE\_REP} & {\bf 3.20} & {\bf 1.35} & {\bf 2.36} & \tiny{UWF\_FIELD\_NOT\_INITIALIZED\_IN\_CONSTRUCTOR} & {\bf 2.29} & {\bf 3.49} & {\bf 1.53} \\
6 & \tiny{SIC\_INNER\_SHOULD\_BE\_STATIC\_ANON} & 3.15 & 8.81 & 0.36 & \tiny{RCN\_REDUNDANT\_NULLCHECK\_OF\_NONNULL\_VALUE} & 5.15 & 3.24 & 0.63 \\
7 & \tiny{EI\_EXPOSE\_REP2} & {\bf 3.14} & {\bf 1.50} & {\bf 2.09} & \tiny{NM\_METHOD\_NAMING\_CONVENTION} & 2.77 & 2.89 & \textbf{\color{green} 1.04} \\
8 & \textbf{\tiny{SE\_BAD\_FIELD}} & \textbf{\color{red}3.04} & \textbf{\color{red}0.29} & \textbf{\color{red}10.36} & \tiny{URF\_UNREAD\_FIELD} & {\bf 1.26} & {\bf 2.88} & {\bf 2.29} \\
9 & \tiny{NM\_METHOD\_NAMING\_CONVENTION} & 2.77 & 2.89 & \textbf{\color{green} 0.96} & \tiny{BC\_UNCONFIRMED\_CAST} & 4.24 & 2.65 & 0.63 \\
10 & \tiny{DM\_DEFAULT\_ENCODING} & 2.49 & 3.52 & 0.71 & \tiny{REC\_CATCH\_EXCEPTION} & 2.43 & 2.42 & \textbf{\color{green} 1.00} \\
11 & \tiny{REC\_CATCH\_EXCEPTION} & 2.43 & 2.42 & \textbf{\color{green} 1.00} & \tiny{BC\_UNCONFIRMED\_CAST\_OF\_RETURN\_VALUE} & 2.02 & 2.25 & \textbf{\color{green} 1.12} \\
12 & \tiny{UWF\_FIELD\_NOT\_INITIALIZED\_IN\_CONSTRUCTOR} & 2.29 & 3.49 & 0.66 & \tiny{SE\_NO\_SERIALVERSIONID} & 8.19 & 2.19 & 0.27 \\
13 & \tiny{PZLA\_PREFER\_ZERO\_LENGTH\_ARRAYS} & \textbf{\color{red}2.19} & \textbf{\color{red}0.63} & \textbf{\color{red} 3.46} & \tiny{UPM\_UNCALLED\_PRIVATE\_METHOD} & {\bf 1.03} & {\bf 2.16} & {\bf 2.10} \\
14 & \tiny{BC\_UNCONFIRMED\_CAST\_OF\_RETURN\_VALUE} & 2.02 & 2.25 & \textbf{\color{green} 0.90} & \textbf{\tiny{VA\_FORMAT\_STRING\_USES\_NEWLINE}} & \textbf{\color{red} 0.25} & \textbf{\color{red} 1.72} & \textbf{\color{red} 6.90} \\
15 & \tiny{RI\_REDUNDANT\_INTERFACES} & \textbf{\color{red}1.91} & \textbf{\color{red}0.62} & \textbf{\color{red} 3.10} & \tiny{MS\_SHOULD\_BE\_FINAL} & 1.72 & 1.71 & \textbf{\color{green} 0.99} \\
16 & \tiny{DM\_CONVERT\_CASE} & 1.89 & 4.04 & 0.47 & \tiny{RV\_RETURN\_VALUE\_IGNORED\_BAD\_PRACTICE} & 0.78 & 1.67 & {\bf 2.15} \\
17 & \tiny{ST\_WRITE\_TO\_STATIC\_FROM\_INSTANCE\_METHOD} & 1.82 & 1.57 & \textbf{\color{green} 1.16} & \tiny{ST\_WRITE\_TO\_STATIC\_FROM\_INSTANCE\_METHOD} & 1.82 & 1.57 & \textbf{\color{green} 0.87} \\
18 & \tiny{SF\_SWITCH\_NO\_DEFAULT} & {\bf 1.81} & {\bf 0.86} & {\bf 2.10} & \tiny{EI\_EXPOSE\_REP2} & 3.14 & 1.50 & 0.48 \\
19 & \tiny{MS\_SHOULD\_BE\_FINAL} & 1.72 & 1.71 & \textbf{\color{green} 1.01} & \tiny{URF\_UNREAD\_PUBLIC\_OR\_PROTECTED\_FIELD} & 1.26 & 1.49 & \textbf{\color{green} 1.18} \\
20 & \tiny{NP\_LOAD\_OF\_KNOWN\_NULL\_VALUE} & 1.40 & 1.14 & 1.22 & \tiny{WMI\_WRONG\_MAP\_ITERATOR} & {\bf 0.72} & {\bf 1.47} & {\bf 2.04} \\
21 & \tiny{URF\_UNREAD\_PUBLIC\_OR\_PROTECTED\_FIELD} & 1.26 & 1.49 & \textbf{\color{green} 0.85} & \tiny{OBL\_UNSATISFIED\_OBLIGATION} & {\bf 0.52} & {\bf 1.40} & {\bf 2.69} \\
22 & \tiny{URF\_UNREAD\_FIELD} & 1.26 & 2.88 & 0.44 & \tiny{EI\_EXPOSE\_REP} & 3.20 & 1.35 & 0.42 \\
23  & \textbf{\tiny{LI\_LAZY\_INIT\_STATIC}} & \textbf{\color{red} 1.25} & \textbf{\color{red} 0.39} & \textbf{\color{red} 3.20} & \tiny{NP\_LOAD\_OF\_KNOWN\_NULL\_VALUE} & 1.40 & 1.14 & \textbf{\color{green} 0.82} \\
24  & \textbf{\tiny{NM\_CLASS\_NAMING\_CONVENTION}} & \textbf{\color{red} 1.23} & \textbf{\color{red} 0.10} & \textbf{\color{red} 12.89} & \tiny{DM\_NUMBER\_CTOR} & 1.10 & 1.04 & \textbf{\color{green} 0.94} \\
25 & \tiny{MS\_PKGPROTECT} & {\bf 1.14} & {\bf 0.55} & {\bf 2.07} & \tiny{SIC\_INNER\_SHOULD\_BE\_STATIC} & 0.83 & 1.00 & 1.22 \\
26 & \tiny{DM\_NUMBER\_CTOR} & 1.10 & 1.04 & \textbf{\color{green} 1.07} & \tiny{SBSC\_USE\_STRINGBUFFER\_CONCATENATION} & {\bf 0.48} & {\bf 0.97} & {\bf 2.03} \\
27 & \tiny{UPM\_UNCALLED\_PRIVATE\_METHOD} & 1.03 & 2.16 & 0.48 & \tiny{OS\_OPEN\_STREAM\_EXCEPTION\_PATH} & {\bf 0.49} & {\bf 0.91} & {\bf 1.87} \\
28 & \textbf{\tiny{FE\_FLOATING\_POINT\_EQUALITY}} & {\bf 1.02} & {\bf 0.48} & {\bf 2.14} & \textbf{\tiny{NP\_NONNULL\_RETURN\_VIOLATION}} & \textbf{\color{red} 0.04} & \textbf{\color{red} 0.87} & \textbf{\color{red} 23.00} \\
29 & \tiny{IS2\_INCONSISTENT\_SYNC} & 0.99 & 0.71 & 1.38 & \tiny{SF\_SWITCH\_NO\_DEFAULT} & 1.81 & 0.86 & 0.48 \\
30  & \makecell[l]{\tiny{NP\_PARAMETER\_MUST\_BE\_NONNULL}\tiny{\_BUT\_MARKED\_AS\_NULLABLE}} & {\bf 0.87} & {\bf 0.50} & {\bf 1.75} & \textbf{\tiny{UWF\_UNWRITTEN\_FIELD}} & \textbf{\color{red} 0.18} & \textbf{\color{red} 0.84} & \textbf{\color{red} 4.73} \\%\hspace{0.3cm}
31 & \textbf{\tiny{SE\_TRANSIENT\_FIELD\_NOT\_RESTORED}} & \textbf{\color{red} 0.85} & \textbf{\color{red} 0.04} & \textbf{\color{red} 22.95} & \tiny{DE\_MIGHT\_IGNORE} & 0.71 & 0.79 & \textbf{\color{green} 1.11} \\
32 & \textbf{\tiny{NP\_METHOD\_PARAMETER\_TIGHTENS\_ANNOTATION}} & \textbf{\color{red} 0.83} & \textbf{\color{red} 0.04} & \textbf{\color{red} 23.16} & \tiny{IS2\_INCONSISTENT\_SYNC} & 0.99 & 0.71 & 0.72 \\
33 & \tiny{SIC\_INNER\_SHOULD\_BE\_STATIC} & 0.83 & 1.00 & \textbf{\color{green} 0.83} & \textbf{\tiny{DM\_BOXED\_PRIMITIVE\_FOR\_PARSING}} & {\bf 0.24} & {\bf 0.71} & {\bf 2.90} \\
34 & \tiny{RV\_RETURN\_VALUE\_IGNORED\_BAD\_PRACTICE} & 0.78 & 1.67 & 0.47 & \textbf{\tiny{RV\_RETURN\_VALUE\_IGNORED\_NO\_SIDE\_EFFECT}} & {\bf 0.27} & {\bf 0.64} & {\bf 2.42} \\
35 & \textbf{\tiny{DLS\_DEAD\_LOCAL\_STORE\_OF\_NULL}} & \textbf{\color{red} 0.72} & \textbf{\color{red} 0.19} & \textbf{\color{red} 3.85} & \textbf{\tiny{ODR\_OPEN\_DATABASE\_RESOURCE}} & {\bf 0.22} & {\bf 0.64} & {\bf 2.86} \\
36 & \tiny{WMI\_WRONG\_MAP\_ITERATOR} & 0.72 & 1.47 & 0.49 & \tiny{PZLA\_PREFER\_ZERO\_LENGTH\_ARRAYS} & 2.19 & 0.63 & 0.29 \\
37 & \tiny{DE\_MIGHT\_IGNORE} & 0.71 & 0.79 & \textbf{\color{green} 0.90} & \tiny{RI\_REDUNDANT\_INTERFACES} & 1.91 & 0.62 & 0.32 \\
38 & \textbf{\tiny{CI\_CONFUSED\_INHERITANCE}} & {\bf 0.62} & {\bf 0.23} & {\bf 2.67} & \textbf{\tiny{NP\_NULL\_ON\_SOME\_PATH\_FROM\_RETURN\_VALUE}} & {\bf 0.31} & {\bf 0.61} & {\bf 1.99} \\
39 & \textbf{\tiny{NM\_CONFUSING}} & 0.61 & 0.43 & 1.44 & \tiny{UCF\_USELESS\_CONTROL\_FLOW} & 0.53 & 0.61 & \textbf{\color{green} 1.15} \\
40 & \textbf{\tiny{EQ\_DOESNT\_OVERRIDE\_EQUALS}} & \textbf{\color{red} 0.56} & \textbf{\color{red} 0.08} & \textbf{\color{red} 6.96} & \tiny{UC\_USELESS\_CONDITION} & 0.39 & 0.59 & 1.49 \\
41 & \tiny{UCF\_USELESS\_CONTROL\_FLOW} & 0.53 & 0.61 & \textbf{\color{green} 0.87} & \textbf{\tiny{NP\_NULL\_ON\_SOME\_PATH}} & {\bf 0.29} & {\bf 0.59} & {\bf 2.07} \\
42 & \tiny{OBL\_UNSATISFIED\_OBLIGATION} & 0.52 & 1.40 & 0.37 & \textbf{\tiny{UC\_USELESS\_OBJECT}} & \textbf{\color{red} 0.15} & \textbf{\color{red} 0.58} & \textbf{\color{red} 3.96} \\
43 & \tiny{ES\_COMPARING\_STRINGS\_WITH\_EQ} & 0.52 & 0.51 & \textbf{\color{green} 1.00} & \tiny{DM\_FP\_NUMBER\_CTOR} & 0.40 & 0.57 & 1.43 \\
44 & \tiny{OS\_OPEN\_STREAM\_EXCEPTION\_PATH} & 0.49 & 0.91 & 0.53 & \tiny{MS\_PKGPROTECT} & 1.14 & 0.55 & 0.48 \\
45 & \tiny{SBSC\_USE\_STRINGBUFFER\_CONCATENATION} & 0.48 & 0.97 & 0.49 & \makecell[l]{\textbf{\tiny{SQL\_PREPARED\_STATEMENT\_GENERATED}}\tiny{\bf\_FROM\_NONCONSTANT\_STRING}} & {\bf 0.27} & {\bf 0.55} & {\bf 2.07} \\% \hspace{0.4cm}
46 & \textbf{\tiny{SF\_SWITCH\_FALLTHROUGH}} & {\bf 0.44} & {\bf 0.15} & {\bf 2.95} & \textbf{\tiny{OBL\_UNSATISFIED\_OBLIGATION\_EXCEPTION\_EDGE}} & {\bf 0.28} & {\bf 0.55} & {\bf 1.96} \\
47 & \textbf{\tiny{RCN\_REDUNDANT\_NULLCHECK\_OF\_NULL\_VALUE}} & {\bf 0.40} & {\bf 0.24} & {\bf 1.68} & \tiny{ES\_COMPARING\_STRINGS\_WITH\_EQ} & 0.52 & 0.51 & \textbf{\color{green} 1.00} \\
48 & \tiny{DM\_FP\_NUMBER\_CTOR} & 0.40 & 0.57 & 0.70 & \textbf{\tiny{OS\_OPEN\_STREAM}} & 0.36 & 0.51 & 1.40 \\
49 & \tiny{UC\_USELESS\_CONDITION} & 0.39 & 0.59 & 0.67 & \textbf{\tiny{RCN\_REDUNDANT\_NULLCHECK\_WOULD\_HAVE\_BEEN\_A\_NPE}} & {\bf 0.24} & {\bf 0.50} & {\bf 2.08} \\
50 & \textbf{\tiny{HE\_EQUALS\_USE\_HASHCODE}} & 0.39 & 0.47 & \textbf{\color{green} 0.82} & \tiny{NP\_PARAMETER\_MUST\_BE\_NONNULL\_BUT\_MARKED\_AS\_NULLABLE} & 0.87 & 0.50 & 0.57 \\
        \bottomrule
        \end{tabular}
    \end{threeparttable}
    }
\end{table*}

Table~\ref{tab:comparisonVTvsFVT} provides comparison data on the occurrence ratios of fixed violation types against detected violation types. We consider two rankings based on the occurred quantities for all detected violations and for only fixed violations respectively, and select top-50 violation types in each ranking for comparison.
If the value of R1/R2 or R2/R1 is close to 1, it means that the violation type has a similar ratio in both fixed instances and detected ones. We refer to this value as {\em Fluctuation Ratio} (hereafter FR).

In the left side of Table~\ref{tab:comparisonVTvsFVT}, there are 12 violation types marked in \textcolor{green}{green}, for which FR values range between 0.80 and 1.20. We consider in such cases that the occurrences are comparable across all violations and fixed violations instances. These 12 violation types have one more type than the types marked in \textcolor{green}{green} in the right side because the last type in the left side is not in the top 50 of the right side.
On the other hand, FR values of 21 violation types are over 1.5, 10 of them are over 3.0, and 4 of them are even over 10: these numbers suggest that the relevant violation types with high recurrences do not appear to have high priorities of being fixed.
Combining FR values and Ratio\_2 values, one can infer that developers make a few efforts to fix violation instances for types  \texttt{\footnotesize SE\_BAD\_FIELD}, \texttt{\footnotesize NM\_CLASS\_NAMING\_CONVENTION}, \texttt{\footnotesize SE\_TRANSIENT\_FIELD\_NOT\_RE} \texttt{\footnotesize STORED},
\texttt{\footnotesize NP\_METHOD\_PARAMETER\_TIGHTENS\_ANNOTATION} or
\texttt{\footnotesize EQ\_} \texttt{\footnotesize DOESNT\_OVERRIDE\_EQUALS}.

In the right side of Table~\ref{tab:comparisonVTvsFVT}, %there are 11 violation types marked in \textcolor{green}{green}, for which FR values range between 0.80 and 1.20. We consider in such cases that the occurrences are comparable across all violations and fixed violations instances. On the other hand, 
FR values of 23 violation types are over 1.5, 4 of them are over 3.0, and one of them is even over 20: these numbers suggest that the relevant violation types with low recurrences do appear to have high priorities of being fixed. Combining FR values and  Ratio\_2 values, which can infer that developers ensure that violations of type \texttt{\footnotesize NP\_NONNULL\_RETURN\_VIOLATION} are fixed with higher priority than others.
Additionally, 13 violation types marked in {\bf bold} in the right side are in the top 50 ranking of fixed violations but not in the top 50 ranking of all detected violations, and vice versa to the types marked in {\bf bold} in the left side of this table.

To sum up, these findings suggest that fixed violation types have different recurrences compared against detected violation types. The order of fixed violation types and the FR values of fixed violation types can provide better criteria to help prioritize violations than the order of all detected violation types, since fixed violations are concerned and resolved by developers. 

%\begin{tcolorbox}
%(Comparing with other similar research work,) The statistic data can provide new and different metrics for prioritizing violations.
%\end{tcolorbox}

\subsection*{Category distributions of fixed violations}

\begin{table}[tp]
    \centering
    \setlength\tabcolsep{2pt}
    \caption{Category distributions of fixed violations.}
    \label{tab:fixedcategories}
    \resizebox{1.0\columnwidth}{!}{
    \begin{threeparttable}
        \begin{tabular}{l|r|ccc|c}
        \toprule
        % Category & \# Violations & \# Violation types & \# Projects\\
        \multirow{2}{*}{Category} & \multirow{2}{*}{\makecell[c]{\# Violation \\instances}} & \multicolumn{3}{c|}{\# Violation types}  & \multirow{2}{*}{\# Projects}
        \\\cline{3-5}
         & & top-50 & top-100 &  All & \\
        \hline
        Dodgy code	&	30,419 & 18 & 31 & 72 & 505 \\
        Performance	&	19,248 & 9 & 13 & 27 & 450 \\
        Bad practice	&	15,640 & 9 & 24 & 71 & 419 \\
        Correctness	&	6,809 & 5 & 16 & 99 & 384 \\
        Internationalization	& 6,719 & 2 & 2 & 2 & 347 \\
        Malicious code vulnerability & 5,505 & 3 & 7 & 16 & 299 \\
        Multithreaded correctness	&	2,018 & 1 & 3 & 34 & 208 \\
        Experimental	& 1,748 & 2 & 2 & 3 & 162 \\
        Security	&	821 & 1 & 2 & 7 & 47 \\
        \bottomrule
        \end{tabular}
    \end{threeparttable}
    }
\end{table}

Table~\ref{tab:fixedcategories} presents the category distributions of fixed violations.
{\tt Dodgy code} is the most common fixed category, and the following two secondary common fixed categories are {\tt Performance} and {\tt Bad practice} in terms of occurrences and spread. Fixed {\tt Security} violations are the least common, although they are found in 10\% of the projects with fixed violations.

In terms of violation types, {\tt Correctness} regroups the largest number of fixed violation types, but they are not among the top occurring. Figure~\ref{fig:fixedCategory} illustrates the detailed distributions of categories. The number of violation types of {\tt Correctness} increases sharply from the ranking 50 to 331, and there are a few correctness-related violation types in the top 50 types. However, the violation types out of top 50 have much lower number of occurrences compared to top 50 types. Therefore, {\tt Correctness} has a low number of overall fixed occurrences, although it contains the largest number of fixed violation types and is seen in many projects. 
The top 50 types are mainly occupied by \texttt{Dodgy code}, \texttt{Performance} and \texttt{Bad practice} categories.

Category {\tt Performance} represents the inefficient memory usage or buffer allocation, or usage of non-static class~\cite{root2009evaluation}. Figure~\ref{fig:performanceExample} presents an example of a fixed {\tt Performance} violation. It is a \texttt{\footnotesize SBSC\_USE\_STRINGBUFFER} \texttt{\footnotesize\_CONCATENATION} violation which denotes that concatenating strings using the {\tt +} operator in a loop~\cite{findbugs:description}. In each iteration, the {\tt String} is converted to a {\tt StringBuffer}  or {\tt StringBuilder}, appended to, and converted back to a {\tt String}, which can lead to a cost quadratic in the number of iterations, as the growing string is recopied in each iteration.

\begin{figure}[!tp]
    \centering
\begin{lstlisting}[linewidth={\linewidth},frame=tb,basicstyle=\footnotesize\ttfamily]
(@{\bf Violation Type:}@) SBSC_USE_STRINGBUFFER_CONCATENATION.

(@{\bf Fixing DiffEntry:}@)
(@{\color{red}-~~~~~String colorStr = "";}@)
(@{\color{codegreen}+~~~~~StringBuilder sb = new StringBuilder();}@)
        for (float f : color) {
(@{\color{red}-~~~~~~~~if (!colorStr.isEmpty())\{ }@)
(@{\color{codegreen}+~~~~~~~~if (sb.length() > 0)\{ }@)
(@{\color{red}-~~~~~~~~~~~colorStr += " "; }@)
(@{\color{codegreen}+~~~~~~~~~~~sb.append(' '); }@)
            }
(@{\color{red}-~~~~~~~~colorStr += String.format("\%3.2f", f); }@)
(@{\color{codegreen}+~~~~~~~~sb.append(String.format("\%3.2f", f)); }@)
        }
\end{lstlisting}
    \caption{Example of a fixed {\tt Performance} violation, taken from {\em Vertex.java} file within Commit \texttt{36e820} in \texttt{Apache pdfbox}\protect\footnotemark{} project.}
    \label{fig:performanceExample}
\end{figure}
~\footnotetext{\url{https://github.com/apache/pdfbox}}

{\tt Internationalization} is also found to be a common fixed category since it has 6,719 fixed violation instances taken from 347 (63.3\%) projects and contains only two violation types (i.e., \texttt{\footnotesize DM\_CONVERT\_CASE} and \texttt{\footnotesize DM\_DEFAULT\_ENCODING}) that are among top-5 most occurring violation types and among top-10 most widespread throughout projects.

To sum up, these findings suggest that developers may prefer to take more efforts on fixing violations of the four categories, i.e., {\tt Dodgy code}, {\tt Performance}, {\tt Bad practice} and {\tt Internationalization}, than others.

%\begin{tcolorbox}
%{\tt Dodgy code}, {\tt Performance}, {\tt Bad practice} and {\tt Internationalization} violations attract more attention from developers than others.
%\end{tcolorbox}

% \vspace{\baselineskip}
% \noindent
\subsection*{Category distributions: fixed Violations VS. all detected ones}

\begin{table}[!tp]
    \centering
    \setlength\tabcolsep{2pt}
    \caption{Comparison of category distributions of all detected violations and fixed violations.}
    \label{tab:comparisonOfCategories}
    \resizebox{1.0\columnwidth}{!}{
    \begin{threeparttable}
        \begin{tabular}{l|c|c|c}
        \toprule
        \multirow{2}{*}{Category} & \multicolumn{2}{c|}{\% Violation Occurrences in} & \multirow{2}{*}{\makecell[c]{FR values\\(F/D)}}\\\cline{2-3}
         & All detected (D) & Fixed (F) & \\
        \hline
        Experimental & 0.8 & 1.97 & 2.46\\
        Correctness	& 3.21  & 7.66 & 2.37\\
        Performance	& 10.77 & 21.64 & 2.01 \\
        Internationalization & 4.38 & 7.56 & 1.73\\
        Security & 0.56 & 0.92 & 1.70\\
        Dodgy code	& 39.82 & 34.21 & 0.86\\
        Bad practice & 26.41 & 17.59 & 0.67\\
        Multithreaded correctness & 3.56 & 2.27 & 0.64\\
        Malicious code vulnerability & 10.49 & 6.19 & 0.59\\
        \bottomrule
        \end{tabular}
    \end{threeparttable}
    }
\end{table}

Table~\ref{tab:comparisonOfCategories} shows the comparing results of category distributions of fixed violations against all detected ones. 
Overall, the ratios of top-5 categories occurrences in fixed violations have increases compared against their ratios in all detected ones. Particularly, the top-3 categories have great increases (more than one fold).

The ratio of {\tt Performance} occurrence in fixed violations has a great increase of $11\%$ compared against its ratio in all ones, which can suggest that developers take many efforts to fix {\tt Performance} violations. 
The ratio of {\tt Internationalization} occurrence in fixed violations also has a great increase compared against its ratio in all detected ones. And the {\tt Internationalization} contains only two violation types that have high rankings in quantity and spread distributions respectively. So that, it implies that developers take many efforts to fix {\tt Internationalization} violations as well. 
Even though {\tt Correctness} {\tt Experimental} and {\tt Security} occurrences in fixed violations and all detected ones do not present good rankings, their occurrence ratios in fixed violations have great increases compared against their ratios in all detected ones.
% \sout{The ratio of {\tt Internationalization} occurrence in fixed violations has a great increase compared against its ratio in all detected ones. And the {\tt Internationalization} contains only two violation types that have high rankings in quantity and spread distributions respectively. So that, it also corroborates that developers take many efforts to fix {\tt Internationalization} violations.}
% \sout{The ratios of {\tt Correctness} {\tt Experimental} and {\tt Security} occurrences in fixed violations have great increases compared against their ratios in all detected ones, but their occurrences in both fixed violations and all detected ones do not present good rankings. }
%Even though the ratios of {\tt Dodgy code} and {\tt Bad practice} occurrences in fixed violations have some decreases compared with their ratios in all detected ones, they occupy the main proportion in fixed violations. 
The ratios of {\tt Dodgy code} and {\tt Bad practice} occurrences in fixed violations have great decreases compared with their ratios in all detected ones, although they occupy the main proportion in fixed violations. 

To sum up, when ranking categories with their FR values, total different priorities of violation categories can be carried out.

%\begin{tcolorbox}
%The distributions of fixed violation categories are different from the distribution of detected violation categories, which can provide different priorities of violation categories. 
%\end{tcolorbox}
%\sout{these findings suggest that developers prefer to take more efforts to fix {\tt Dodgy code}, {\tt Bad practice}, {\tt Performance} and {\tt Internationalization} violations than other category ones.  }

% \subsubsection*{Violation Prioritization\kui{R1.1}}
% \textcolor{red}{}

\section{Violation code patterns}
\label{appendix:VCP}

\subsection*{Example of mined violation code patterns which are consistent with \findbugs documentation.}
We note that identified common code patterns of many violation types are consistent with the bug descriptions by \findbugs.
We consider in the following 10 example cases of violation types to investigate the possibility for mining patterns.  

\texttt{\footnotesize DM\_CONVERT\_CASE} is converting a string variable or literal to an upper or lower case with the platform's default encoding~\cite{findbugs:description}. It may result in improper conversions when used with international characters. The two patterns are method invocations of \texttt{\footnotesize String.toUpperCase()} and \texttt{\footnotesize String.toLowerCase()}.

\texttt{\footnotesize RCN\_REDUNDANT\_NULLCHECK\_OF\_NONNULL\_VALUE} represents that the current statement contains a redundant check of a known non-null value against the constant null~\cite{findbugs:description}. Four kinds of common patterns are found in this study, which are shown in Table~\ref{tab:unfixedPatterns}.

\texttt{\footnotesize BC\_UNCONFIRMED\_CAST} denotes that the current cast is unchecked with an \texttt{instanceof} test, and not all instances can be cast from their type to the target type that is being cast to~\cite{findbugs:description}. In the three patterns, {\tt T1} is the target type, and {\tt v2} or {\tt exp1} are the value or expression being cast. \texttt{\footnotesize BC\_UNCONFIRMED\_CAST\_OF\_RETURN\_VALUE} has similar patterns, which denotes an unchecked cast of the return value of a method invocation.

\texttt{\footnotesize RV\_RETURN\_VALUE\_IGNORED\_BAD\_PRACTICE} means that the current statement does not check the return value of a method invocation which could indicate an unusual or unexpected function execution~\cite{findbugs:description}. Its patterns consist of a file's creation, a file's deletion and a method invocation with a return value.

\texttt{\footnotesize DM\_NUMBER\_CTOR} is using a number constructor to create a number object, which is inefficient~\cite{findbugs:description}.
For example, using \texttt{\footnotesize new Integer(...)} is guaranteed to always result in a new \texttt{\footnotesize Integer} object whereas \texttt{\footnotesize Integer.valueOf(...)} allows caching of values to be done by the compiler, class library, or JVM. Using cached values can avoid object allocation and the code will be faster. Our mined patterns are the five types of number creations with number constructors. \texttt{\footnotesize DM\_FP\_NUMBER\_CTRO} has the similar patterns with it.

% \texttt{\footnotesize SBSC\_USE\_STRINGBUFFER\_CONCATENATION} is concatenating strings using `+' operator in a loop~\cite{findbugs:description}.
% In each iteration of a loop, the String is converted to a \texttt{\footnotesize StringBuffer} or \texttt{\footnotesize StringBuilder}, appended to, and converted back to a String. This can lead to a cost quadratic in the number of iterations, as the growing string is recopied in each iteration.

\texttt{\footnotesize DM\_BOXED\_PRIMITIVE\_FOR\_PARSING} denotes that a boxed primitive value is created from a \texttt{\footnotesize String} value without using an effective static \texttt{\footnotesize parseXXX} method~\cite{findbugs:description}. The two common patterns are \texttt{\footnotesize Integer.valueOf(str)} and \texttt{\footnotesize Long.valueOf(str)}.

\texttt{\footnotesize PZLA\_PREFER\_ZERO\_LENGTH\_ARRAYS} means that an array-returned method returns a {\tt null} reference which is not an explicit presentation of an empty list of results~\cite{findbugs:description}. It leads to the clients needing a null check for this return value.

\texttt{\footnotesize ES\_COMPARING\_STRINGS\_WITH\_EQ} denotes the comparison of two strings using \texttt{\footnotesize ==} or \texttt{\footnotesize !=} operator~\cite{findbugs:description}. Unless both strings either were constants in a source file or had been interned using the {\tt \footnotesize String.intern()} method, the same string value might be represented by two different {\tt \footnotesize String} objects.

%\begin{tcolorbox}
%Our proposed violation pattern mining approach can cluster violation instances effectively.
%\end{tcolorbox}

\section{Reasons for failure to resolve unfixed violations}
\label{appendix:RFU}
We have identified 23 violation types where we could not successfully resolve the associated unfixed violations. According to our observation, it might be caused by the following reasons:

{\bf Reason 1.} It is difficult to match effective fix patterns for specific violations. For example, \texttt{\footnotesize DE\_MIGHT\_IGNORE} violations are fixed by replacing the \texttt{\footnotesize Exception} with a specific exception class. Therefore, it is challenging to match an appropriate specific exception class for this kind of violations in terms of syntax without any semantic information or test cases.
%REC_CATCH_EXCEPTION, NP_LOAD_OF_KNOWN_NULL_VALUE, NP_PARAMETER_MUST_BE_NONNULL_BUT_MASRKED_AS_NULLLABLE, EI_EXPOSE_REP, IS2_INCONSISTENT_SYNC

{\bf Reason 2.} It is challenging to identify common fix patterns from the source code changes of some violations without an exact position.
For example, \texttt{\footnotesize UWF\_FIELD\_NOT} \texttt{\footnotesize \_INITIALIZED\_IN\_CONSTRUCTOR} means that non-null fields are not initialized in any constructors~\cite{findbugs:description}. Our observation shows that the positions of this kind of violations are located in one constructor. So that, it is impossible to obtain any information about these violations. Even if some information of these violations could be identified, which are the specific information, it is still a challenge to match any effective fix patterns for them.

{\bf Reason 3.} It is unable to fix \texttt{\footnotesize NM\_METHOD\_NAMING\_CONVEN}-\texttt{\footnotesize TION} violations which do not comply the method naming convention. Even if violated method names can be fixed by matched fix patterns, the changed name may cause compilation errors or API changes that may break client programs.
% because it is impossible to locate and change all invocations of this method with our technique.

{\bf Reason 4.} It is challenging to fix all related violations just by deleting the violated source code. For example, the common fix pattern of \texttt{\footnotesize EI\_EXPOSE\_REP} is deleting the violated source code. When the fix pattern is used to fix related violations, the changed source code may not be correctly compiled.

{\bf Reason 5.} There might be a lack of effective fix patterns for some violation types. The fix patterns of some violation types are deleting the violated source code. We do not adopt this kind of fix patterns, even though the violation can be fixed or removed by deleting the violated source code, which removes the feature of original source code and many of them failed to pass compile or checkstyle.
\end{appendices}

\end{document}